\PassOptionsToPackage{hyphens}{url}
\documentclass[lettersize,journal]{IEEEtran}
\usepackage{amssymb}
\usepackage{times}
\usepackage{amsfonts}

\usepackage{subfig}
\usepackage{wrapfig}
\usepackage{float}
\usepackage{multirow, graphicx}

\usepackage{xspace}

\usepackage{url}

\usepackage{sty/widetext} 

\usepackage{stackrel}

\let\emptyset\varnothing

\ifdefined\TTSTYLETARGETreport
\usepackage[
    backend=bibtex,
    maxbibnames=100,
    firstinits=true,
    bibstyle=numeric,
    citestyle=numeric
]{biblatex}
\fi

\usepackage{pifont}
\newcommand{\cmark}{\ding{51}}%
\newcommand{\xmark}{\ding{55}}%

\usepackage{todonotes}
\usepackage{marginnote}
\newcommand{\ignore}[1]{}

\usepackage{enumitem}
\setlist[itemize]{leftmargin=*}

\usepackage{graphicx, multirow}

\usepackage{tabularx, multirow, rotating} 
\usepackage{subfig} 
\usepackage{algorithm} 
\usepackage{algpseudocode} 
\usepackage{ulem} 
\usepackage{color} 

\usepackage{listings} 
\lstdefinestyle{Oracle}{basicstyle=\ttfamily,
                        keywordstyle=\lstuppercase,
                        emphstyle=\itshape,
                        showstringspaces=true,
                        }
\makeatletter
\newcommand{\lstuppercase}{\uppercase\expandafter{\expandafter\lst@token
                           \expandafter{\the\lst@token}}}
\newcommand{\lstlowercase}{\lowercase\expandafter{\expandafter\lst@token
                           \expandafter{\the\lst@token}}}
\makeatother

\usepackage{tikz}

\newcommand{\ballnumber}[1]{\tikz[baseline=(myanchor.base)] \node[circle,fill=.,inner sep=1pt] (myanchor) {\color{-.}\bfseries\footnotesize #1};}

\newif\ifboldnumber

\algrenewcommand\alglinenumber[1]{%
  \footnotesize\ifboldnumber\bfseries\fi\global\boldnumberfalse#1:}

\usepackage{color, colortbl}
\definecolor{Gray}{gray}{0.9}
\definecolor{LightCyan}{rgb}{0.88,1,1}
\usepackage[first=0,last=9]{lcg}

\usepackage{colortbl}
\usetikzlibrary{calc}
\usepackage{zref-savepos}

\newcounter{NoTableEntry}
\renewcommand*{\theNoTableEntry}{NTE-\the\value{NoTableEntry}}

\renewcommand{\thesection}{\arabic{section}}
\usepackage{cite}
\usepackage{amssymb,amsfonts}
\usepackage{graphicx}
\usepackage{textcomp}
\usepackage{xcolor}

\newcommand*{\renameenviron}[1]{%
  \expandafter\let\csname exam-#1\expandafter\endcsname
      \csname #1\endcsname
  \expandafter\let\csname endexam-#1\expandafter\endcsname
      \csname end#1\endcsname
  \expandafter\let\csname #1\endcsname\relax
  \expandafter\let\csname end#1\endcsname\relax
}
\renameenviron{rcases}
\renameenviron{rbrace}
\usepackage{mathtools}
\usepackage{tikz}

\usepackage{makecell}
\usepackage{soul}

\newtheorem{theorem}{Theorem}[section]

\newtheorem{definition}[theorem]{Definition}
\usepackage{caption}
\usepackage{hyperref}
\usepackage{url}

\newenvironment{reduce}
 {\hbox\bgroup\scriptsize$\displaystyle}
 {$\egroup}

\renewcommand{\marginpar}{\marginnote}

\newsavebox{\mpfootsave}

\makeatletter

\newif\iflogvar
\logvarfalse

\newcommand{\tangSide}[1]{%
\iflogvar
\setbox\mpfootsave=\vbox{\unvbox\@mpfootins}%
\todo[caption={},color=cyan!20]{\tiny #1}%
\setbox\@mpfootins=\vbox{\unvbox\mpfootsave}%
\else
\fi
}

\begin{document}
\title{Towards Automated Discovery of Asymmetric Mempool DoS in Blockchains}

\author{\IEEEauthorblockN{Yibo~Wang\IEEEauthorrefmark{1},
        Yuzhe~Tang\IEEEauthorrefmark{1}~\IEEEmembership{IEEE Member},
        Kai~Li\IEEEauthorrefmark{2},
        Wanning Ding\IEEEauthorrefmark{1},
        Zhihua Yang\IEEEauthorrefmark{1}
}
\IEEEauthorblockA{\IEEEauthorrefmark{1}Syracuse University,
    \{ywang349,ytang100,wding04,zyang47\}@syr.edu}
\IEEEauthorblockA{\IEEEauthorrefmark{2}San Diego State University,
    kli5@sdsu.edu}
\thanks{Yuzhe Tang is the corresponding author.}
}
\markboth{Journal of \LaTeX\ Class Files,~Vol.~14, No.~8, August~2021}%
{Shell \MakeLowercase{\textit{et al.}}: A Sample Article Using IEEEtran.cls for IEEE Journals}


\maketitle

\begin{abstract}
  {\colorlet{blue}{black} \colorlet{red}{black}
\tangSide{R10}
In blockchains, mempool controls transaction flow before consensus, denial of whose service hurts the health and security of blockchain networks. This paper presents \textsc{mpfuzz}, the first mempool fuzzer to find asymmetric DoS bugs by 
{\color{blue}
exploring the space of symbolized mempool states} 
and optimistically estimating the promisingness of an intermediate state in reaching bug oracles. Compared to the baseline blockchain fuzzers, \textsc{mpfuzz} achieves a $>100\times$ speedup in finding known DETER exploits. Running \textsc{mpfuzz} on major Ethereum clients leads to discovering new mempool vulnerabilities, which exhibit a wide variety of sophisticated patterns, including stealthy mempool eviction and mempool locking. Rule-based mitigation schemes are proposed against all newly discovered vulnerabilities.

\ignore{
This paper presents \textsc{mpfuzz}, the first blockchain fuzzer that automatically and systematically finds known and unknown mempool DoS bugs in Ethereum. Fuzzing mempools, unlike fuzzing other blockchain components tackled in the existing literature, poses new design problems. To solve them, \textsc{mpfuzz} explores the previously unexplored transaction space, namely invalid transactions with varying fees, and defines a bug oracle specifically for asymmetric mempool DoS. \textsc{mpfuzz} mutates transactions and explores mempool states, {\it symbolically}, that is, based on the transaction symbols tailored to Ethereum protocols. To search deep-state bugs, \textsc{mpfuzz} optimistically estimates mempool costs. 

Running \textsc{mpfuzz} leads to the discovery of new mempool DoS bugs in leading Ethereum clients of the latest versions, all of which have been patched against the known bugs. In particular, all Ethereum clients are found vulnerable to at least one of two DoS patterns:  mempool eviction or locking.


This paper also presents a defensive mempool framework, \textsc{saferAd}, that provides provable security against the newly discovered DoS patterns. \textsc{saferAd} upper-bounds the victim transaction fees under mempool locking and lower-bounds the attacker transaction fees under mempool evictions.
}
}



\end{abstract}

\providecommand{\ssssp}{{\sc SS\_SSP}\xspace}
\newcommand{\tremark}[1]{\footnote{\textcolor{red}{(Ting's comment: #1)}}}
\newcommand{\xremark}[1]{\footnote{\textcolor{red}{(Xin's comment: #1)}}}
\newcommand{\jj}[1]{\footnote{\textcolor{blue}{(Jiyong: #1)}}}
\newcommand{\yz}[1]{\footnote{\textcolor{red}{(Yuzhe: #1)}}}
\newcommand{\yw}[1]{\footnote{\textcolor{green}{(Yibo's comment: #1)}}}

\definecolor{mygreen}{rgb}{0,0.6,0}
\lstset{ %
  backgroundcolor=\color{white},   
  basicstyle=\scriptsize\ttfamily,        
  breakatwhitespace=false,         
  breaklines=true,                 
  captionpos=b,                    
  commentstyle=\color{mygreen},    
  deletekeywords={...},            
  escapeinside={\%*}{*)},          
  extendedchars=true,              
  keepspaces=true,                 
  keywordstyle=\color{blue},       
  language=Java,                 
  morekeywords={*,...},            
  numbers=left,                    
  numbersep=5pt,                   
  numberstyle=\scriptsize\color{black}, 
  rulecolor=\color{black},         
  showspaces=false,                
  showstringspaces=false,          
  showtabs=false,                  
  stepnumber=1,                    
  tabsize=2,                       
  title=\lstname,                  
  moredelim=[is][\bf]{*}{*},
}

{\colorlet{blue}{black} \colorlet{red}{black} 
\colorlet{violet}{black} 
  \section{Introduction}

In Ethereum, a mempool buffers unconfirmed transactions from web3 users before they are included in the next blocks. Mempool provides the essential functionality to bridge the gap between varying rates of submitted transactions and rates of produced blocks, regardless of public or private transactions it serves. As shown in recent studies~\cite{DBLP:conf/ccs/LiWT21}, denying a mempool service can force the blockchain to produce blocks of low or even zero (Gas) utilization, undermining validators' incentives and shrinking the blockchain networks in the long run, re-introducing the $51\%$ attacks. Besides, a denied mempool service can prevent normal transactions from block inclusion, cutting millions of web3 users off the blockchain and failing the decentralized applications relying on real-time blockchain access.


\noindent{\bf Problem}: Spamming the mempool to deny its service has been studied for long~\cite{DBLP:conf/fc/BaqerHMW16,me:unconfirmed,DBLP:conf/icbc2/SaadNKKNM19,me:backlogged}. Early designs by sending spam transactions at high prices burden attackers with high costs and are of limited practicality. What poses a real threat is Asymmetric DeniAl of Mempool Service, coined by ADAMS, in which the mempool service is denied at an asymmetrically low cost. That is, the attack costs, in terms of the fees of adversarial transactions, are significantly lower than those of normal transactions victimized by the denied mempool. In the existing literature, DETER~\cite{DBLP:conf/ccs/LiWT21} is the first  ADAMS attack, and it works by sending invalid transactions to {\it directly evict} normal transactions in the mempool. MemPurge~\cite{cryptoeprint:2023/956} is a similar mempool attack that finds a way to send overdraft transactions into Geth's pending transaction pool and causes eviction there. 
These known attacks are easy to detect (i.e., following the same direct-eviction pattern). In fact, the DETER bugs reported in 2021 have been successfully fixed in all major Ethereum clients as of Fall 2023, including Geth, Nethermind, Erigon, and Besu. Given this state of affairs, we pose the following research question: 
Are there new ADAMS vulnerabilities in the latest Ethereum clients already patched against direct-eviction based attacks? 

This work takes a systematic and semi-automated approach to discovering ADAMS vulnerabilities, unlike the existing DETER bugs that are manually found. 
Fuzzing mempool implementations is a promising approach but also poses unique challenges: 
Unlike the consensus implementation that reads only valid confirmed transactions, the mempool, which resides in the pre-consensus phase, needs to handle various unconfirmed transactions, imposing a much larger input space for the fuzzer. For instance, a mempool can receive \textit{invalid transactions under legitimate causes},\footnote{For instance, future transactions can be caused by out-of-order information propagation in Ethereum.} and factors such as fees or prices are key in determining transaction admission outcomes. Existing blockchain fuzzers including Fluffy~\cite{DBLP:conf/osdi/YangKC21}, Loki~\cite{DBLP:conf/ndss/MaCRZ00L023} and Tyr~\cite{me:tyr} all focus on fuzzing consensus implementation and don't explore the extra transaction space required by mempool fuzzing. As a result, directly re-purposing a consensus fuzzer to fuzz mempool would be unable to detect the DETER bugs, let alone discover more sophisticated new ADAMS bugs.


\noindent{\bf 
Proposed methods}:
 To efficiently fuzz mempools, our key observation is that real-world mempool implementation admits transactions based on abstract ``symbols'', such as pending and future transactions, instead of concrete value. Thus, {\it sending multiple transactions under one symbol would trigger the same mempool behavior repeatedly}, and it suffices to explore just one transaction per symbol during fuzzing without losing the diversity of mempool behavior.

{\tangSide{R9} \color{blue} We propose symbolized-stateful mempool fuzzing or \textsc{mpfuzz}. 
To begin with, \textsc{mpfuzz} is set up and run in a three-step workflow: It is first manually set up against a size-reduced mempool under test, which induces a much smaller search space for fuzzing. Second, \textsc{mpfuzz} is iteratively run against the reduced mempool to discover short exploits. 
{\color{violet}Third, \textsc{mpfuzz} automatically extends the short exploits to actual ones that are functional on real mempools and Ethereum clients.}
Internally, the design of \textsc{mpfuzz} is based on seven transaction symbols we design out of Ethereum semantics: future, parent, overdraft, latent overdraft, and replacement transactions. Under these symbols, in each iteration of fuzzing, \textsc{mpfuzz} explores one concrete transaction per symbol, sends the generated transaction sequence to the target mempool, observes the mempool end state, and extracts the feedback of symbolized state coverage and state promisingness in reaching bug oracles to guide the next round of fuzzing. In particular, \textsc{mpfuzz} employs a novel technique to evaluate state promisingness, that is, by {\it optimistically} estimating the costs of unconfirmed transactions whose validity is subject to change in the future. {\color{violet}After finding a short exploit against the reduced mempool, \textsc{mpfuzz} automatically extends the small exploit to a more extensive exploit that is effective within a larger mempool. Specifically, \textsc{mpfuzz} extracts all the unique admission event patterns in the small exploit and aligns these patterns to construct the attack transaction sequence on the larger-size mempool. }}



\noindent{\bf 
Found attacks}:
{\tangSide{R5} \color{blue} 
We run \textsc{mpfuzz} on six leading execution-layer Ethereum clients deployed on the mainnet's public-transaction path (Geth~\cite{me:geth}, Nethermind~\cite{me:nethermind}, Erigon~\cite{me:erigon}, Besu~\cite{me:besu}, Reth~\cite{me:reth}, and OpenEthereum~\cite{me:parity}), three PBS builders (proposer-builder separation) on the mainnet's private-transaction and bundle path (Flashbot $v1.11.5$~\cite{me:flashbotbuilder}, EigenPhi~\cite{me:eigenphibuilder} and bloXroute~\cite{me:bloXroutebuilder}), and three Ethereum-like clients (BSC $v1.3.8$~\cite{me:bscclient} deployed on Binance Smart Chain, go-opera $v1.1.3$~\cite{me:goopera} on Fantom, and core-geth $v1.12.19$~\cite{me:coregeth} on Ethereum Classic). 
}
On Ethereum clients of historical versions, \textsc{mpfuzz} can rediscover known DETER bugs. On the latest Ethereum clients where DETER bugs are fixed, \textsc{mpfuzz} can find new ADAMS attacks, described next.

We summarize the newly discovered ADAMS exploits into several patterns: 1) Indirect eviction by valid-turned-invalid transactions. Unlike the direct-eviction pattern used in DETER, the indirect-eviction attack works more stealthily in two steps: The attacker first sends normal-looking transactions to evict victim transactions from the mempool and then sends another set of transactions to {\it turn} the admitted normal-looking transactions into invalid ones, bringing down the cost. 2) Locking mempool, which is to occupy the mempool to decline subsequent victim transactions. Mempool locking does not need to evict existing transactions from the mempool as the eviction-based DETER attacks do.
3) Adaptive attack strategies where the attack composes adversarial transactions in an adaptive way to the specific policies and implementations of the mempool under test. Example strategies include composing transactions of multiple patterns in one attack, re-sending evicted transactions to evict ``reversible'' mempools, locking mempool patched against turning, etc. 

All newly found ADAMS bugs are reported to the developer communities, including Ethereum Foundation, BSC, Fantom, PBS builders, etc. with $15$ bugs confirmed and $4$ fixed in recent client releases~\cite{me:gethfix11}. 
Bug reporting is documented on the webpage~\cite{me:mpfuzz:report}, which also includes demos of \textsc{mpfuzz} on the tested Ethereum clients. \textsc{mpfuzz} will be open-sourced to facilitate vulnerability discovery in more and future clients.

{\color{violet}
\noindent{\bf Systematic evaluation}: We conducted an ablation study to evaluate the performance of \textsc{mpfuzz}, which incorporates three distinct techniques. We aim to identify the contribution and significance of each component to the overall \textsc{mpfuzz}. We implement four baseline fuzzers to compare against \textsc{mpfuzz}. The first baseline is a stateless fuzzer built within the GoFuzz framework~\cite{me:gofuzz}. For the remaining three baselines, each one selectively disables a single technique from \textsc{mpfuzz}: symbolized state, symbolized input, and promising feedback and energy. The experimental result shows that the stateless fuzzer is the least effective in discovering ADAMS exploits. Among the three distinct techniques, the symbolized input technique improves efficiency more compared to that of the promising feedback and symbolized state coverage, and the state promisingness in feedback is more beneficial for performance than symbolized state coverage. 

We systematically evaluate the attack success rate and cost of all the exploits found by \textsc{mpfuzz} on the Ethereum testnet and a local network. Specifically, we evaluate a newly discovered eviction attack on Goerli testnet. The result of the eviction attack shows the Gas used by normal transactions significantly drops to $3.19\%$ after the launch of the attack, which indicates that only a few normal transactions are included in the blocks following the attack. The locking attack evaluated on Goerli testnet shows the normal transactions received from the victim node drop essentially to zero, which indicates the locking attack is highly successful. Additionally, we evaluate other exploits on a local network. The evaluation results of all attacks on the Ethereum clients, PBS and Ethereum-like clients show that the attack success rates are all higher than $84.63\%$, and the attack costs are all lower than $1.172$ Ether per block.

}
\noindent{\bf Contributions}: The paper makes contributions as follows:

\vspace{2pt}\noindent$\bullet$\textit{ 
New fuzzing problem}: This paper is the first to formulate the mempool-fuzzing problem to automatically find asymmetric DoS vulnerabilities as bug oracles. Fuzzing mempools poses new unique challenges that existing blockchain fuzzers don't address and entails a larger search space including invalid transactions and varying prices. 

\tangSide{R10}

\vspace{2pt}\noindent$\bullet$\textit{ 
New fuzzing method}: The paper presents the design and implementation of \textsc{mpfuzz}, 
{\color{blue} 
a symbolized-stateful mempool fuzzer. Given a mempool implementation, \textsc{mpfuzz} defines the search space by the mempool states covered under symbolized transaction sequences and efficiently searches this space using the feedback of symbolized state coverage and the promising-ness of an intermediate state in triggering bug oracles. } 
With a Python prototype and runs on real Ethereum clients, \textsc{mpfuzz} achieves a $>100\times$ speedup in finding known DETER exploits compared to baselines.

\vspace{2pt}\noindent$\bullet$\textit{ 
New discovery of mempool vulnerabilities}: \textsc{mpfuzz} discovers new asymmetric-DoS vulnerabilities in six major Ethereum clients in the mainnet. By evaluation under real transaction workloads, all found attacks achieve $84.6-99.6\%$ success rates and low costs such as adversarial transaction fees $100\times$ lower than the victim transaction fees.

{\color{violet}
\vspace{2pt}\noindent$\bullet$\textit{ 
New discovery of mempool vulnerabilities}: \textsc{mpfuzz} discovers new asymmetric-DoS vulnerabilities in six major Ethereum clients, PBS and Ethereum-like clients. We found and reported $24$ newly discovered ADAMS bugs. After our reporting, $15$ bugs are confirmed and $3$ bugs are fixed.

\vspace{2pt}\noindent$\bullet$\textit{ 
New found attack evaluation}: This paper presents the evaluation results of the attack success rate and cost
of all the exploits found by \textsc{mpfuzz} on the Ethereum testnet and a local network. By evaluation under real transaction workloads, all found attacks achieve $84.6-99.6\%$ success rates and low costs such as adversarial transaction fees $100\times$ lower than the victim transaction fees.

}

  \section{Related Works}
\label{sec:rw}

\noindent{\bf 
Consensus fuzzers}:
Fluffy~\cite{DBLP:conf/osdi/YangKC21} is a code-coverage based differential fuzzer to find consensus bugs in Ethereum Virtual Machines (EVM). 
Specifically, given an EVM input, that is, a transaction sequence, Fluffy sends it to multiple nodes running different Ethereum clients (e.g., Geth and OpenEthereum) for execution. 
1) The test oracle in Fluffy is whether the EVM end states across these nodes are different, implying consensus failure.
2) Fluffy mutates ordered transaction sequences at two levels: it reorders/adds/deletes transactions in the sequence, and for each new transaction to generate, it randomly selects values in Gas limits and Ether amounts. For the data field, it employs semantic-aware strategies to add/delete/mutate bytecode instructions of pre-fixed smart contract templates. 
3) Fluffy uses code coverage as feedback to guide the mutation.

Loki~\cite{DBLP:conf/ndss/MaCRZ00L023} is a stateful fuzzer to find consensus bugs causing crashes in blockchain network stacks.
Specifically, Loki runs as a fuzzer node interacting with a tested blockchain node through sending and receiving network messages.
1) The test oracle in Loki is whether the tested blockchain node crashes. 
2) The fuzzer node uses a learned model of blockchain network protocol to generate the next message of complying format, in which the content is mutated by randomly choosing integers and bit-flipping strings. In other words, the mutation in Loki is unaware of application level semantics. For instance, if the message is about propagating an Ethereum transaction, the transaction's nonce, Ether amount, and Gas price are all randomly chosen; the sender, receiver, and data are bit flipped.
3) Loki records messages sent and received from the test node. It uses them as the state. New states receive positive feedback.

Tyr~\cite{me:tyr} is a property-based stateful fuzzer that finds consensus bugs causing the violation of safety, liveness, integrity, and other properties in blockchain network stacks.
Specifically, in Tyr, a fuzzer node is connected to and executes a consensus protocol with multiple neighbor blockchain nodes. 
1) The test oracle Tyr uses is the violation of consensus properties, including safety (e.g., invalid transactions cannot be confirmed), liveness (e.g., all valid transactions will be confirmed), integrity, etc. The checking is realized by matching an observed end state with a ``correct'' state defined and run by the Tyr fuzzer.
2) Tyr mutates transactions on generic attributes like randomly selecting senders and varying the amount of cryptocurrencies transferred (specifically, with two values, the full sender balance, and balance plus one). When applied to Ethereum, Tyr does not mutate Ethereum-specific attributes, including nonce or Gas price. It generates one type of invalid transaction, that is, the double-spending one, which in account-based blockchains like Ethereum are replacement transactions of the same nonce. 
3) Tyr uses state divergence as the feedback, that is, the difference of end states on neighbor nodes receiving consensus messages in the same iteration. In Tyr, the states on a node include both the messages the node sent and received, as well as blocks, confirmed transactions, and other state information.

These existing blockchain fuzzers~\cite{me:tyr,DBLP:conf/ndss/MaCRZ00L023,DBLP:conf/osdi/YangKC21} cannot detect mempool DoS bugs and are different from this work. Briefly, mempool DoS does not trigger system crashes and can not be detected by Loki. Mempool content difference across nodes does not mean insecurity, and the mempool DoS bugs can not be detected by Fluffy (which detects consensus state difference). 
While Tyr aims to detect liveness and safety violations, their model of invalid transactions is rudimentary. Specifically, they only model double-spending transactions and cannot detect DETER bugs~\cite{DBLP:conf/ccs/LiWT21} that rely on more advanced invalid transactions, like future transactions and latent overdrafts.

\noindent{\bf
Blockchain DoS}: Blockchain DoS security has been examined at different system layers, including P2P networks~\cite{DBLP:conf/uss/HeilmanKZG15,DBLP:journals/iacr/MarcusHG18,DBLP:conf/sp/ApostolakiZV17,tran2020stealthier}, mining-based consensus~\cite{mirkin2019bdos,me:finney}, and application-level smart contracts~\cite{DBLP:conf/ndss/0002L20,me:eip150,me:dosblockgas} and DApp (decentralized application) services~\cite{DBLP:conf/ndss/LiCLT0L21}. These DoSes are not related to mempools and are orthogonal to this work.

For mempool DoS, a basic attack is by sending spam transactions with high fees to evict victim transactions of normal fees~\cite{DBLP:conf/fc/BaqerHMW16,me:unconfirmed,DBLP:conf/icbc2/SaadNKKNM19,me:backlogged}, which incur high costs to attackers. DETER~\cite{DBLP:conf/ccs/LiWT21} is the first asymmetric DoS on Ethereum mempool where the adversary sends invalid transactions of high fees (e.g., future transactions or latent overdrafts) to evict victim transactions of normal fees. Concurrent to this work is MemPurge~\cite{cryptoeprint:2023/956} posted online in June 2023; in it, the attacker reconnects her adversarial future transactions and makes them latent-overdraft transactions. 
The DETER bugs have been fixed on the latest Ethereum clients such as Geth $v1.11.4$, as we tested them in July 2023. Both DETER and MemPurge are manually discovered. 

{\color{red}
  \section{Background}
\tangSide{R12}

\noindent{\bf Ethereum mempools}:
In Ethereum, users send their transactions to the blockchain network, which are propagated to reach validator nodes. On every blockchain node, unconfirmed transactions are buffered in the mempool before they are included in the next block or evicted by another transaction.
In Ethereum 2.0, transactions are propagated in two fashions: public transactions are broadcast among all nodes, and private transactions are forwarded to selected nodes, more specifically, selected builders and proposers (as in PBS or proposer-builder separation).
The mempool is present on both paths, and it is inside Ethereum clients such as Geth~\cite{me:geth} and Nethermind~\cite{me:nethermind} (handling public transactions), and Flashbot~\cite{me:flashbotbuilder} (private transactions).
On these clients, a mempool also serves many other downstream modules, including MEV searchers or bots (in PBS), Gas stations, RPC queries, etc.

Mempools on different nodes are operated independently and don't need to synchronize as the consensus layer does. For instance, future transactions are not propagated, and the set of future transactions in the mempool on one node is different from another node.

The core mempool design is transaction admission: Given an initial state, whether and how the mempool admits an arriving transaction. In practice, example policies include those favoring admitting transactions of higher prices (and thus evicting or declining the ones of lower prices), transactions arriving earlier, certain transaction types (e.g., parent over child transactions), and valid transactions (over future transactions). See different types of Ethereum transactions in \S~\ref{sec:notations}.
}

\noindent{\bf 
Transactions and fees}: In Ethereum, Gas measures the amount of computations caused by smart contract execution. When preparing for transaction $tx$, the sender needs to specify \textit{Gas} and \textit{GasPrice}; the former is the maximal amount of computations allowed for running smart contracts under $tx$, and the latter is the amount of Ether per Gas the sender is willing to pay. 
After $tx$ is included in a block, the actual amount of computations consumed by contract execution is denoted by \textit{GasUsed}. The fees of transaction $tx$ are the product of \textit{GasUsed} and \textit{GasPrice}, that is, $tx.fee=\textit{GasUsed}*\textit{GasPrice}$.

After EIP-1559~\cite{me:eip1559}, \textit{GasPrice} is divided into two components: $BasePrice$ and $PriorityPrice$, that is, $\textit{GasPrice}=BasePrice+PriorityPrice$. Blockchain nodes follow the Ethereum protocol to derive $BasePrice$ from Gas utilization in recent blocks. The fees associated with the $BasePrice$ are burnt upon transaction inclusion. $PriorityPrice$ is set by the transaction sender.

{\color{blue}
\ignore{
\noindent{\bf 
Transactions and fees}: Given an Ethereum transaction $tx$, its fee is the product of \textit{GasUsed} and \textit{GasPrice}, that is, $tx.fee=\textit{GasUsed}*\textit{GasPrice}$. 
\textit{tx.GasUsed} is the sum of a fixed $21000$ Gas (for computation done for basic transaction validation) and a variable amount proportional to the computations caused by smart-contract execution in $tx$. \textit{GasUsed} is determined only after the execution of $tx$. 
Typically, a transaction is executed when selected to be included in the next block, and the fees of the transaction are charged when the next block is finalized with sufficient confirmations.

In Ethereum, the sender of $tx$ may also set $tx.Gas$, which is the upper bound of \textit{GasUsed}.
After EIP-1559, $\textit{GasPrice}=BasePrice+PriorityPrice$, where the $BasePrice$ is decided by the blockchain and its associated fees are burnt upon transaction inclusion. The validator collects the rest of the fees, namely, $\textit{GasUsed}*(PriorityPrice)$.
}
}

  \section{Threat Model and Bug Oracle}
\label{sec:threatmodel}

{\color{red}
\tangSide{R12} 
In the threat model, an attacker controls one or few nodes to join an Ethereum network, discovers and neighbors critical nodes (e.g., top validators, backends of an RPC Service, MEV searcher) if necessary, and sends crafted adversarial transactions to their neighbors. The attacker in this work has the same networking capacity as in DETER~\cite{DBLP:conf/ccs/LiWT21}.

The attacker's goal is to disable the mempools on the critical nodes or all nodes in an Etheurem network and cause damage such as deployed or dropped transactions in block inclusion, produced blocks of low or zero (Gas) utilization, and disabled other downstream services like MEV searchers. In practice, this damage is beneficial to competing block validators, MEV searchers, and RPC services. Besides, the attacker aims to cause this damage at asymmetrically low costs, as will be described. 
}

\ignore{
In the threat model, an attacker controls one or a few nodes joining a target Ethereum network. The attacker's goal is to disable some critical functions in the entire network, including the block production (or mining) and relay of submitted transactions through RPC services. Specifically, in either case of disabling network-wide block production or RPC Service, a successful ADAMS attack can result in the blockchain appended with empty blocks. 

The attacker has the capability of joining nodes in the target Ethereum network,  propagating transactions in the network, and passively observing the traffic of messages that are received by or sent by the attack nodes. Using these capabilities, the attacker may be able to discover critical service nodes running behind the popular RPC infrastructures and top mining pools, which is feasible in the pre-merge Ethereum network~\cite{DBLP:conf/ccs/LiWT21}. 

We describe the attack design goals: 1) Against a given victim node, the attacker aims to find crafted transactions to be sent to the victim and to deny its mempool service to the local miner (or block producer). So, the local miner reads zero unconfirmed transactions and proposes empty blocks. 2) Optionally, the attacker aims at finding the craft transactions that will be propagated. That is, the victim node receiving the crafted transactions would propagate them to its neighbor nodes, and so do the neighbors until the crafted transactions are propagated to the entire network.

While causing the DoS damage to the Ethereum network, an ADAMS attacker aims at achieving asymmetrically low cost, that is, the fees of crafted transactions sent in an attack must be significantly lower than those of victim transactions, either evicted or declined upon admission. 
}

  \subsection{Notations}
\label{sec:notations}

\noindent{\bf 
Transactions}:
In Ethereum, we consider two types of accounts: a benign account of index $i$ denoted by $B_i$, and an adversarial account of index $i$ denoted by $A_i$. A (concrete) Ethereum transaction $tx$ is denoted by $tx{\begin{reduce}\begin{bsmallmatrix}{s} & {v}\\ {n} & {f}\end{bsmallmatrix}\end{reduce}}$, where the sender is $s$, the nonce is $n$, the Gas price is $f$, and the transferred value in Ether is $v$.
This work does not consider the ``data'' field in Ethereum transactions or smart contracts because smart contracts do not affect the validity of the Ethereum transaction. The notations are summarized in Table~\ref{tab:notations}.

\begin{table}[!htbp] 
\caption{Notations}
\label{tab:notations}\centering{\small
\begin{tabularx}{0.475\textwidth}{ |l|X|l|X| }
  \hline
Symbol & Meaning & Symbol & Meaning  \\ \hline
$m$ & Mempool length 
&
$ops$ & Tx sequence
 \\ \hline
$st$ & State of residential txs in mempool
&
$dc$ & History of declined txs from mempool
 \\ \hline
$A$ & Adversarial tx sender
&
$B$ & Benign tx sender
 \\ \hline
&\multicolumn{3}{p{2in}|}{}\\[-0.8em]
  $tx\begin{bsmallmatrix}{s} & {v}\\ {n} & {f}\end{bsmallmatrix}$
  & 
  \multicolumn{3}{p{2.5in}|}{
A transaction from sender $s$, of nonce $n$, transferring Ether value $v$, and with Gas Price $f$
}
 \\ \hline
\end{tabularx}
}
\end{table}

We use each of the following symbols to represent a disjoint group of transactions. Informally, $\mathcal{N}$ is a transaction sent from a benign account, $\mathcal{F}$ a future transaction, $\mathcal{P}$ a parent transaction, $\mathcal{C}$ a child transaction, $\mathcal{O}$ an overdraft transaction, $\mathcal{L}$ a latent overdraft transaction, and $\mathcal{R}$ a replacement transaction. These symbols are formally specified in \S~\ref{sec:symbolize}.

Particularly, we use the notion of latent overdraft from the existing work~\cite{DBLP:conf/ccs/LiWT21}, which indicates a child transaction that by itself does not overdraft but does overdraft when taking into account its parent transactions. Suppose Alice has a balance of $5$ Ether and sends $tx_1$ of nonce $1$ spending $2$ Ether and $tx_2$ of nonce $2$ spending $4$ Ether. $tx_2$ is a latent overdraft.

\noindent{\bf 
States and transitions}: We recognize two mempool-related states in Ethereum, the collection of transactions stored in mempool $st$, and the collection of declined transactions $dc$. 

Suppose a mempool of state $\langle{}st_i, dc_i\rangle{}$ receives an arriving transaction $tx_i$ and transitions to state $\langle{}st_{i+1}, dc_{i+1}\rangle{}$. 

An arriving transaction can be admitted with eviction, admitted without eviction, or declined by a mempool. The three operations are defined as follows. 1) An arriving transaction $tx_i$ is admitted with evicting a transaction $tx_i'$ from the mempool, if 
$st_{i+1}=st_{i}\setminus{}tx_i'\cup{}tx_i,dc_{i+1}=dc_i$.
2) $tx_i$ is admitted without evicting any transaction in the mempool, if 
$st_{i+1}=st_{i}\cup{}tx_i,dc_{i+1}=dc_i$.
3) $tx_i$ is declined and does not enter the mempool, if 
$st_{i+1}=st_{i},dc_{i+1}=dc_i\cup{}tx_i$.

\begin{figure}[!bthp]
\centering
\includegraphics[width=0.41\textwidth]{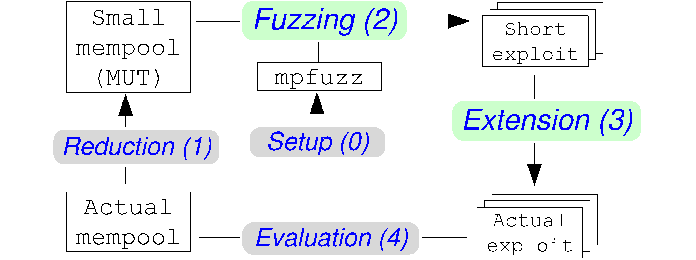}
\caption{Overview of exploit discovery and evaluation workflow: 0) \textsc{mpfuzz} setup, 1) mempool reduction, 2) fuzzing on reduced mempool under test (MUT) to discover short exploits, and 3) exploit extension. The extended exploits are 4) evaluated on the actual mempool of the original size. Green means automated tasks, and gray requires manual effort.}
\label{fig:overview}
\end{figure}

\tangSide{R11} 
{\color{blue} 
  \subsection{Exploit Discovery Workflow}

To set up the stage, we present the workflow overview in this work. The workflow is depicted in Figure~\ref{fig:overview}. To discover vulnerabilities and exploits in an actual mempool, one has to 0) set up \textsc{mpfuzz} by various parameters (as will be introduced), 1) reduce the mempool from its original configuration to a much smaller version, named mempool under test (MUT). 2) \textsc{mpfuzz} is run on the reduced MUT deployed locally (with settings described in \S~\ref{sec:mut:setting}). Fuzzing leads to the discovery of short exploits. 3) Short exploits are extended to the longer ones, that is, actual exploits. 4) The success of the actual exploit is evaluated on the actual mempools deployed in close-to-operational settings, such as using popular Ethereum testnets like Goerli (see \S~\ref{sec:evaluate:testnet}) or a network we set up locally running unmodified Ethereum clients (see \S~\ref{sec:evaluate:locking}, \S~\ref{sec:evaluate:locking2} and \S~\ref{sec:setup:singlenode} ).
Fuzzing a reduced mempool (MUT) instead of an original one is necessary to ensure the efficient execution of the \textsc{mpfuzz}, which entails many iterations of re-initiating and populating the mempool. 

In this workflow, fully automated is Step 2) {\color{violet} and Step 3), that is, the discovery of short exploits on a given MUT and extension of the short exploit to an actual exploit} (colored green in Figure~\ref{fig:overview}). Other steps colored in gray require manual efforts and are described below:
In {\bf Step 0)}, \textsc{mpfuzz} setup entails setting parameters in bug oracles (e.g., $\epsilon$ and $\lambda$) and fixed symbols in guiding fuzzing. The setup is usually one-time.
In {\bf Step 1)}, mempool reduction entails reconfiguring the mempool to a much smaller capacity (e.g., MUT length $m=6$ or $m=16$, compared to the original size like $m'=6144$ as in Geth). In addition, policy-specific parameters in the actual mempool are tuned down proportionally. For instance, the Geth mempool buffers up to $py_1'=1024$ future transactions, and when buffering more than $py'_3=5120$ pending transactions, the mempool triggers a protective policy to limit the number of transactions sent from the same sender by $py_2'=16$ . In a reduced MUT of $m=6$, these parameters are reduced while retaining the same proportion, such as $py_1=1, py_3=5, py_2=2$. The mempool reduction requires expert understanding and manual efforts, and it is by nature the best effort.
\ignore{
In {\bf Step 3)}, it first de-duplicates the raw short exploits found by \textsc{mpfuzz}. Then, for each distinct short exploit, it extends it by repeating the transaction-admission events to fit into the actual mempool, during which one may practice strategies like increasing nonce, switching sender accounts, and measuring normal transactions in the actual mempool. In exploit extension, one could also consider the factors not modeled in MUT or \textsc{mpfuzz} such as $BasePrice$, as seen in the case in \S~\ref{sec:eval:locking:xt8a}.}
}

{\color{blue}
  \subsection{Bug Oracle: Definitions \& Rationale}
\label{sec:oracle}

\begin{definition}[Tx admission timeline]
\label{def:timeline}
In a transaction admission timeline, a mempool under test is initialized at state $\langle{}st_0, dc_0=\emptyset\rangle{}$, receives a sequence of arriving transactions $ops$, and ends up with state $\langle{}st_n, dc_n\rangle{}$. Then, a validator continually builds blocks by selecting and clearing transactions in the mempool until the mempool is empty. This transaction timeline is denoted by $\langle{}st_0, dc_0=\emptyset\rangle{}, ops\Rightarrow{}\langle{}st_n, dc_n\rangle{}$. 
\end{definition}

The transaction admission timeline is simplified from the timeline that an operational mempool experiences where transaction arrival and block building can be interleaved. 

We use two mempool-attack templates to define two bug oracles: an eviction-based DoS where adversarial transactions evict existing victim transactions in the mempool, and a locking-based DoS where existing adversarial transactions in the mempool decline arriving victim transactions.

\begin{definition}[Eviction bug oracle]
\label{def:oracle:evict}
A transaction admission timeline, $\langle{}st_0, dc_0=\emptyset\rangle{}, ops \Rightarrow{}\langle{}st_n, dc_n\rangle{}$, is a successful \textsc{ADAMS} eviction attack, if.f. 1) the admission timeline causes full damage, that is, initial state $st_0$ contains only benign transactions, 
$ops$ are adversarial transactions, and the result end state $st_n$ contains only adversarial transactions (in Equation~\ref{eqn:adams:evict:1}), and 2) the attack cost is asymmetrically low, that is, the total adversarial transaction fees in the end state $st_n$ to be charged are smaller, by a multiplicative factor $\epsilon$, than the attack damage measured by the fees of evicted transactions in the initial state $st_0$ (in Equations~\ref{eqn:adams:evict:2} and~\ref{eqn:adams:evict:3}). Formally, 

\begin{eqnarray}
\label{eqn:adams:evict:1}
st_0\cap{}st_n &=& \emptyset 
\\
\label{eqn:adams:evict:2}
asym_E(st_0, ops) &\stackrel[]{def}{=}& \frac{\sum_{tx\in{}st_n}tx.fee}{\sum_{tx\in{}st_0}tx.fee} 
\\
\label{eqn:adams:evict:3}
asym_E(st_0, ops) &<& \epsilon
\end{eqnarray}
\end{definition}

\begin{definition}[Locking bug oracle]
\label{def:oracle:lock}
A transaction admission timeline, $\langle{}st_0=\emptyset, dc_0=\emptyset\rangle{}, ops \Rightarrow{}\langle{}st_n, dc_n\rangle{}$, is a successful \textsc{ADAMS} locking attack, if.f. 1) the admission timeline causes full damage, that is, the mempool is first occupied by adversarial transactions (i.e., $st_n$ contains only adversarial transactions) and then declines the arriving normal transactions (i.e., $dc_n$ contains only normal transactions); the normal transactions and adversarial transactions are sent by different accounts (i.e., Equation~\ref{eqn:adams:decline:1}), and 2) the attack cost is asymmetrically low, that is, the average adversarial transaction fees in the mempool are smaller, by a multiplicative factor $\lambda$, than the attack damage measured by the average fees of victim transactions declined (i.e., Equations~\ref{eqn:adams:decline:2} and~\ref{eqn:adams:decline:3}).
Formally, 

\begin{eqnarray}
\cup_{tx\in{}dc_n}{tx.sender}  &\cap &
\cup_{tx\in{}st_n}{tx.sender} = \emptyset
\label{eqn:adams:decline:1}  
\\ 
asym_D(ops) &\stackrel[]{def}{=}& 
\frac{\sum_{tx\in{}st_n} tx.fee / \|st_n\|}{\sum_{tx\in{}dc_n}tx.fee / \|dc_n\|}
\label{eqn:adams:decline:2}
\\
asym_D(ops) &<& \lambda 
\label{eqn:adams:decline:3}
\end{eqnarray}
\end{definition}

\tangSide{R1}
\noindent{\bf Design rationale}: Both definitions use the normal transaction fees to measure the attack damage because these fees are not collected by the validator in the timeline under attack and are collectible without attack. Specifically, the damage in Definition~\ref{def:oracle:evict} is measured by the fees of normal transactions in $st_0$ which are evicted in end state $st_n$ and which would not have been evicted had there been no adversarial transactions in $ops$. The damage in Definition~\ref{def:oracle:lock} is measured by the fees of normal transactions in $ops$ which are declined from end state $st_n$ and which would not have been declined had there been no adversarial transactions in $ops$.

Both definitions are strict and require causing full damage, such as evicting or declining {\it all} normal transactions. Strict definitions are necessary to ensure that what is found as exploits on the small MUT is true positive and works on the actual mempool. 

The targeted attacks by this work are denial of mempool service in feeding downstream validators with valid transactions sent from benign users. They are different from other forms of DoS in blockchains, including resource exhaustion via executing malicious smart contracts~\cite{cryptoeprint:2023/956,DBLP:conf/ndss/LiCLT0L21}.

Particularly, in the threat model of mempool DoS, we assume the validators are benign and functional in building blocks. Because in all Ethereum clients we know, transactions are admitted into mempool without execution (i.e., executing the smart contracts they invoke), \textit{GasUsed} is not a factor in transaction admission.\footnote{The further design rationale for admitting transactions without execution is that an Ethereum transaction $tx$'s \textit{GasUsed} is non-deterministic unless the ordering of $tx$ w.r.t. other transactions in the block is fixed and $tx$ is executed.} Thus, our bug oracle does not vary \textit{GasUsed}. Instead, all the transactions are fixed at $21000$ Gas (i.e., as if they don't invoke smart contracts).\footnote{We may use price and fees interchangeably.} This design significantly reduces the search space without loss in covering different mempool behaviors.
Besides, our bug oracle does not model the effect of EIP1559 or $BasePrice$. Without modeling $BasePrice$, our bug oracle can still capture the alternative transaction fee (the product of \textit{GasPrice} and $21000$ Gas), which is still proportional to actual $tip$ or damage on validator revenue. }

  \section{Stateful Mempool Fuzzing by \textsc{mpfuzz}}
\label{sec:symbolize}

  \subsection{Transaction Symbolization}

The key challenge in designing \textsc{mpfuzz} is the large input space of transaction sequences. Recall that an Ethereum transaction consists of multiple attributes (sender, nonce, \textit{GasPrice}, value, etc.), each defined in a large domain (e.g., 64-bit string). Exploring the raw transaction space is hard; the comprehensive search is inefficient, and randomly trying transactions as done in the state-of-the-art blockchain fuzzers (in \S~\ref{sec:rw}) is ineffective.

\noindent{\bf
Intuition}: We propose symbolizing transactions for efficient and effective mempool stateful fuzzing. Our key idea is to map each group of concrete transactions, triggering an equivalent mempool behavior into a distinct symbol so that searching for one transaction is sufficient to cover all other transactions under the same symbol. Transaction symbolization is expected to reduce the search space from possible concrete transactions to the symbol space.

{\tangSide{R7} \color{blue} 
In this work, we manually design seven symbols to represent transactions based on the Ethereum ``semantics'', that is, how different transactions are admitted by the mempool. 
}


\begin{table}[!htbp] 
  \caption{Symbols, transactions, and associated costs. Tx refers to the instantiated transaction under a given symbol.}
 \label{tab:symbols:2}\centering{\scriptsize
 \begin{tabularx}{0.49\textwidth}{ X|l|l|l|l }
Symbol        & Description & Tx  & $cost()$ & $opcost()$ \\ \hline 
&&&&\\[-0.8em]
$\mathcal{N}$ & Benign & $tx{\begin{reduce}\begin{bsmallmatrix}{B*} & {1}\\ {*} & {3}\end{bsmallmatrix}\end{reduce}}$ & $3$ & $3$ \\ [3pt]\hline 
&&&&\\[-0.8em]
$\mathcal{F}$ & Future & $tx{\begin{reduce}\begin{bsmallmatrix}{A*} & {1}\\ {m+1} & {m+4}\end{bsmallmatrix}\end{reduce}}$ & $0$ & $0$ \\ [3pt]\hline 
&&&&\\[-0.8em]
$\mathcal{P}$ & Parent & $tx{\begin{reduce}\begin{bsmallmatrix}{A>r} & {1}\\ {1} & {[4,m+3]}\end{bsmallmatrix}\end{reduce}}$& $[4, m+3]$ & $[4,m+3]$ \\ [4pt]\hline 
&&&&\\[-0.8em]
$\mathcal{C}$ & Child & $tx{\begin{reduce}\begin{bsmallmatrix}{A[1,r]} & {1}\\ {\geq{}2} & {m+4}\end{bsmallmatrix}\end{reduce}}$ & $m+4$ & {\color{red} $1$} \\ [4pt]\hline 
&&&&\\[-0.8em]
$\mathcal{O}$ & Overdraft & $tx{\begin{reduce}\begin{bsmallmatrix}{A[1,r]} & {m+1}\\ {\geq{}2} & {m+4}\end{bsmallmatrix}\end{reduce}}$ & $0$ & $0$ \\ [4pt]\hline 
&&&&\\[-0.8em]
$\mathcal{L}$ & Latent overdraft & $tx{\begin{reduce}\begin{bsmallmatrix}{A[1,r]} & {m-1}\\ {\geq{}2} & {m+4}\end{bsmallmatrix}\end{reduce}}$ & $0$ & $0$ \\ [4pt] \hline 
&&&&\\[-0.8em]
$\mathcal{R}$ & Replacement & $tx{\begin{reduce}\begin{bsmallmatrix}{A[1,r]} & {m-1}\\ {1} & {m+4}\end{bsmallmatrix}\end{reduce}}$ & $m+4$ & $0$  \\ [4pt]\hline
\end{tabularx}}
\end{table}

\noindent{\bf 
Specification}: Suppose the current state contains adversarial transactions sent from $r$ accounts $A_1,\dots, A_r$. 
These accounts have an initial balance of $m$ Ether, where $m$ equals the mempool capacity (say $m=1000$).
{\tangSide{R13} \color{blue}
The rationale is that these accounts can send at most $m$ adversarial transactions, each minimally spending one Ether, to just occupy a mempool of $m$ slots. A larger value of attacker balance is possible but increases the search space.
}

Symbol $\mathcal{N}$ defines a transaction subspace covering any normal transactions sent from any benign account and of any nonce, namely $tx\begin{bsmallmatrix}{B*} & {*}\\ {*} & {*}\end{bsmallmatrix}$
. In \textsc{mpfuzz}, Symbol $\mathcal{N}$ is instantiated to concrete transactions of fixed \textit{GasPrice} $3$ wei and of value $1$ Ether. That is, symbol $\mathcal{N}$ is instantiated to transaction $tx\begin{bsmallmatrix}{B*} & {1}\\ {*} & {3}\end{bsmallmatrix}$
, as shown in Table~\ref{tab:symbols:2}. Among all symbols, $\mathcal{N}$ is the only symbol associated with benign sender accounts.

Symbol $\mathcal{F}$ defines the transaction subspace of any future transaction sent from an adversarial account; the future transaction is defined w.r.t. the current state. \textsc{mpfuzz} instantiates symbol $\mathcal{F}$ to concrete transactions of nonce $m+1$, value $1$ Ether, \textit{GasPrice} $m+4$ wei, and any adversarial account $A*$. The instantiating pattern for symbol $\mathcal{F}$ is $tx\begin{bsmallmatrix}{A*} & {1}\\ {m+1} & {m+4}\end{bsmallmatrix}$. 

Symbol $\mathcal{P}$ defines the transaction subspace of any parent transaction from an adversarial account w.r.t. the current state. \textsc{mpfuzz} instantiates $\mathcal{P}$ to a transaction of Pattern $tx{\begin{reduce}\begin{bsmallmatrix}{A\geq{r}} & {1}\\ {1} & {[4,m+3]}\end{bsmallmatrix}\end{reduce}}$. Here, the transaction is fixed with a \textit{GasPrice} in the range $[4,m+3]$ wei, value at $1$ Ether, and nonce at $1$. 
{\tangSide{R13} \color{blue}
The \textit{GasPrice} of a future transaction ($\mathcal{F}$) is $m+4$ wei, which is higher than the price of a parent transaction ($\mathcal{P}$) in $[4,m+3]$ wei. The purpose is to ensure that a future transaction can evict any parent transaction during fuzzing.
}

Symbol $\mathcal{C}$ defines the transaction subspace of any adversarial child transaction w.r.t. the state. \textsc{mpfuzz} instantiates $\mathcal{C}$ to a transaction of Pattern $tx{\begin{reduce}\begin{bsmallmatrix}{A[1,r]} & {1}\\ {\geq{}2} & {m+4}\end{bsmallmatrix}\end{reduce}}$. The instantiated child transaction is fixed at \textit{GasPrice} of $m+4$.

Symbol $\mathcal{O}$ defines the transaction subspace of any adversarial overdraft transaction w.r.t. state $st$. \textsc{mpfuzz} instantiates $\mathcal{O}$ to a transaction of Pattern $tx{\begin{reduce}\begin{bsmallmatrix}{A[1,r]} & {m+1}\\ {\geq{}2} & {m+4}\end{bsmallmatrix}\end{reduce}}$. Recall that all adversarial accounts have an initial balance of $m$ Ether.

Symbol $\mathcal{L}$ defines the transaction subspace of any adversarial latent-overdraft transaction w.r.t. state $st$. \textsc{mpfuzz} instantiates $\mathcal{L}$ to a transaction of Pattern $tx{\begin{reduce}\begin{bsmallmatrix}{A[1,r]} & {m-1}\\ {\geq{}2} & {m+4}\end{bsmallmatrix}\end{reduce}}$. All adversarial accounts have an initial balance of $m$ Ether.

Symbol $\mathcal{R}$ defines the transaction subspace of any adversarial ``replacement'' transaction w.r.t., the state. Unlike the symbols above, the transactions under Symbol $\mathcal{R}$ must share the same sender and nonce with an existing transaction in the current state $st$. \textsc{mpfuzz} instantiates $\mathcal{R}$ to a transaction of Pattern $tx{\begin{reduce}\begin{bsmallmatrix}{A[1,r]} & {m-1}\\ {1} & {m+4}\end{bsmallmatrix}\end{reduce}}$. Here, we simplify the problem and only consider replacing the transaction with nonce $1$.

\begin{figure}[!bthp]
\centering
\includegraphics[width=0.38\textwidth]{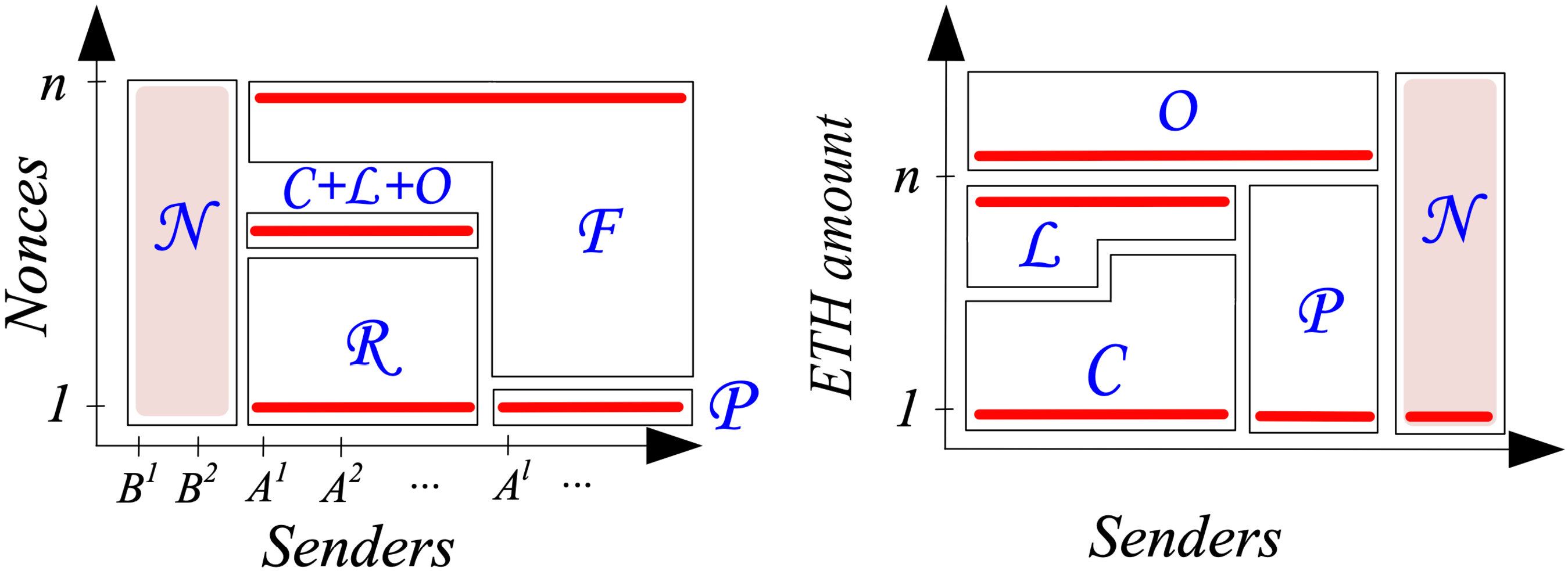}
\caption{Symbols and transaction space reduction.}
\label{fig:opsymbols}
\end{figure}

Given that a concrete transaction in this work consists of four attributes, transactions are defined in a four-dimensional space. We visualize transaction symbols (with an incomplete view) in two two-dimensional spaces in Figure~\ref{fig:opsymbols}, that is, one of transaction sender and nonce and the other of sender and value. 
For each symbol, the figures depict the transaction subspace covered by the symbol (in white shapes of black lines) and the transaction pattern instantiated by \textsc{mpfuzz} under the symbol (in red lines or shapes). In our design, the instantiated transactions are a much {\it smaller subset} of the defining transaction space. 
For instance, given state $st$ of transactions sent from account $A_1,\dots,A_r$, 
transaction $tx{\begin{reduce}\begin{bsmallmatrix}{A_{r+1}} & {*}\\ {3} & {*}\end{bsmallmatrix}\end{reduce}}$
is within the defining space of symbol $\mathcal{F}$, but \textsc{mpfuzz} does not instantiate $\mathcal{F}$ by transaction $tx{\begin{reduce}\begin{bsmallmatrix}{A_{r+1}} & {*}\\ {3} & {*}\end{bsmallmatrix}\end{reduce}}$ (but instead to transactions $tx{\begin{reduce}\begin{bsmallmatrix}{A*} & {1}\\ {m+1} & {m+4}\end{bsmallmatrix}\end{reduce}}$).

  \subsubsection{State Search Algorithm}

\begin{algorithm}[h]
\caption{\textsc{mpfuzz}(SeedCorpus $sdb$, Mempool $mp$)}
\small \label{alg:fuzz3}
\begin{algorithmic}[1]
\State $sdb$.init($mp$);
\While{!$sdb$.is\_empty() or timeout}
\State $st, ops$ = $sdb$.next(); \Comment{Selection by energy}
\label{alg:fuzz2:next}
\ForAll{$ops', st'$ = mutateExec($ops, st$)}
\label{alg:fuzz2:exec}
  \If{is\_ADAMS($st'$)} \Comment{Test oracle}
    \State emit($ops'$, $st'$, "Found an exploit");
  \Else
    \If{!$sdb$.feedback($st'$, $sdb$)} 
        \State $sdb$.add($st'$, $ops'$);
    \EndIf
  \EndIf
\EndFor
\EndWhile
\end{algorithmic}
\end{algorithm}

\noindent{\bf
Algorithm overview}:
We propose a stateful fuzzing algorithm, listed in Algorithm~\ref{alg:fuzz3}, to selectively explore the input space in a way that prioritizes new and promising mempool states towards triggering ADAMS oracle. 

The core data structure is a seed corpus or $sdb$ which stores a list of input-state pairs or seeds. Upon running the algorithm, the corpus maintains the mempool states covered so far by the algorithm execution.
The algorithm runs an outer loop that continues until the corpus is empty, or only the states of zero energy are left, or timeout. In each iteration, the algorithm retrieves the next seed from the corpus based on how promising the seed is in reaching a state triggering the oracle (a.k.a., the energy~\cite{DBLP:conf/uss/BaBMR22}). 
A seed consists of a symbolized input $ops$, which consists of transaction symbols, and a symbolized state $st$ reached by running a transaction sequence instantiated from the input $ops$ (Line~\ref{alg:fuzz2:next}). The algorithm runs an inner loop, in each iteration of which it mutates the input in the current seed ($\langle{}st, in\rangle{}$) and executes the mutated input against a reinitialized empty mempool, producing end state $st'$ (Line~\ref{alg:fuzz2:exec}). If the end state satisfies ADAMS conditions, it emits the mutated input-state pair $ops',st'$ as a newly found exploit. The algorithm further checks the feedback: The feedback is positive if the mutated input $ops'$ brings the mempool state $st'$ closer to triggering the test oracle than $st$. This entails mutated state $st'$ to be different from state $st$ (i.e., increased state coverage) and state $st'$ to achieve larger damage or lower attack cost. In case of positive feedback, the algorithm would add mutated input-state pair $ops',st'$ to the corpus. 

This algorithm assumes the mempool is deterministic. That is, given a seed $\langle{}ops, st\rangle{}$, running the same input $ops$ against an empty mempool multiple times always results in the same end state $st$. In practice, we generate inputs to avoid the non-deterministic behavior of real mempool implementations. 

The algorithm can be configured with initial seeds and input-mutation strategies. By default, we use one initial seed whose input fills up the mempool with normal transactions. The default input-mutation strategy is to append the current input with a newly generated transaction. 

\begin{algorithm}[h]
\caption{mutateExec(SymbolInput $ops$, SymbolState $st$)}
\small \label{alg:fuzz3:input}
\begin{algorithmic}[1]
\State $ops'$=mutateSymbol($ops$);
\label{alg:fuzz3:input:1}
\State $st_c$=executeUnappended($ops, ops'$);
\label{alg:fuzz3:input:2}
\State $st_c'$=executeAppended($ops', st_c, st$);
\label{alg:fuzz3:input:3}
\State \Return $st_c'$
\end{algorithmic}
\end{algorithm}

\noindent{\bf 
Input mutation}:
Given a symbolized input $ops$ and symbolized state $st$, the mutation algorithm is to explore each ``slightly'' different input $ops'$ and its associated state $st'$, such that the next stage can find the inputs producing positive feedback.

Internally, the algorithm proceeds at two levels, that is, symbols and concrete value. Specifically, as shown in Algorithm~\ref{alg:fuzz3:input}, 1) it appends the symbolized input $ops$ with a previously untried symbol, generating the ``mutated'' symbolized input $ops'$ (Line~\ref{alg:fuzz3:input:1} in Algorithm~\ref{alg:fuzz3:input}). In this step, the algorithm tries different symbols in the following order: $\mathcal{P}, \mathcal{L}, \mathcal{C}$. 
2) It instantiates symbolized $ops'$ to a transaction sequence and executes the sequence in the tested mempool to obtain the concrete end state (Line~\ref{alg:fuzz3:input:2} in Algorithm~\ref{alg:fuzz3:input}). This step is necessary for instantiating and executing the mutation in the next step. Specifically, in this step, the algorithm instantiates symbolized input $ops$ to a concrete input $in_c$, which is a sequence of transactions. It then drives the transactions to a reinitialized mempool for execution. It returns the resulting concrete state $st_c$.
3) The algorithm instantiates the appended symbols under the context of the previous state $st$ and $st_c$. At last, it executes the appended transaction on mempool $st_c$ to obtain the concrete end state $st'_c$ under the mutated input (Line~\ref{alg:fuzz3:input:3} in Algorithm~\ref{alg:fuzz3:input}).

\noindent{\bf 
Mutation feedback}: We describe how mempool states are symbolized before presenting state-based feedback.
In \textsc{mpfuzz}, a symbolized state $st$ is a list of transaction symbols, ordered first partially as follows: $\mathcal{N}\preceq{}\mathcal{E}\preceq{}\mathcal{F}\preceq{}\{\mathcal{P},\mathcal{L},\mathcal{C}\}$. $\mathcal{E}$ refers to an empty slot. State slots of the same symbols are independently instantiated into transactions, except for $\mathcal{P}, \mathcal{C}, \mathcal{L}$. Child transactions (i.e., $\mathcal{C}, \mathcal{L}$) are appended to their parent transaction of the same sender (i.e., $\mathcal{P}$). Across different senders, symbols are ordered by the parent's \textit{GasPrice}. For instance, suppose a mempool stores four concrete transactions, $tx\begin{bsmallmatrix}{A_1} & {*}\\ {1} & {4}\end{bsmallmatrix}$, $tx\begin{bsmallmatrix}{B} & {*}\\ {100} & {100}\end{bsmallmatrix}$, $tx\begin{bsmallmatrix}{A_2} & {*}\\ {1} & {5}\end{bsmallmatrix}$, $tx\begin{bsmallmatrix}{A_1} & {*}\\ {2} & {10001}\end{bsmallmatrix}$
, they are mapped to four symbols, $\mathcal{P}, \mathcal{F}, \mathcal{P}, \mathcal{C}$, and are further ordered in a symbolized state by $\mathcal{F}\mathcal{P}\mathcal{C}\mathcal{P}$.

In \textsc{mpfuzz}, the feedback of an input $ops'$ is based on the symbolized end state $st'$. Positive feedback on state $st'$ is determined conjunctively by two metrics: 1) State coverage that indicates state $st'$ is not covered in the corpus, and 2) state promising-ness that indicates how promising the current state is to reach a state satisfying bug oracles.

\begin{eqnarray}
\nonumber
feedback(st',st,sdb) &=& st\_coverage(st',sdb)==1 \land{} 
\nonumber
\\ &&
st\_promising(st',st)==1 
\label{eqn:feedback}
\end{eqnarray}

Specifically, state coverage is determined by straightforwardly comparing symbolized state $st'$, as an ordered list of symbols, with all the symbolized states in the corpus. For instance, state $\mathcal{F}\mathcal{P}\mathcal{C}\mathcal{P}$ is different from $\mathcal{F}\mathcal{P}\mathcal{P}\mathcal{C}$. This also implies concrete transaction sequence $tx{\begin{reduce}\begin{bsmallmatrix}{A_1} & {*}\\ {1} & {4}\end{bsmallmatrix}\end{reduce}}$, $tx{\begin{reduce}\begin{bsmallmatrix}{B} & {*}\\ {100} & {100}\end{bsmallmatrix}\end{reduce}}$, $tx{\begin{reduce}\begin{bsmallmatrix}{A_2} & {*}\\ {1} & {5}\end{bsmallmatrix}\end{reduce}}$, $tx{\begin{reduce}\begin{bsmallmatrix}{A_1} & {*}\\ {2} & {10001}\end{bsmallmatrix}\end{reduce}}$ is the same with $tx{\begin{reduce}\begin{bsmallmatrix}{A_1} & {*}\\ {1} & {4}\end{bsmallmatrix}\end{reduce}}$, $tx{\begin{reduce}\begin{bsmallmatrix}{B} & {*}\\ {100} & {100}\end{bsmallmatrix}\end{reduce}}$, $tx{\begin{reduce}\begin{bsmallmatrix}{A_2} & {*}\\ {1} & {6}\end{bsmallmatrix}\end{reduce}}$, $tx{\begin{reduce}\begin{bsmallmatrix}{A_1} & {*}\\ {2} & {10}\end{bsmallmatrix}\end{reduce}}$, as they are both mapped to the same symbolized state.

How promising a state $st'$ is (i.e., $st\_promising(st')$) is determined as follows: A mutated state $st'$ is more promising than an unmutated state $st$ if any one of the three conditions is met: 1) State $st'$ stores fewer normal transactions (under symbol $\mathcal{N}$) than state $st$, implying more transactions evicted and higher damage (i.e., $evict\_normal(st', st) = 1$). 
2) State $st'$ declines (speculatively) more incoming normal transactions than state $st$, also implying higher damage (i.e., $decline\_normal(st',st) = 1$). 
3) The total fees of adversarial transactions in state $st'$ are lower than those in state $st$, implying lower attack costs. 
We consider two forms of costs: concrete cost and symbolized cost. The former, denoted by $cost(st')$, is simply the total fees of adversarial transactions instantiated from a symbolized state $st'$. The latter, denoted by $opcost(st')$, optimistically estimates the cost of transactions in the current state $st'$ that contributes to a future state triggering test oracle. We will describe the symbolized cost next. Formally, state promising-ness is calculated by Equation~\ref{eqn:promising}.  

\begin{eqnarray}
\nonumber
st\_promising(st',st) & = & evict\_normal(st',st) == 1 \lor{} 
\\ \nonumber & & 
decline\_normal(st',st) == 1 \lor{} 
\\ \nonumber & & 
cost(st') < cost(st) \lor{} 
\\ & & 
opcost(st') < opcost(st)
\label{eqn:promising}
\end{eqnarray}

\noindent{\bf
Seed energy}: Recall that in \textsc{mpfuzz}, the next seed is selected from the corpus based on energy. By intuition, the energy of a state is determined based on how promising the state is in triggering the test oracle. In addition to the state promising-ness used in deciding feedback, state energy incorporates fuzz runtime information, such as how many times the state has been selected.

Specifically, the energy of a seed $\langle{}ops, st\rangle{}$ is determined by Equation~\ref{eqn:energy}.

\begin{eqnarray}
Energy(in,st) &=& b / opcost(st)
\label{eqn:energy}
\end{eqnarray}

First, each seed in the corpus records how many times it has been mutated. The more symbols it has mutated in the past, the less energy the seed currently has and the lower priority it will be selected next. 
$b$ can be configured differently to traverse the state tree differently.
In particular, breadth-first search (BFS) is by the following configuration: $b=1$ if at least one symbol has not been tried (for mutation) in the current seed. Otherwise, $b=0$. 

Second, 
we use the symbolized cost to estimate how promising a state is and use it in seed energy. 

\noindent{\bf Estimate state cost}: When \textsc{mpfuzz} determines how promising a state is, it needs to look beyond the current state and into all possible descendant states. We propose a heuristic that optimistically estimates the descendant-state costs given a current state. The key intuition is the following: In Ethereum, the validity of a child transaction (under symbol $\mathcal{C}$) depends on its parent/ancestor transaction. Thus, even though a transaction in the current state is valid, the transaction can be ``turned'' into an invalid one in subsequent states. We thus attribute, optimistically, the cost of transaction $\mathcal{C}$ to value $1$, so that it is preferable to a parent transaction $\mathcal{P}$ in input mutation and seed selection, and turning $\mathcal{C}$ into an invalid one like $\mathcal{L}$ or $\mathcal{F}$ is also encouraged. The cost profiles used in \textsc{mpfuzz} are summarized in Table~\ref{tab:symbols:2}.

{\color{violet}
\subsection{Exploit Extension}
\label{sec:extension}
The initial short exploit identified on MUT necessitates further automatic development to create a more extensive exploit that is effective within a larger mempool. The key idea in extending the small exploit to a more comprehensive one is to ensure the same admission event observed in the smaller MUT. We first define the admission event as given an arriving transaction $tx_i$ at a mempool of initial state $st_0$, the admission of $tx_i$ transitions the mempool into an end state $tx_n$. The admission event is denoted by $f(st_0,tx_i)\Rightarrow{}st_n$. We extract the admission event patterns, representing a state transition from the initial state to the end state. In the pattern, we only symbolize the mempool slots that are updated during the state transition. A pattern is denoted as $\langle{}st_0', tx_i, st_n'\rangle{}$. For example, in the state transition from $\mathcal{NNPC}$ to $\mathcal{NPPC}$ by admitting a transaction $\mathcal{P}_0$, the symbolized state of a mempool slots transitions from $\mathcal{N}$ to $\mathcal{P}$. Thus, we extract the pattern as $\langle{}\mathcal{N}, \mathcal{P}, \mathcal{P}\rangle{}$.

we extract all the unique admission event patterns in the small exploit and store them into an admission event pattern set ($aps$). We then construct the attack transaction sequence on the larger-size mempool by aligning the patterns identified from the small exploit with corresponding admission events. This entails validating that the admission event pattern of a given event $f(st_0,tx_i)\Rightarrow{}st_n$ on the larger-size mempool exists within the $aps$. Specifically, \textsc{mpfuzz} first selects $ap_i$ from the set $aps$ whose initial state $st_0$ matches the state of the large mempool. We then instantiate $tx_i$ of the $ap_i$ and send it to the large mempool. If the state transition of the large mempool $\langle{}st_0', tx_i, st_n'\rangle{}$ aligns with the pattern $ap_i$, this positive outcome prompts \textsc{mpfuzz} to proceed with the extension process accordingly. Otherwise, \textsc{mpfuzz} backtracks by rolling back the mempool state to its state prior to the admission of $tx_i$ and tries the next pattern. \textsc{mpfuzz} tries the patterns in the decreasing order of $opcost()$, which means if multiple admission event patterns in $aps$ yield positive results,  \textsc{mpfuzz} proceeds the extension process by applying the pattern with the lower $opcost()$.

}

  \section{Found Exploits}
\label{sec:exploits}
\newcolumntype{C}{>{\centering\let\newline\\\arraybackslash\hspace{0pt}}m{2cm}}
\definecolor{darkolivegreen}{rgb}{0.33, 0.42, 0.18}
\definecolor{ferngreen}{rgb}{0.31, 0.47, 0.26}
\definecolor{forestgreen}{rgb}{0.0, 0.27, 0.13}

\begin{table*}[!htbp]
  \caption{ADAMS exploit patterns found by \textsc{mpfuzz} across Ethereum clients; $XT_{1-7}$ are eviction based, and $XT_{8-9}$ are locking based. $XT_{1-3}$ are known patterns in DETER~\cite{DBLP:conf/ccs/LiWT21}, while others are new. \xmark{} indicates the presence of a bug,  {\color{green} \cmark{}} indicates the fixing of a bug after our reporting, and \cmark{} indicates the fixing of a bug by previous works.}
  \label{tab:bug-report}
  \centering{\footnotesize
  \begin{tabularx}{0.85\textwidth}{ l|ccccccc|cX }
  & $XT_1$ &$XT_2$&$XT_3$&$XT_4$&$XT_{5}$&$XT_6$&$XT_7$ &$XT_8$&$XT_9$\\ \hline 
   \makecell{Geth $\geq{}v1.11.4$, {\color{blue} Flashbot $\le{}v1.11.5$ , bloXroute,} \\ {\color{blue}BSC $\le{}v1.3.8$, core-geth$\le{}v1.12.18$}} & \cmark{} & \cmark{} & & {\color{green}\cmark} & \xmark{} & \xmark{} &  &&\\ 
   \rowcolor{Gray} Geth $<v1.11.4$, {\color{blue} EigenPhi} & \xmark{} & \xmark{} & \xmark{} & \xmark{} & \xmark{} & \xmark{} & && \\ 
   {\color{blue} go-opera $\le{}v1.1.3$} &  & \xmark{} & \xmark{} & \xmark{} & \xmark{} & \xmark{} & && \\ 
   \rowcolor{Gray} Erigon $\le{}v2.42.0$ &  &  & & \xmark{} & & & &&\\
   Besu $\geq{}v22.7.4$ & \cmark{} & \xmark{} &  & \xmark{} &  & & &&\\ 
   \rowcolor{Gray} Besu $<v22.7.4$ &\xmark{} & \xmark{} &  & \xmark{} &  & & &&\\ 
   Nethermind $\geq{}v1.18.0$ & \cmark{} & &  & \xmark{} & & & {\color{green}\cmark} &&\\ 
   \rowcolor{Gray} Nethermind $<v1.18.0$& \xmark{} &  &  & \xmark{} &  &  & \xmark{} &&\\ 
   Reth $\geq{}v0.1.0-alpha.6$ &  &  &  & \xmark{} & & & & {\color{green} \cmark} &\\
   \rowcolor{Gray} Reth $<v0.1.0-alpha.6$ &  &  & &  & & & &\xmark&\\
   OpenEthereum $\le{}v3.3.5$ &  &  & & \xmark{} & & & &&\xmark
 \end{tabularx}}
\end{table*}

{\tangSide{R5} \color{blue} 
We have run \textsc{mpfuzz} across a variety of Ethereum clients, including six leading execution-layer clients on the public-transaction path of Ethereum mainnet (Geth, Besu, Nethermind, Erigon, Reth, and OpenEthereum), three PBS clients (proposer-builder separation) on the mainnet's private-transaction and bundle path (Flashbot builder $v1.11.5$~\cite{me:flashbotbuilder}, EigenPhi builder~\cite{me:eigenphibuilder} and bloXroute builder-ws~\cite{me:bloXroutebuilder}), and the clients deployed on three operational Ethereum-like networks (BSC $v1.3.8$~\cite{me:bscclient} deployed on Binance Smart Chain, go-opera $v1.1.3$~\cite{me:goopera} on Fantom, and core-geth $v1.12.19$~\cite{me:coregeth} on Ethereum Classic). 

On the six public-transaction clients, \textsc{mpfuzz} leads to the discovery of $22$ bugs, as listed in Table~\ref{tab:bug-report}, including $7$ conforming to the known DETER attacks on clients of historical versions and $15$ new bugs on the clients of the latest versions. 

Other clients, including the three PBS clients and three clients on Ethereum-like networks, are mostly forks of the Geth clients, and on them, \textsc{mpfuzz} discovered $13$ bugs of a similar nature to those found on Geth (of the historical and latest versions), as seen in Table~\ref{tab:bug-report}.

We describe the patterns of these bugs as follows.
}

  \subsection{Found Eviction Attacks}
\label{sec:foundevictionattacks}
\noindent{\bf 
Exploit $XT_1$: Direct eviction by future transactions}:
In this attack, given a mempool's initial state storing normal transactions, the attacker sends future transactions at high fees to evict the normal transactions. This exploit is essentially the DETER-X attack~\cite{DBLP:conf/ccs/LiWT21}.

In practice, Geth ($\leq{}v1.11.4$), Nethermind, Besu,{\tangSide{R5} \color{blue} and EigenPhi} are vulnerable under this exploit, as in Table~\ref{tab:bug-report}.

\noindent{\bf 
Exploit $XT_2$: Direct eviction by latent overdraft transactions}:
In this eviction attack, the attacker sends latent-overdraft transactions at high fees to evict the normal transactions initially stored in the target mempool. These transactions are sent from $k$ accounts, each of which sends $l$ transactions, denoted by $k\times{}l$. The intention is to evade the limit on the number of transactions per sender. The evasion increases the attacker cost from one pending transaction to multiple. 
This exploit is essentially the DETER-Z attack~\cite{DBLP:conf/ccs/LiWT21}.

In practice, Geth ($\leq{}v1.11.4$), Besu,{\tangSide{R5} \color{blue} 
EigenPhi, and go-opera}
are found vulnerable under this exploit. 

\noindent{\bf 
Exploit $XT_3$: Compositional direct eviction} (by combining $XT_1$ and $XT_2$):
In some Ethereum clients, notably Geth, the limit on the number of transactions per sender is triggered under the condition that the mempool stores enough pending transactions (e.g., more than 5120 transactions in Geth). 
This eviction attack combines $XT_1$ and $XT_2$ to avoid triggering the condition and evade the protection. It works by maximizing the eviction of mempool under $XT_1$ until it is about to trigger the condition. It then conducts $XT_2$ by sending latent overdraft transactions under one sender, that is, $1\times{}l$. Compared to $XT_2$, the combined exploit $XT_3$ achieves lower costs.

In practice, Geth ($\leq{}v1.11.4$) is vulnerable under this exploit where its mempool of capacity of 6144 slots triggers the limit of 16 transactions per sender when there are more than 5120 pending transactions. That is, $XT_3$ is configured with $l=5120$.
Other clients including{\tangSide{R5} \color{blue} 
EigenPhi and go-opera} are also vulnerable.

\noindent{\bf 
Exploit $XT_4$: Indirect eviction by valid-turned-overdraft transactions}: 
This eviction attack works in two steps: 1) The attacker first sends valid transactions at high fees to evict normal transactions initially stored in the mempool. These transactions are sent from $k$ accounts, each of which sends $l$ transactions. 2) She then sends $k$ transactions; each of the transactions is of nonce $1$, at a high fee, of high value $v_1$, and from the same $k$ accounts in Step 1). The transactions would replace the transaction of the same sender and nonce sent in Step 1). Once the replacement is finished, they turn their child transactions into latent overdraft. Specifically, if a sender's balance is $bal$ and the Ether value of transaction of nonce 2 is $v_2$, then $v_1$ is carefully crafted to enable turned latent overdraft, that is, $v_1<bal$ and $v_1+v_2>bal$.

In practice, many clients are found vulnerable under this exploit, including Geth $<v1.11.4$, Erigon, Besu, Nethermind, OpenEthereum,{\tangSide{R5} \color{blue} 
EigenPhi, and go-opera}. 

\noindent{\bf 
Exploit $XT_5$: Indirect eviction by valid-turned-future transactions}:
This eviction attack works in two steps: 1) The attacker first sends valid transactions at high fees to evict normal transactions initially stored in the mempool. These transactions are sent by $k\times{}l$, that is, from $k$ accounts, each with $l$ transactions. For each sender, the transaction of nonce 1 has a fee, say $f_1$, slightly lower than the transactions of other nonces, that is, $f_1<f_2$. 
2) She then sends $k$ transactions; each of the transactions is from a distinct sender from those used in Step 1) and of fee $f'$ that $f_1<f'<f_2$. The intent is that the $k$ transactions in Step 2) evict the transactions of nonce $1$ sent in Step 1), turning other child transactions sent in Step 1) into future transactions. 

\noindent{\bf 
Exploit $XT_6$: Compositional indirect eviction} (by multi-round valid-turned-future transactions): 
{\color{blue}
This exploit is an adaptive attack to the Geth $\geq{}v1.11.4$. Geth $\geq{}v1.11.4$ is patched against $XT_{1-2}$ and adopts the following admission policies: The mempool of $m'=6144$ slots admits up to $py_1'=1024$ future transactions. When containing more than $py_3'=5120$ pending transactions, the mempool starts to limit that no more than $py_2'=16$ pending transactions from the same sender can be admitted.

Exploit $XT_6$ works in three steps: 1) it first evicts the initial mempool of normal transactions with $m'/py_2'=384$ sequences, each of $16$ transactions sent from one unique sender, 2) it then sends $py_1'/py_2'=64$ transactions to evict the parent transactions and creates $py_1'=1024$ future transactions in the mempool, 3) it sends one sequence of $5120$ transactions from one sender to evict the normal transactions sent in Step 1), and 4) it sends one transaction to evict the parent transaction in Step 3, leaving the mempool of just one valid transaction. Step 3) can succeed because the precondition (w.r.t. $py_3'=5120$) of limiting transactions of the same sender does not hold.}

In practice, Geth of all versions and the Geth forks, including 
{\tangSide{R5} \color{blue} 
Flashbot, bloXroute, BSC, and Ethereum Classic} are found vulnerable under $XT_6$.

\noindent{\bf 
Exploit $XT_7$: Reversible evictions}: 
Suppose a mempool of state $st$ admits an arriving transaction $tx$ by evicting an existing transaction $tx'$, transitioning its state to $st'$. A mempool is reversible if one sends $tx'$ to state $st'$, and the mempool admits $tx'$ by evicting $tx$, transitioning its state back to $st$.

The mempool on Nethermind $<v1.18.0$ can be reversible: Minimally, suppose a two-slot mempool stores $tx_1$ of low fee from one sender and $tx_2$ of medium fee from another sender $A2$ and receives $tx_3$ as a child of $tx_1$ and of high fee. For instance, $tx_1\begin{bsmallmatrix}{A} & {1}\\ {1} & {1}\end{bsmallmatrix}$, $tx_2\begin{bsmallmatrix}{B} & {1}\\ {1} & {3}\end{bsmallmatrix}$,  $tx_3\begin{bsmallmatrix}{A} & {1}\\ {2} & {5}\end{bsmallmatrix}$. The Nethermind $<v1.18.0$ mempool would admit $tx_3$ and evict $tx_2$. After that, if one resends the evicted $tx_2$ back to the mempool storing $tx_1$ and $tx_3$, the Nethermind  $<v1.18.0$ mempool would admit $tx_2$ and evict $tx_3$, looping back to its initial state.

An attacker observing reversible mempool mounts Exploit $XT_7$ to evict the mempool while bringing down attack costs. On Nethermind $<v1.18.0$, the attack works in two steps: 1) She first sends $k$ transaction sequences from $k$ senders,\footnote{Nethermind does not limit transactions per sender, thus $k=1$ in actual attacks.} each of $l$ transactions (i.e., $k\times{}l$). In each $l$-transaction sequence, the nonce-$1$ transaction has a low fee, say $f_1$, while all its child transactions are of high fees, evicting normal transactions initially stored in the mempool. 2) The attacker sends $k\times(l-1)$ transactions from $k\times(l-1)$ new senders, all with fees slightly higher than $f_1$. The mempool would admit these transactions to the mempool by evicting all child transactions sent in Step 1), bringing down the attack costs.

  \subsection{Found Locking Attacks}

\noindent{\bf
Exploit $XT_8$: Locking FIFQ}: Certain mempool design prohibits eviction by admitting transactions as a FIFQ. Transactions are admitted to a mempool only when there are empty slots. A full mempool always declines an arriving transaction. In practice, Reth $v0.1.0-alpha.4$ adopts the FIFQ mempool design.

While a FIFQ mempool can not be vulnerable to eviction attacks, it can be easily locked by Exploit $XT_8$ as follows: Whenever the attacker observes an empty slot present in the mempool, she sends a pending transaction of a minimally necessary fee to occupy the slot. Any transactions that arrive after will encounter a full mempool and are declined.

\noindent{\bf
Exploit $XT_9$: Locking mempool of no turning}: Because of the risk of evicting a parent transaction that could lead to turning, certain mempool is designed to restrict eviction victims to child transactions; that is, only the transaction of maximal nonce under a given sender can be evicted. For instance, given an arriving transaction $tx$, the OpenEthereum mempool finds an existing child transaction $tx'$ with a lower fee than $tx$ and evicts it to admit $tx$. 

In Exploit $XT_9$, the attacker observes empty slots in the mempool (e.g., created by arriving blocks) and sends the same number of transactions to occupy them. The transactions are sent from one account where the child transaction has a higher fee $f_1$ than normal transactions, while all its parent transactions have minimal fees. A normal transaction that arrives subsequently is declined because $f_1$ is higher than the fee of a normal transaction. Because each of its parent transactions has a low fee, the mempool is locked at a low cost.

\section{Attack Evaluation}
  \subsection{Evaluation of Found Eviction  Exploits on Testnet}
\label{sec:evaluate}
\label{sec:evaluate:turning}

We design experiments to measure the success of ADAMS attacks in real Ethereum networks.

  \subsubsection{Evaluation in Testnet}
\label{sec:eval:multi:turning}

\ignore{
\begin{center}
\fbox{\parbox{0.90\linewidth}{
RQ1. What's the success rate and attack cost in mounting a turning/eviction attack against a testnet? Particularly, can the crafted transactions in a turning attack be propagated through the testnet?
}}\end{center}
}

\label{sec:evaluate:testnet}
\noindent{\bf Experimental settings}: 
We set up our experiment platform by connecting the attacker node to the Goerli testnet. The attacker node runs an instrumented Geth client at
the execution layer which sends crafted transactions to the testnet.

\ignore{
\begin{figure}[!bthp]
\centering
\includegraphics[width=0.425\textwidth]{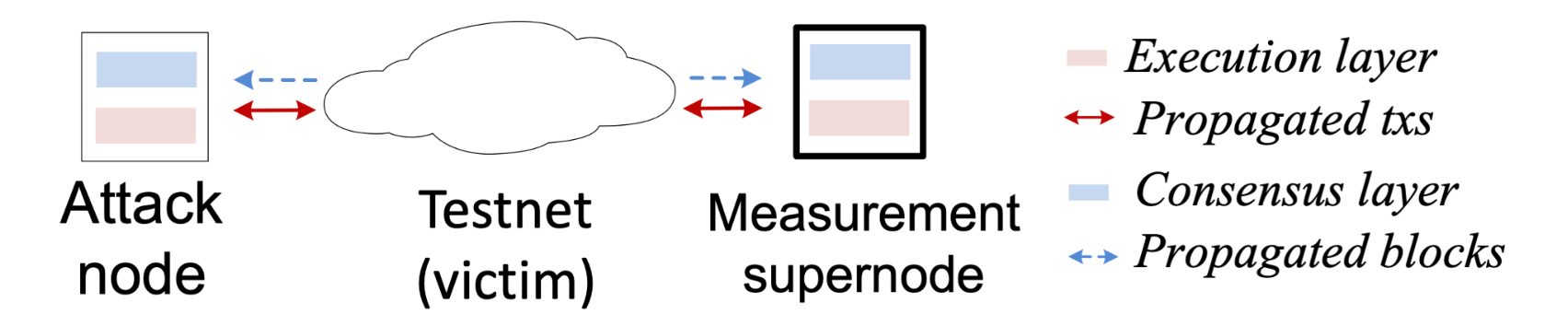}
\caption{Experimental setup for evaluation of turning attacks on Goerli testnet}
\label{fig:exp:setup5}
\end{figure}
}

To monitor transaction propagation, we launch an independent supernode (called measurement node) in the same testnet. The measurement node runs a reconfigured Geth client where the limit of peers/neighbors is removed and the measurement node can connect to as many neighbors as possible, hence a super node. 
The measurement node is not (directly) connected to the attacker node. When setting up our experiments, we ran the measurement node in Goerli for seven days and found the node is stabilized at $290$ neighbors. 

\ignore{
\begin{figure*}[!htbp]
 \centering
  \subfloat[Transaction propagation in six messages]{%
 \includegraphics[width=0.295\textwidth]{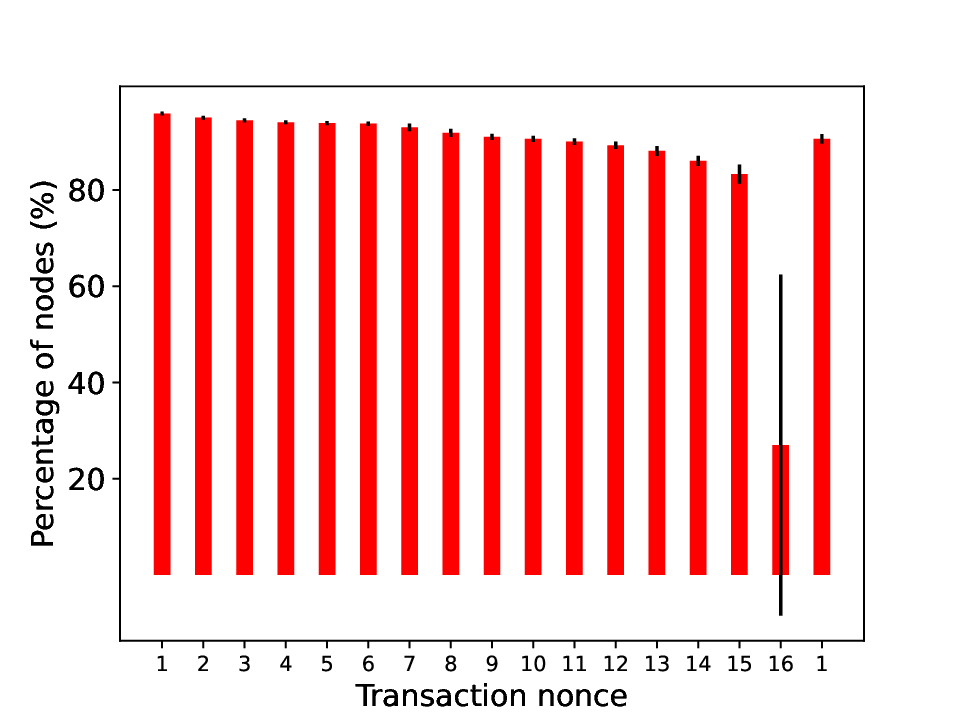}
 \label{fig:testnet:rinkeby_ed4_percentage}}
  \subfloat[Transaction propagation in two messages]{%
 \includegraphics[width=0.295\textwidth]{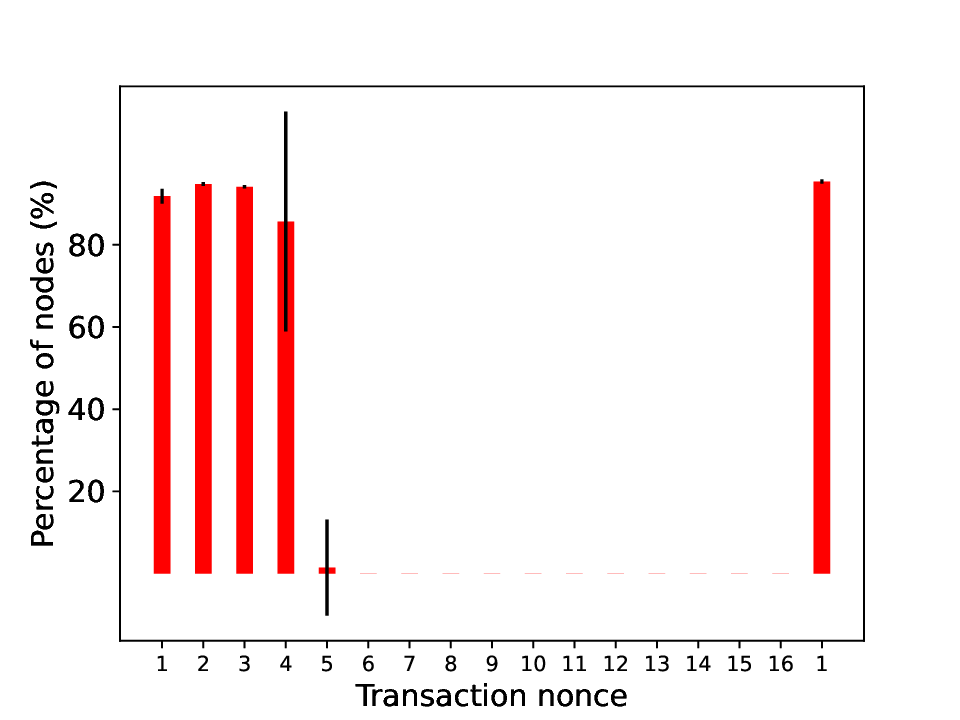}
 \label{fig:testnet:rinkeby_ed4_percentage_1msg}}
  \subfloat[Etherscan screenshot of the blocks generated during the attack on Goerli (txs in six messages)]{%
 \includegraphics[width=0.325\textwidth]{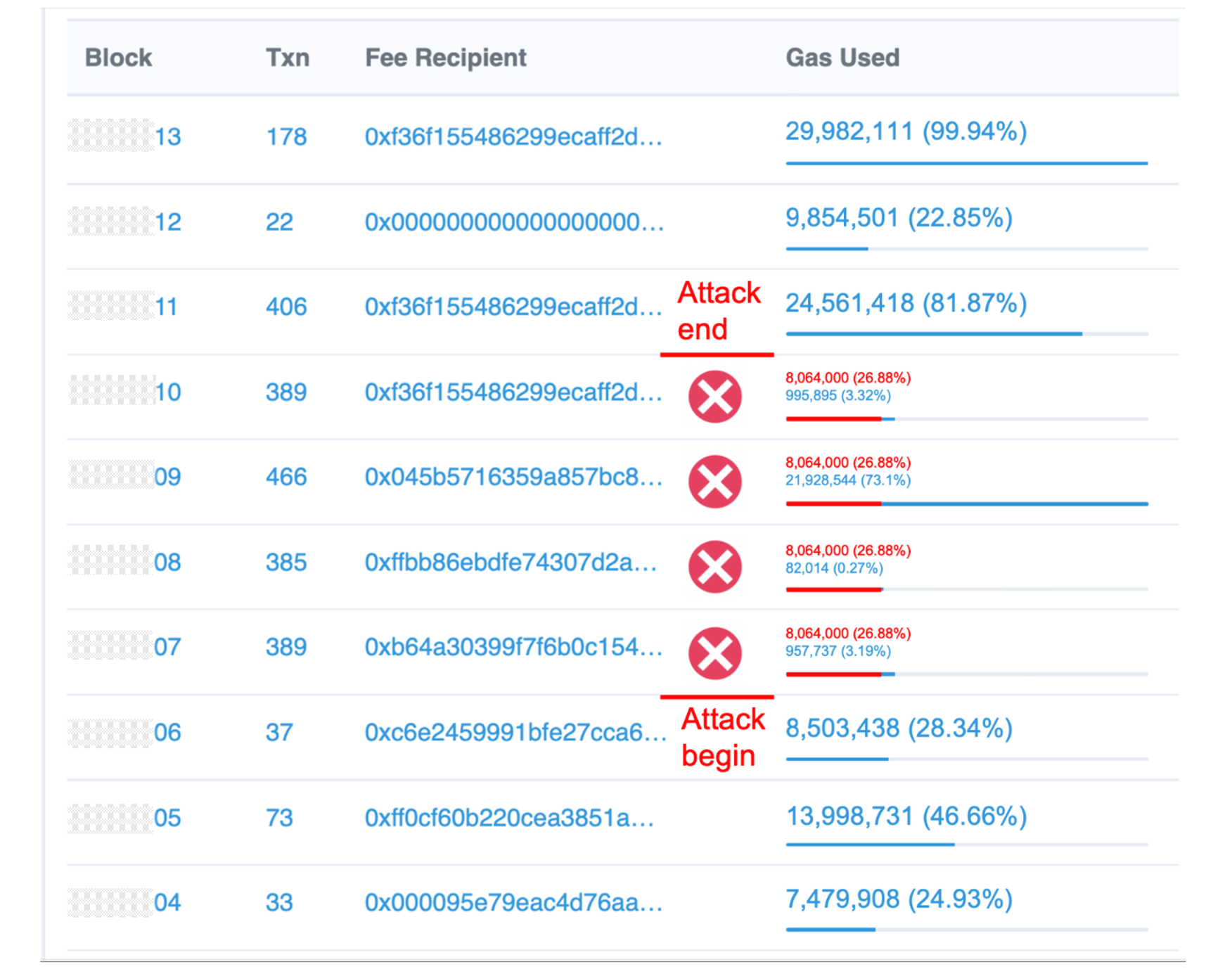}
 \label{fig:testnet:rinkeby_ed4}
  }%
 \caption{Mounting $XT_4$ attacks on Goerli: Emptying blocks and propagating malicious transactions to the entire network.}
\end{figure*}

\begin{table}[!htbp]
  \caption{Types of transactions included in the four blocks during the attack in Figure~\ref{fig:testnet:rinkeby_ed4}. `\# private tx' means the number of private transactions.}
  \label{tab:privatetransactions2}
  \centering{\small
  \begin{tabularx}{0.425\textwidth}{ |l|X|X|X| }
  \hline
  \# block & \# private txs & \# public txs & \# attack txs \\ \hline 
  $8040010$ & $5$ & $0$ & $384$ \\ \hline 
  $8040009$ & $78$ & $4$ & $384$ \\ \hline 
  $8040008$ & $1$ & $0$ & $384$ \\ \hline 
  $8040007$ & $5$ & $0$ & $384$ \\ \hline 
\end{tabularx}
}
\end{table}
}

\begin{figure}[!bthp]
\centering
\includegraphics[width=0.30\textwidth]{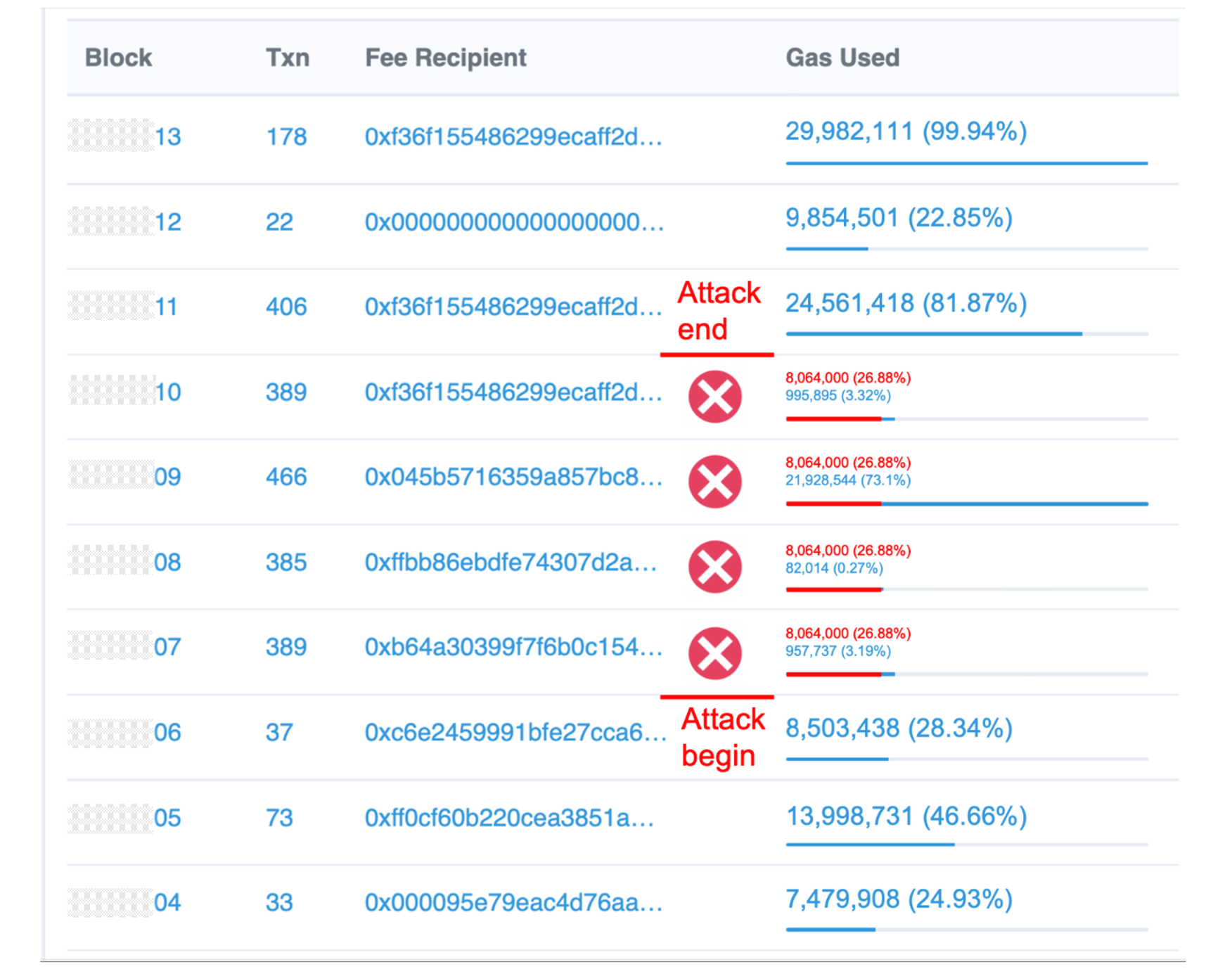}
\caption{Mounting $XT_4$ attacks on Goerli: Etherscan screenshot of the blocks generated during the attack}
\label{fig:testnet:rinkeby_ed4}
\end{figure}

\noindent{\bf Experiment results}: The instrumented measurement node is able to log the received messages from different neighbors; these messages include those of transactions and of transaction hashes. The measurement node could receive the same transaction from different neighbors, and the log stores the transaction-neighbor pairs. We started logging the received messages one month before the experiment.

To do an experiment, we make the attacker node send $XT_4$ transactions using $384$ accounts. Each account sends $16$ valid pending transactions, followed by a replacement transaction. In total, the attacker node sends $6144$ valid transactions and $384$ replacement transactions. The \textit{GasPrice} of the valid and replacement transactions are set to be $8$ Gwei and $130$ Gwei, respectively. Finally, we wrap up the experiment by waiting after the attacker node sends all messages and all replacement transactions are included in the blockchain. 

\ignore{
\begin{table}[!htbp]
      \caption{Types of transactions included in the four blocks during the attack in Figure~\ref{fig:testnet:rinkeby_ed4}. `\# private tx' means the number of private transactions.}
      \label{tab:privatetransactions}
      \centering{\small
      \begin{tabularx}{0.425\textwidth}{ |l|X|X|X| }
      \hline
      \# block & \# private txs & \# public txs & \# attack txs \\ \hline 
      $8040010$ & $5$ & $0$ & $384$ \\ \hline 
      $8040009$ & $78$ & $4$ & $384$ \\ \hline 
      $8040008$ & $1$ & $0$ & $384$ \\ \hline 
      $8040007$ & $5$ & $0$ & $384$ \\ \hline 
    \end{tabularx}
    }
\end{table}
}


Figure~\ref{fig:testnet:rinkeby_ed4} shows the generated blocks in the experiment. We took the screenshot from \url{etherscan.io} and labeled it in red with information regarding the attack. Before the attack begins, the testnet normally utilizes $24-47\%$ of the Gas in a block. For ethics, our attacks are short (lasting four blocks). 

Right after the launch of the attack, in block $xx07$, the Gas used by normal transactions drops to $3.19\%$ and the Gas used by adversarial transactions (denoted by red bars in Figure~\ref{fig:testnet:rinkeby_ed4}) is $26.88\%$. The included adversarial transactions are $384$ replacement transactions sent in the second round of $XT_4$. The $6144-384=5760$ child transactions sent in the first round are not included in the block. Other blocks during the attack, namely $xx08$ and $xx10$, are similar. In the third block $xx09$, $73.1\%$ Gas is spent on including $82$ normal transactions.
To explain it, we inspect the raw history of transactions arriving at and logged by our measurement supernode and found that out of the $82$ included normal transactions, only $4$ are present in the log, implying the other $78$ normal transactions are private ones that were not broadcast to the measurement node. 
%
Notice that our attacks require no discovery of critical nodes as needed in~\cite{DBLP:conf/ccs/LiWT21}. 

  \subsection{Evaluation of Found Locking Exploits on Reth}
\label{sec:eval:locking:xt8a}
Given Exploit $XT_8$ found on Reth, we manually extended it to two actual exploits on unmodified Reth's mempool, $XT_{8a}$, and $XT_{8b}$. 

  \subsubsection{Exploit Extension}

\noindent{\bf Mempool in Reth $v0.1.0-alpha.4$}: Recall that the mempool is a FIFO queue: Any transaction residential in the mempool is never evicted, no matter how high an arriving transaction's fee is. Besides, when there is an empty slot in its mempool, it requires that the transaction admitted must have fees higher than the latest block's base fee. 

\noindent{\bf 
Actual exploit $XT_{8a}$}: In $XT_8$, the attack cost increases with the block base fee. We manually propose a method to decrease the block base fee in Ethereum by mounting an eviction attack on the previous block. Specifically, in Ethereum (after EIP1559), given a recently produced block $i$, the base block fee of block $i+1$ is calculated dynamically as follows: 

{\fontsize{9.2pt}{9.2pt}\selectfont
\begin{align}
BasePrice(i+1)&=&BasePrice(i) * [\frac{7}{8} + \frac{1}{4} * \frac{\textit{GasUsed}(i)}{BlockLimit}]
\end{align}
}

Therefore, if an eviction attack can persistently lower the Gas utilization of recent blocks, the current block fees can be reduced, which decreases the costs of locking attacks on the current block. In Exploit $XT_{8a}$, we mount a series of eviction attacks first and then a locking attack against the Reth node. 

Specifically, suppose in a network of a Reth node and a block validator node, the $XT_{8a}$ attacker first keeps sending eviction attacks directly to the validator node for several consecutive blocks. When observing the block base fees drop sufficiently, the attacker mounts the regular locking attack $XT_{8}$ against the Reth node.

  \subsubsection{Attack Evaluation on Reth}
\label{sec:evaluate:locking}
\noindent{\bf Experimental settings}: 
We set up an experiment platform for evaluating locking attacks, on which an attacker node is connected to a victim non-validator node, which is also connected to a workload generator node and a victim validator node. The attacker node also maintains a direct connection to the victim validator. The network topology is depicted in Figure~\ref{fig:exp:setup3}. The non-validator node runs the Geth $v1.11.4$ client.

\begin{figure}[!bthp]
\centering
\includegraphics[width=0.425\textwidth]{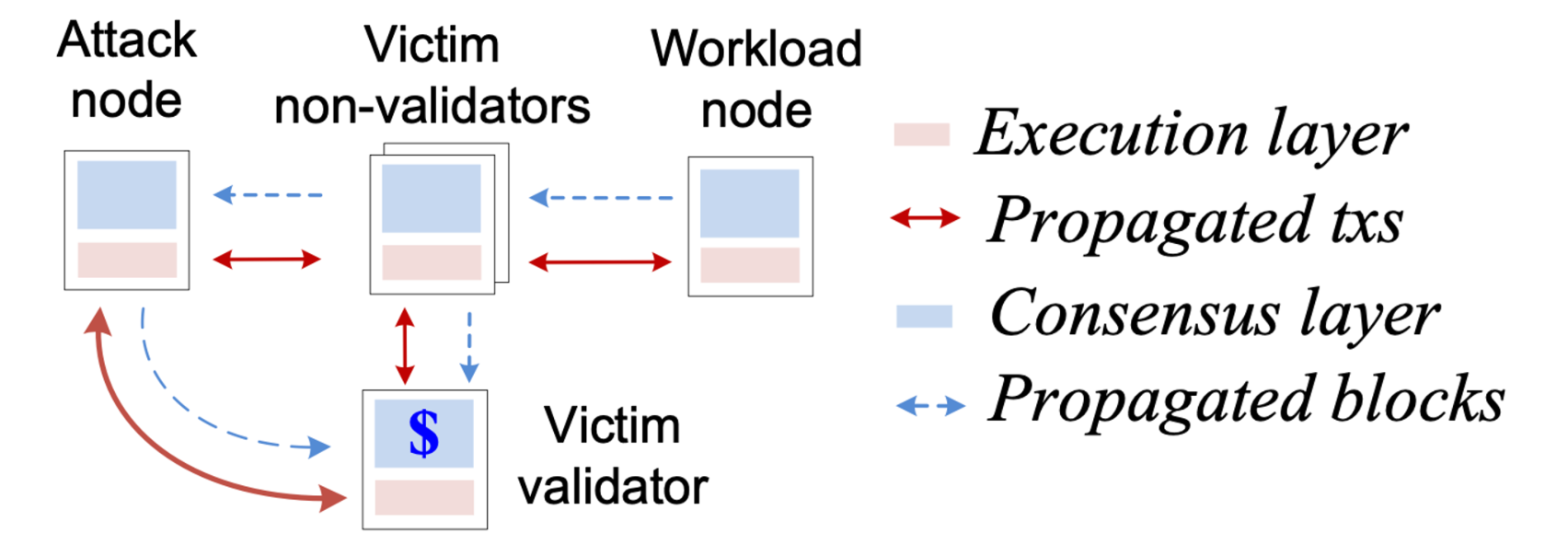}
\caption{Experimental setup for locking attacks on Reth}
\label{fig:exp:setup3}
\end{figure}


\begin{figure}[!htb]
  \centering
  \includegraphics[clip,width=0.8\columnwidth]{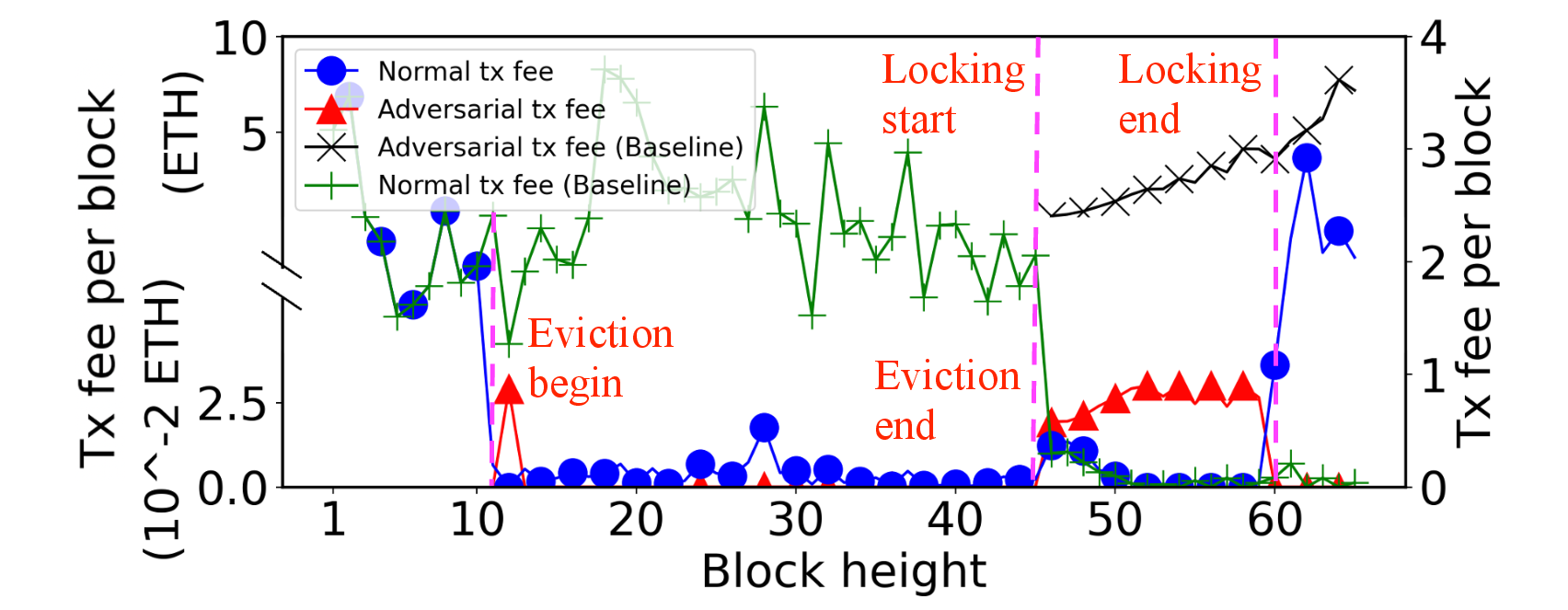}%
  \caption{Evaluation of locking attack $XT_{8a}$ on Reth}%
  \label{fig:reth-lock}
\end{figure}

\noindent{\bf 
Evaluation of $XT_{8a}$}: To evaluate $XT_{8a}$, we set up the workload generator that sends normal transactions to the validator Geth node. The non-validator Reth node has not yet joined the network. We let the attacker node first send eviction attacks ($XT_6$) directly to the victim validator for $35$ consecutive blocks until it observes the block base fees drop sufficiently to $1$ Gwei, which occurs at the $45$-th in our experiment. The Reth node then joins the network with an empty mempool. The attacker node mounts a locking attack $XT_8$ with transactions of price $5$ Gwei to occupy the Reth node's mempool.

We report the total fees of normal transactions included in the blocks – the lower the fees are, the more successful the locking attack is. We also report the adversarial transaction fees as the attack cost shown in Figure~\ref{fig:reth-lock}. When the locking attack begins at the $45$-th block, the normal transaction fees increase because of the empty mempool on the Reth node which accepts some normal transactions in addition to the adversarial locking transactions. Then, after three blocks, the normal transaction fees quickly drop to near-zero Ether, showing the success of locking attacks. Interestingly, unlike the eviction attacks that cause low Gas utilization per block, the locking attacks just reduce the total Ether fees without reducing Gas utilization (i.e., the blocks produced under locking attacks from height $45$ to $60$ use the Gas of almost $100\%$ block limit as shown in Figure~\ref{fig:reth-gas}). Due to the high Gas utilization, block base fees keep increasing during the locking attack. When reaching the $60$-th block, the base fee is higher than the maximal transaction price tolerable by the attacker. In our experiment, the locking attacker stops at $60$-th block, and the normal transaction fees immediately recover to the ``normal'' level. In practice, the attacker can repeat sending eviction attacks to reduce the base fee before mounting the locking attack again.

{\color{violet}

\begin{figure}
  \centering
  \includegraphics[width=0.3\textwidth]{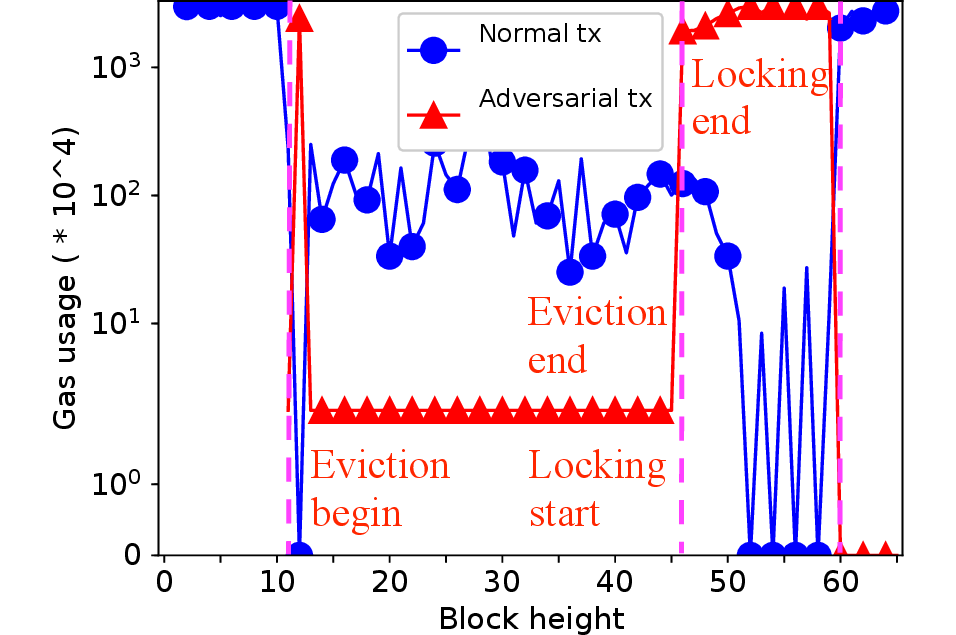}
  \caption{Gas usage of attacks $XT_{8a}$}
  \label{fig:reth-gas}
\end{figure}
Figure~\ref{fig:reth-gas} shows the Gas usage of attack transactions and normal transactions in $XT_{8a}$. During the locking attack from block height $45$-th to $60$-th, the attacker sends locking transactions to occupy all the empty slots. It makes the Gas usage of the attacker almost 100\% block limit while the Gas usage of the normal transactions drops to around $0$. When empty slots are created by arriving blocks, normal transactions compete with the attack transactions to admit into the empty slots. Thus, some normal transactions are included in blocks during the locking attack.

We also manually designed another locking variant for Reth. Suppose the Reth node does not run block validation, and its mempool is initially synchronized with the mempool of a block validator. The attacker first evicts $10,000$ normal transactions from the validator's mempool. Then, the Reth node's mempool is permanently locked by the $10,000$ normal transactions evicted from the validator mempool. Because these 10,000 transactions will never be included in blocks, they would exist in the Reth node's mempool forever, permanently locking it in its current state.

\begin{figure}[!htb]
  \centering
  \includegraphics[clip,width=0.72\columnwidth]{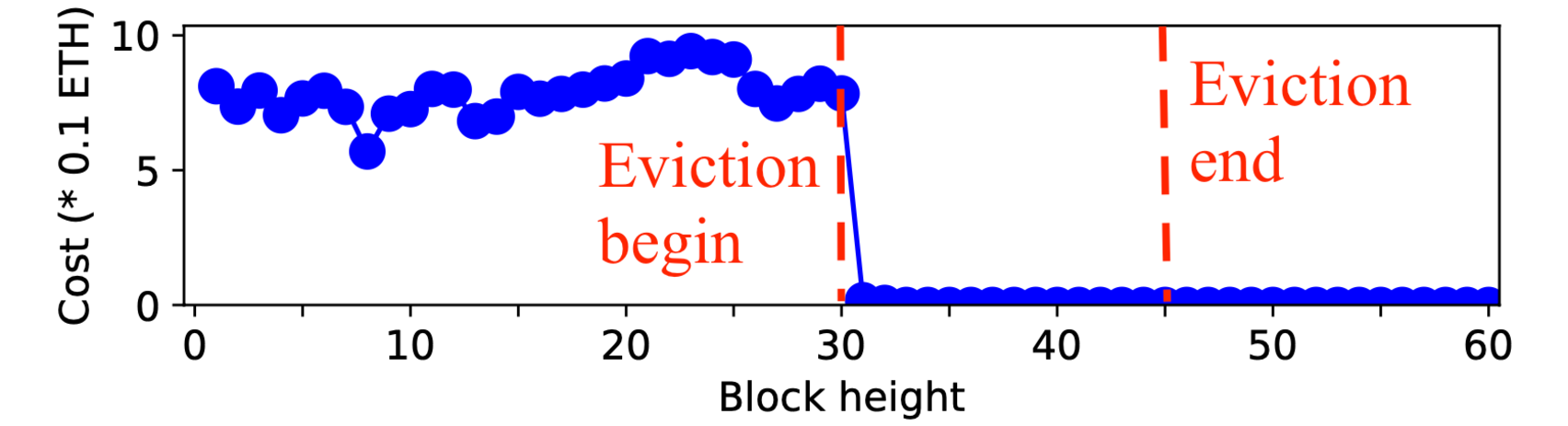}%
\caption{Evaluation of locking attack $XT_{8b}$ on Reth}%
\label{fig:side-effect-locking}
\end{figure}

\noindent{\bf 
Evaluation of $XT_{8b}$}: To evaluate $XT_{8b}$, we initially set up the attacker node connected to the validator node running Geth, the workload generator node connected to the victim non-validator Reth node, and the Reth node connected to the validator. The system is run long enough to produce a certain number of blocks and the normal transactions are propagated to both validator and on-validator nodes' mempools. We then mount the eviction attack ($XT_6$) directly to the validator Geth node that evicts the mempool transactions on the validator node and prevents them from being included in the blocks. 

By mounting eviction attacks between the $30$-th to $45$-th blocks in Figure~\ref{fig:side-effect-locking}, the Reth mempool is successfully locked afterward: From the figure, after the $45$-th block, the total fees of normal transactions included in blocks remain at zero Ether per block until the experiment ends on the $60$-th block.

\subsection{Evaluation of Locking on Geth}
\label{sec:evaluate:locking2}
\begin{figure}[!bthp]
\centering
\includegraphics[width=0.375\textwidth]{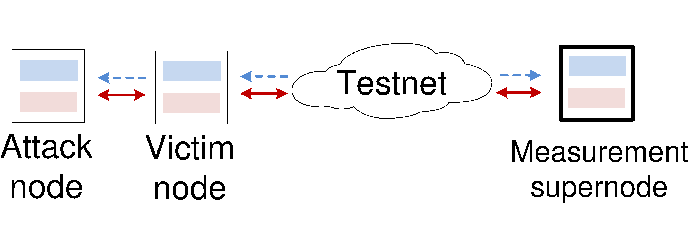}
\caption{Experimental setup for locking attacks on Geth/OpenEthereum}
\label{fig:exp:setup4}
\end{figure}

\ignore{
\begin{center}
\fbox{\parbox{0.90\linewidth}{RQ2. What's the success rate and attack cost in locking a remote mempool secured under turning attacks?
}}\end{center}
}

\begin{figure}[!ht]
 \centering
  \subfloat[Attack damage]{%
  \includegraphics[width=0.245\textwidth]{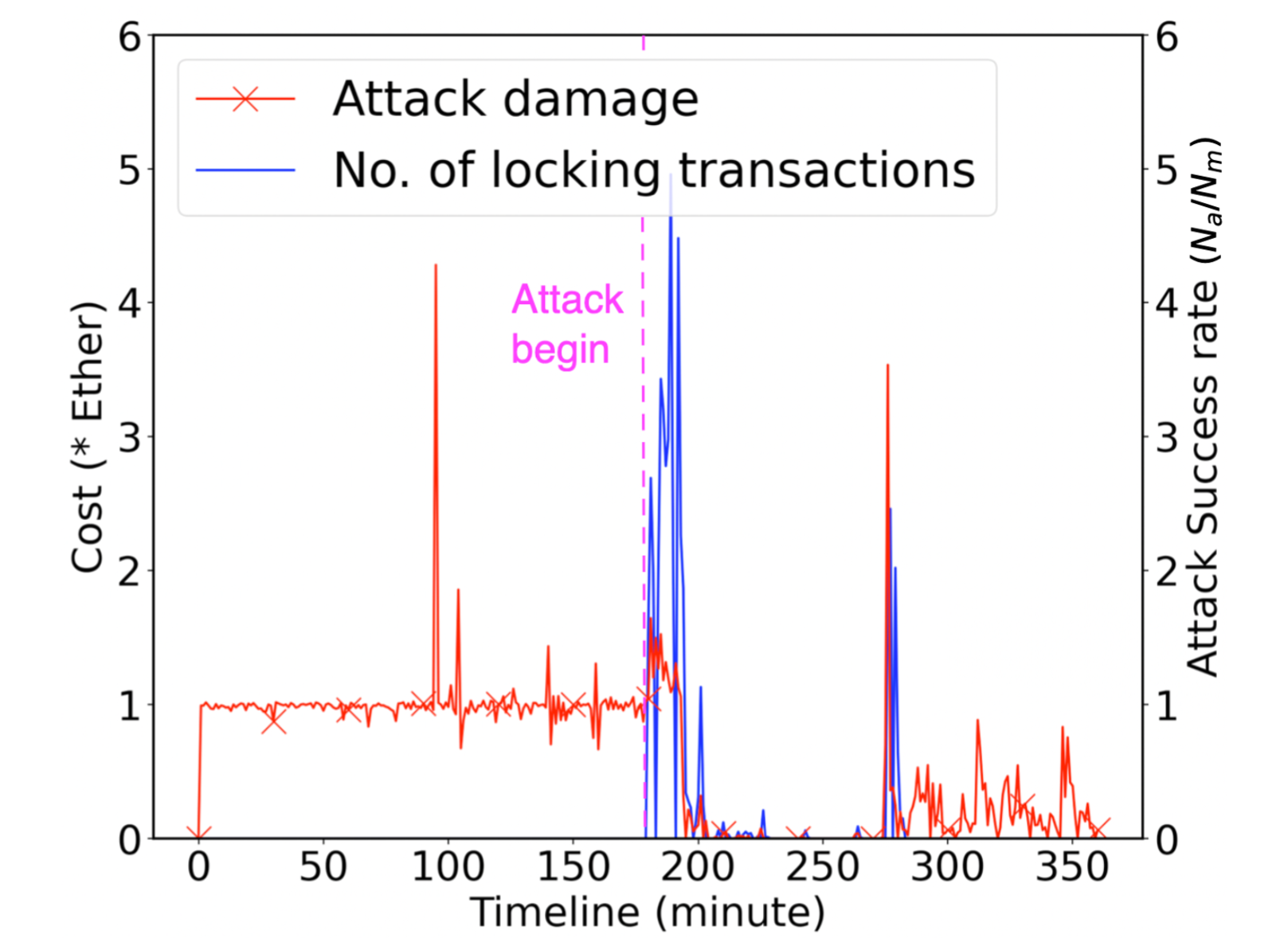}
 \label{fig:ld5:x}}
  \subfloat[Attack cost]{%
 \includegraphics[width=0.245\textwidth]{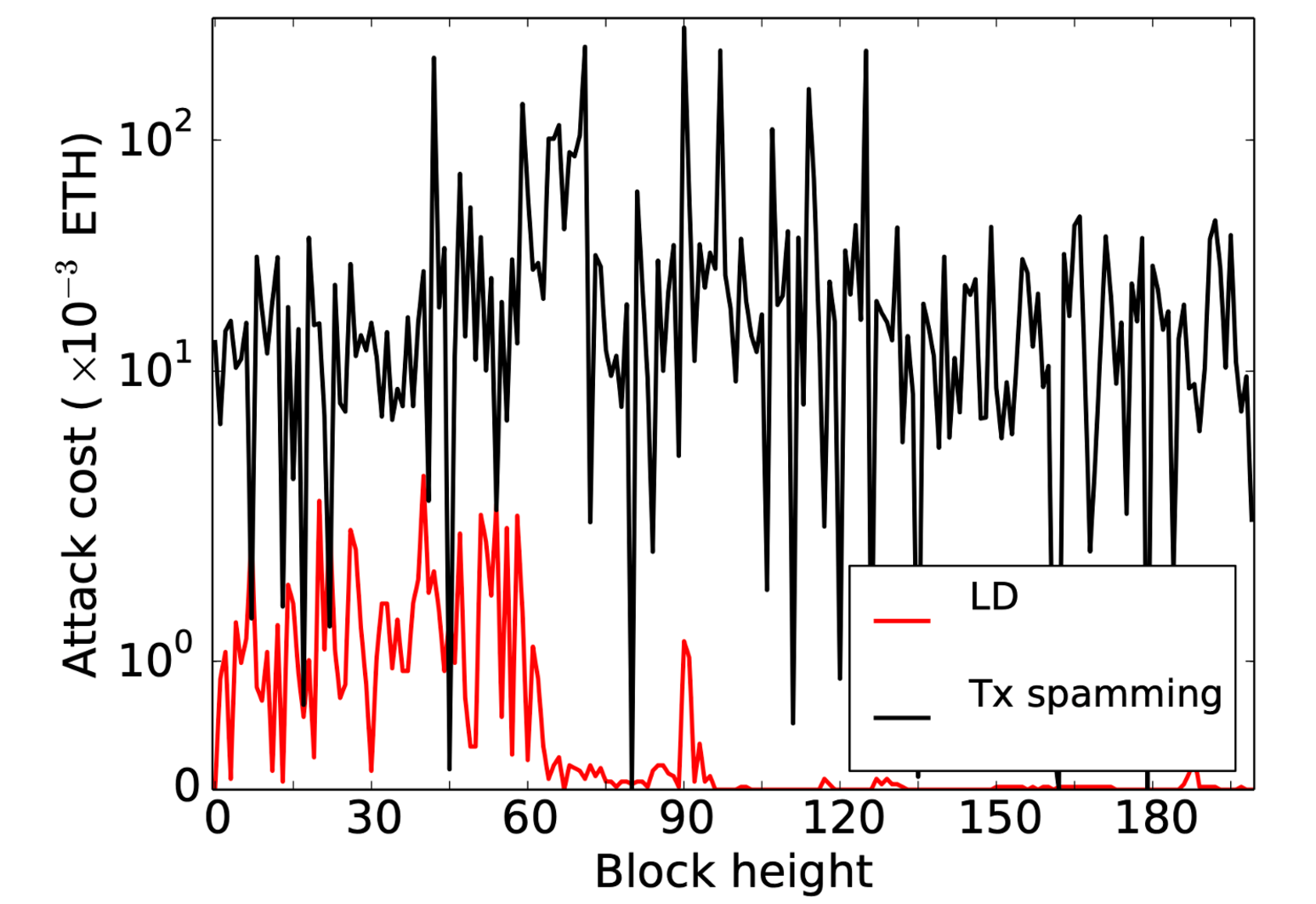}
 \label{fig:ld5:3}
  }%
 \caption{Mounting locking attacks on a Geth node in testnet.}
\end{figure}
\ignore{}

\noindent{\bf 
Design rational}: Recall that OpenEthereum adopts an overly strict policy that determines eviction by only considering the fees of childless transactions. While this policy is intended to prevent turning attacks ($XT_5$), it is vulnerable to locking attacks. Because OpenEthereum has been deprecated from the operational Ethereum 2.0 networks, we re-implement its admission policy on Geth $v1.11.4$ and run the turning-hardened Geth on a victim under locking attacks.

\noindent{\bf Settings}: 
We set up an experimental platform for evaluating locking attacks on Goerli testnet. The victim node runs the turning-hardened Geth client (based on $v1.11.4$) described above. The attacker node runs an instrumented Geth client (based on $v1.11.4$) at the execution layer such that it can mount locking attacks. Specifically, the Geth instrumentation enables the attacker node to monitor the local mempool and send a crafted transaction to the victim node upon each new empty slot created (by block arrival). The network topology is depicted in Figure~\ref{fig:exp:setup4}.

In the experiment, we first run our three nodes for $350$ minutes and then turn on the locking attack on the attacker node. We collect the benign transactions received by the attacker node (the number of which is denoted by $N_a$), the crafted transactions the attacker sends (the number denoted by $N_c$), and the benign transactions received by the measurement node (the number denoted by $N_m$). We use Metric $\frac{N_a}{N_m}$ to report the (inverse) of the attack success rate -- Under the same transactions received by the measurement node, the fewer transactions the attacker node receives from the victim node, the more successful the victim mempool is locked. 

\noindent{\bf 
Evaluation of locking attacks}: 
We report in Figure~\ref{fig:ld5:x} the locking attack's success rate over the period of $350$ minutes. Each tick on the $X$ axis represents $1$ minute; that is, $N_a$/$N_m$ is the number of benign transactions received by the attacker/measurement node every $1$ minute. It is clear that the locking attack is highly successful -- Right after the attack is launched at the $180$-th minute, the rate of normalized transactions received from the victim node drops essentially to zero. And it persists over the period where the attack lasts.

We also calculate the attack cost by the total fees of the attacker's transactions included in the blockchain. In the experiment, we wait three hours after the attack stops and consider all the blocks produced until then. Not a single attacker's transaction sent in our experiment is included in the blockchain. This may be due to the low fee with which crafted transactions are sent.  

Understanding crafted transaction inclusion is nondeterministic and dependent on benign transactions, we estimate the worst-case attack cost, which is the total fees of crafted transactions sent by the attacker (i.e., when all the crafted transactions are included in the blockchain). Figure~\ref{fig:ld5:3} reports the worst-case locking attack cost over time; on average, the attack, as successful as reported in Figure~\ref{fig:ld5:x}, consumes $0.0004$ Ether per block. By comparison, it is $0.028$ Ether of transactions included per block, which is the minimal cost of the baseline transaction-spamming attack~\cite{DBLP:conf/fc/BaqerHMW16}.

}

{\color{violet}

\subsection{Evaluation of Found Exploits on Single Victim Node}
\label{sec:setup:singlenode}

\begin{figure}[!bthp]
\centering
\includegraphics[width=0.375\textwidth]{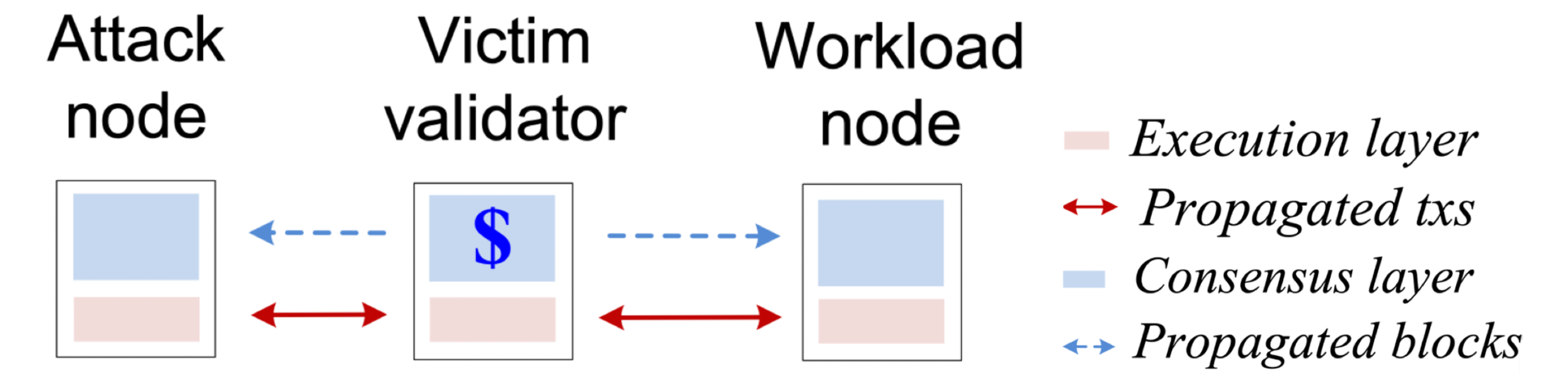}
\caption{Experimental setup}
\label{fig:exp:setup}
\end{figure}

\label{appdx:sec:eval:local:1}
\noindent{\bf Evaluation settings}:
Our goal is to evaluate the success rate and cost of different ADAMS attacks on a single victim node. Because some ADAMS attacks are sensitive to the workload of normal transactions, we first collect transaction workloads from the mainnet. Specifically, we instrumented a Geth client (denoted by Geth-m) to log every message it receives from every neighbor. The logged messages contain transactions, transaction hashes (announcements), and blocks. When the client receives the same message from multiple neighbors, it logs it as multiple message-neighbor pairs. 
We also log the arrival time of a transaction or a block.

{\it Workload collection}: 
We launched a Geth-m node in the mainnet on May 17, 2023, turned on logging for $5$ hours, and collected the transactions propagated to it. We make the collected transactions replayable as follows: We use the account balances and nonces on the mainnet to set up the initial state locally. We then replace the original sender in the collected transactions with the public keys that we generated. By this means, we know the secret keys of transaction senders and are able to send the otherwise same transactions for experiments. 

For experiments, we set up three nodes, an attack node sending crafted transactions, a workload node sending normal transactions collected, and a victim node receiving transactions from the other two nodes. The victim node is connected to both the attack and workload nodes. There is no direct connection between the attack node and the workload node. The attacker node runs an instrumented Geth $v1.11.4$ client (denoted by Geth-a) that can propagate invalid transactions to its neighbors. The victim node runs the tested Ethereum client. The workload node runs a vanilla Geth $v1.11.4$ client. On each node, we also run a Prysm $v3.3.0$ client at the consensus layer. The experiment platform is denoted in Figure~\ref{fig:exp:setup}.
Among the three nodes, we stake Ether to the consensus-layer client on the victim node, so that only the victim node would propose or produce blocks. 

In each experiment, we first run the above ``attacked'' setup (i.e., with victim, attack and workload nodes). We then run a ``regular'' setup that excludes the attack node. Under the regular setup, the workload node sends the normal transactions and blocks to the victim node. We compare the experiment results under the attacked setup and regular setup to show the success of the attack.


\ignore{
\begin{center}
\fbox{\parbox{0.90\linewidth}{RQ1. What's the success rate and attack cost in mounting turning-based exploits against a victim node running leading clients such as Geth, Erigon and Nethermind? 
}}
\end{center}
}

\noindent{\bf 
Evaluation of eviction/turning attacks on Single Node}:
We set up the experiment platform described in \S~\ref{sec:setup:singlenode}. In each experiment, we drive benign transactions from the workload node. Note that the collected workload contains the timings of both benign transactions and produced blocks. On the $30$-th block, we start the attack. Recall that each attack is configured by delay $d$; the attack node observes the arrival of a produced block and waits for $d$ seconds before sending a round of crafted transactions. 

The attack phase lasts for $40$ blocks; after the $70$-th block, we stop the attack node from sending crafted transactions. We keep running workload and victim nodes for another $50$ blocks and stop the entire process at the $120$-th block. We collect the blocks produced and, given a block, we report two metrics: 1) total fees of benign transactions included, 2) total fees of attack transactions included.

In each experiment, we also re-run the workload and victim nodes with the same setup. In this ``no-attack'', we don't run the attack node. We collect the blocks produced, and, given a block, we report two metrics: 3) total fees of transactions included (denoted as ``Benign ops - no attack''), and 4) the cost of a baseline spamming attack with 100\% success rate (denoted as ``Spamming - analytical''). For the latter, given a block, we select the transaction of the highest price, and report the price multiplied by the Ethereum block Gas limit. 


\begin{figure*}[!ht]
  \centering
  \subfloat[$XT_6$ w. $8$-sec. delay (single node)]{%
    \includegraphics[width=0.245\textwidth]{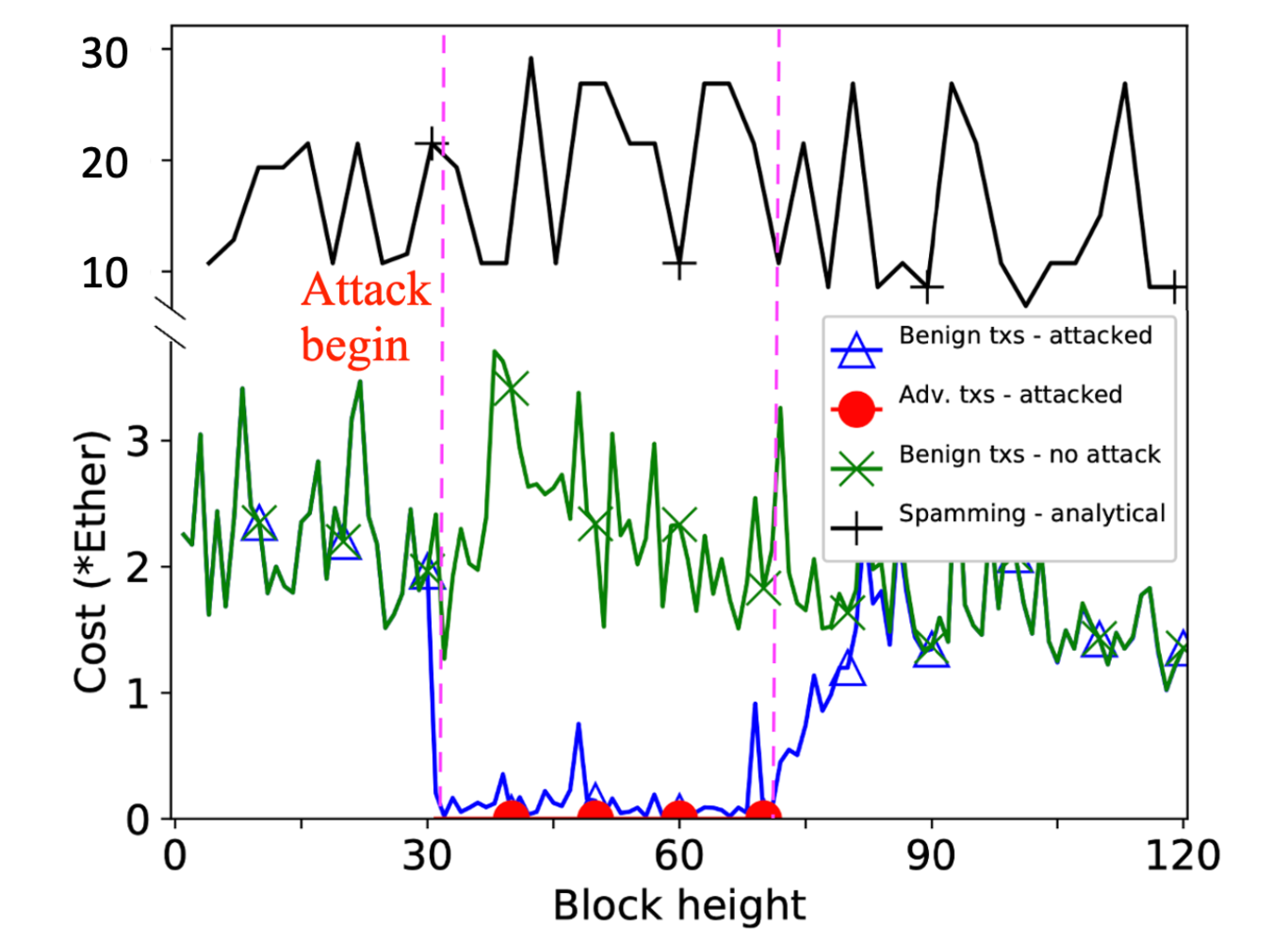}
    \label{fig:successrate-geth}}%
  \subfloat[$XT_6$ w. varying delay (single node)] {%
   \includegraphics[width=0.245\textwidth]{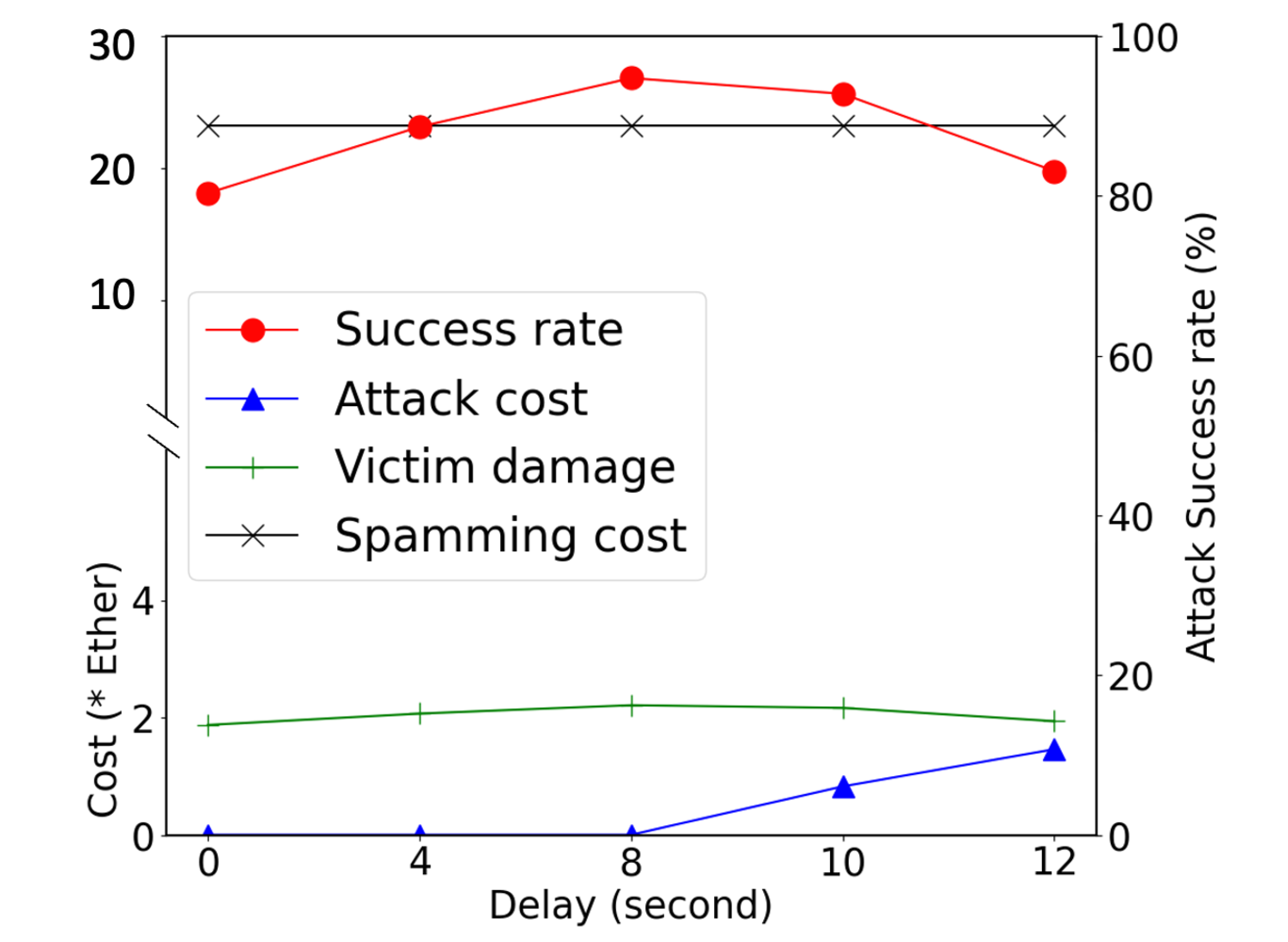}
  \label{fig:delay-geth}}
\ignore{
  \subfloat[Attack with $0$-second delay]{%
   \includegraphics[width=0.245\textwidth]{figures/nethermind-local-new2.eps}
  \label{fig:successrate-nethermind}}%
}
  \subfloat[$XT_4$ on Nethermind w. varying delay (single node)] {%
   \includegraphics[width=0.245\textwidth]{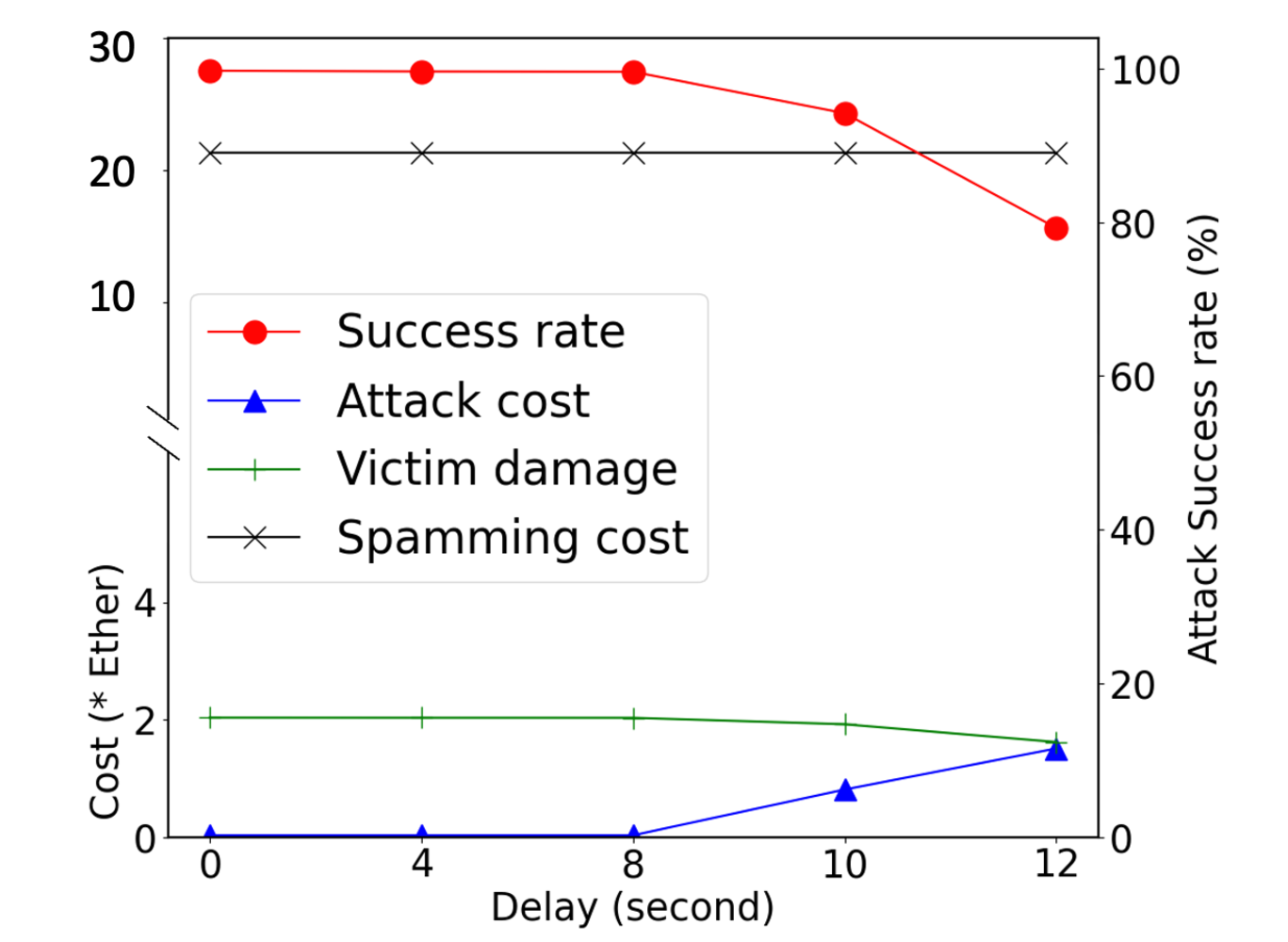}
  \label{fig:delay-nethermind}}
\ignore{
  \subfloat[Propagation Attack with $8$ seconds delay]{%
    \includegraphics[width=0.237\textwidth]{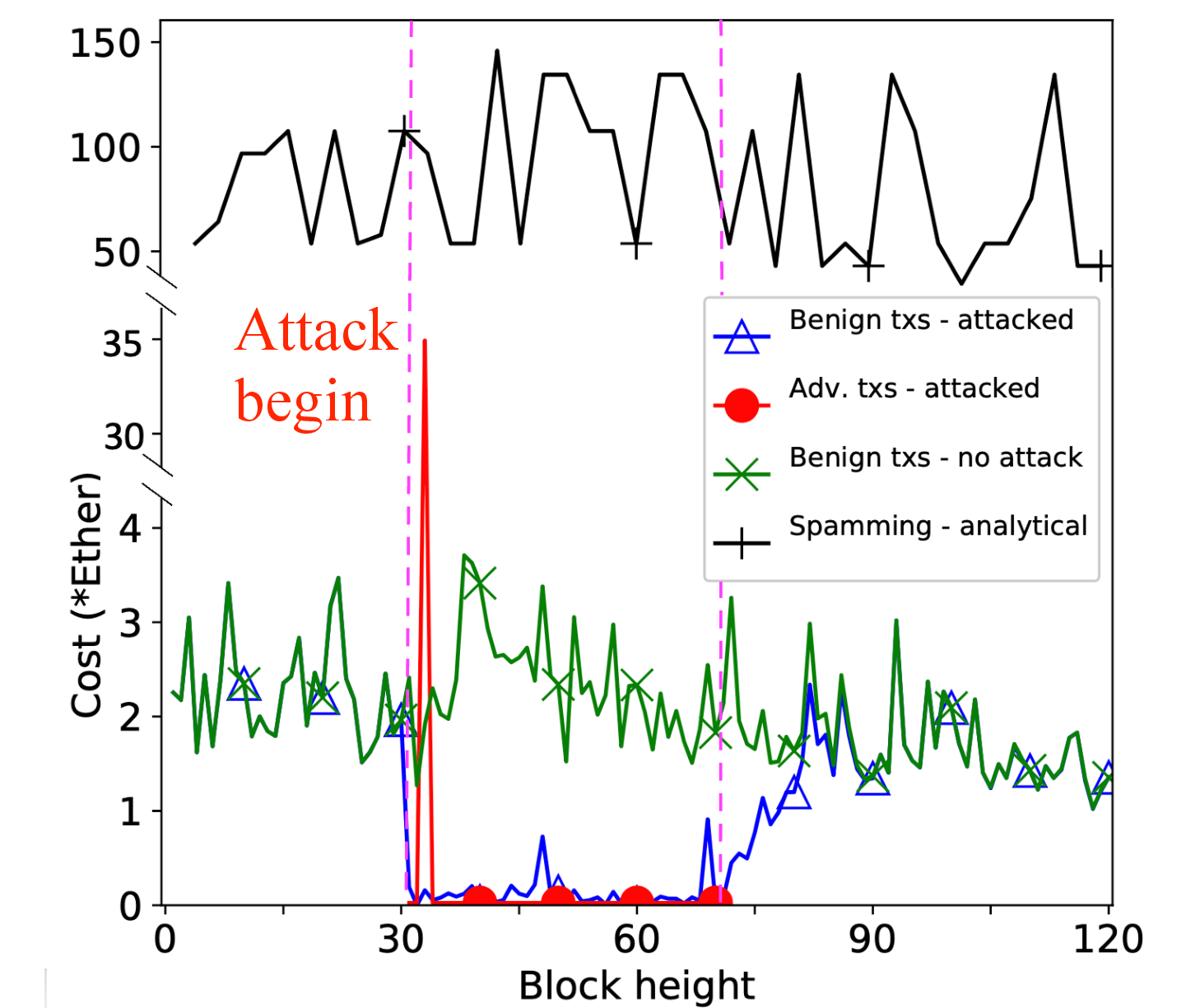}
    \label{fig:successrate-geth-propagate}}%
}
  \subfloat[$XT_6$ w. varying delay ($6$ nodes)] {%
   \includegraphics[width=0.245\textwidth]{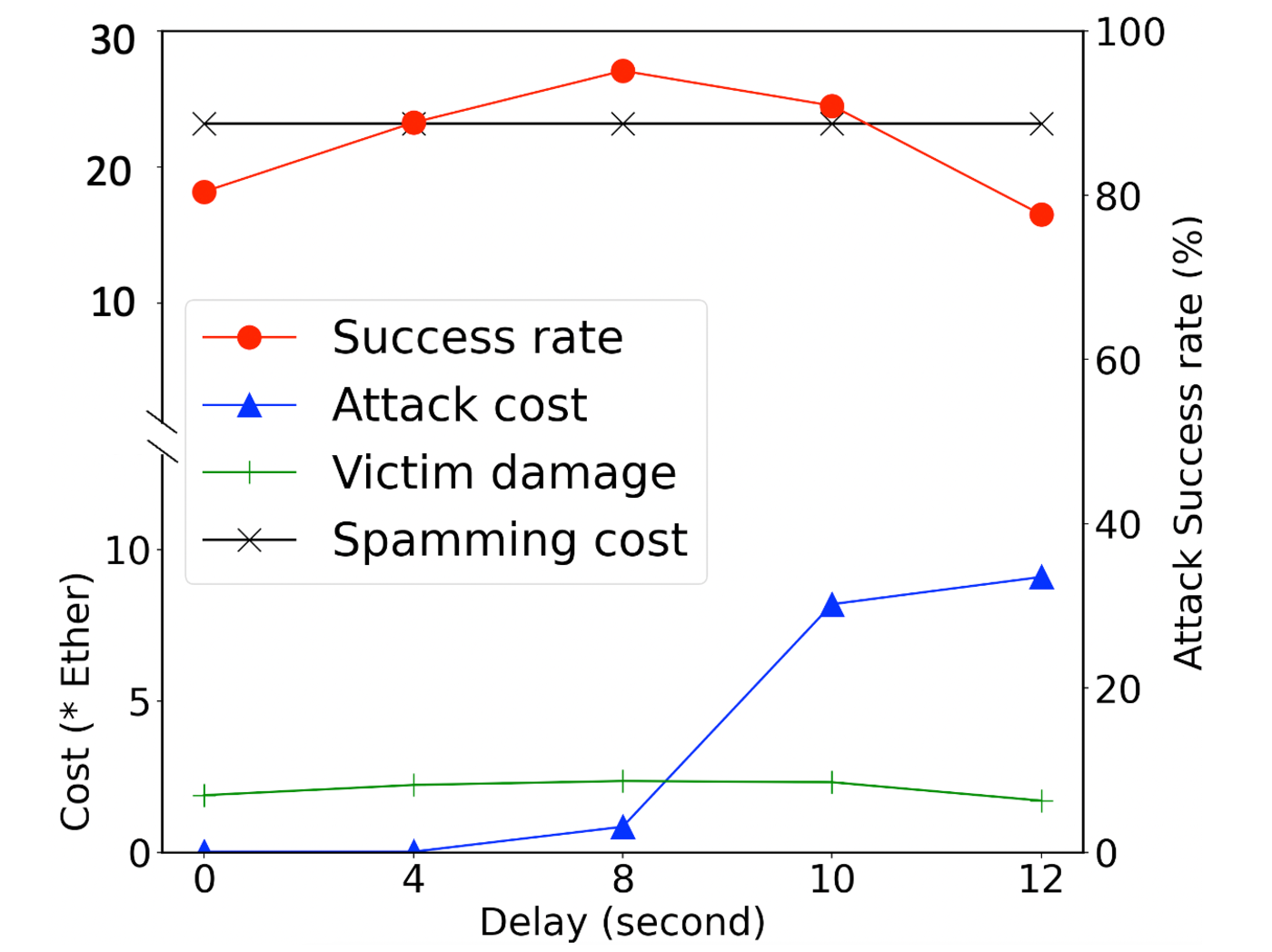}
  \label{fig:delay-geth-propagate}
  \label{fig:turning:multi}
}
\hfill
\bigskip
  \subfloat[$XT_4$ on Geth w. varying delay\\\hspace{\textwidth} (single node)]{%
    \includegraphics[width=0.24\textwidth]{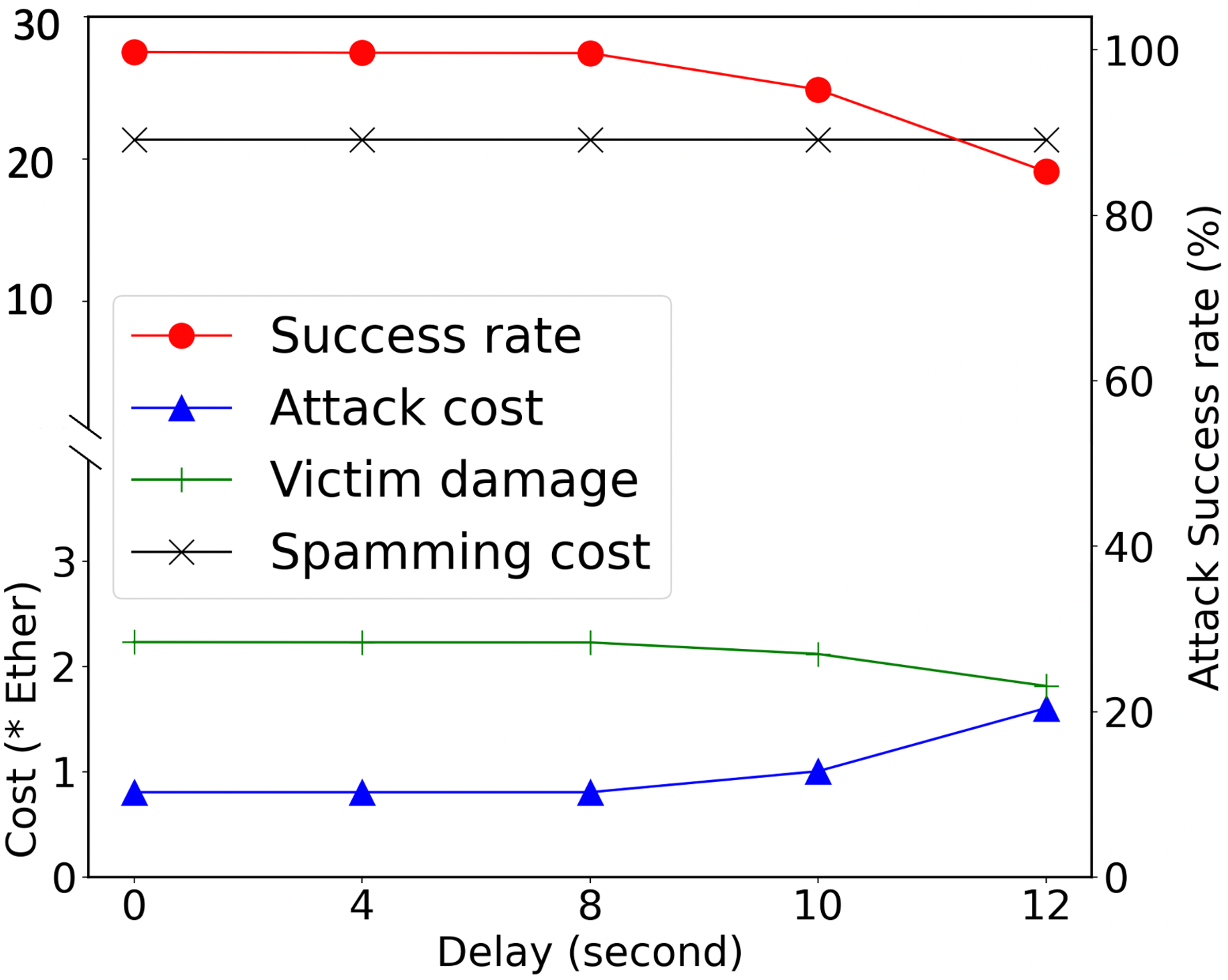}
    \label{fig:trade-off-xt4-geth}}%
  \subfloat[$XT_4$ on Erigon w. varying delay \\\hspace{\textwidth}(single node)] {%
   \includegraphics[width=0.24\textwidth]{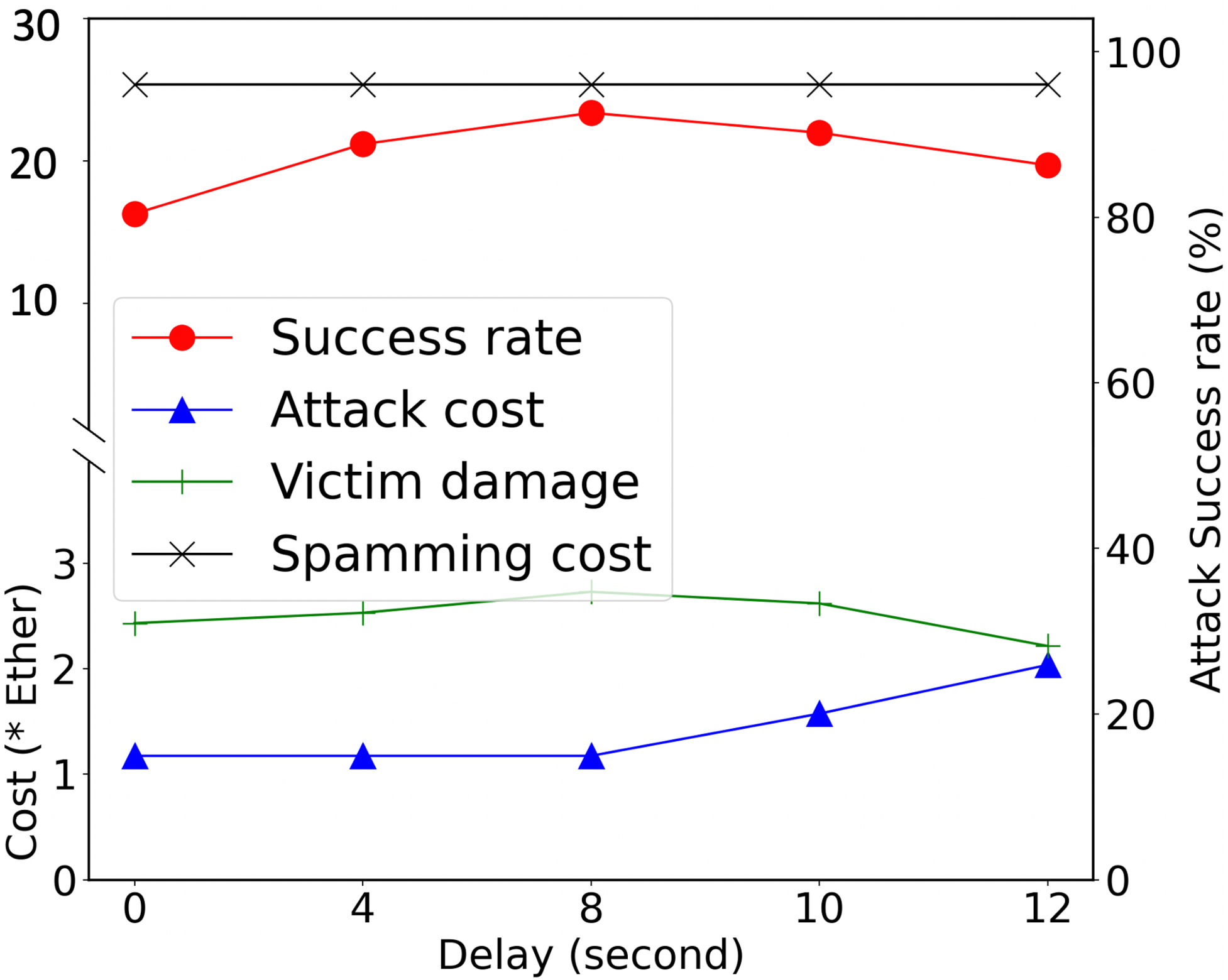}
  \label{fig:trade-off-xt4-erigon}}
  \subfloat[$XT_{8a}$ on Reth w. varying delay\\\hspace{\textwidth} (single node)] {%
   \includegraphics[width=0.24\textwidth]{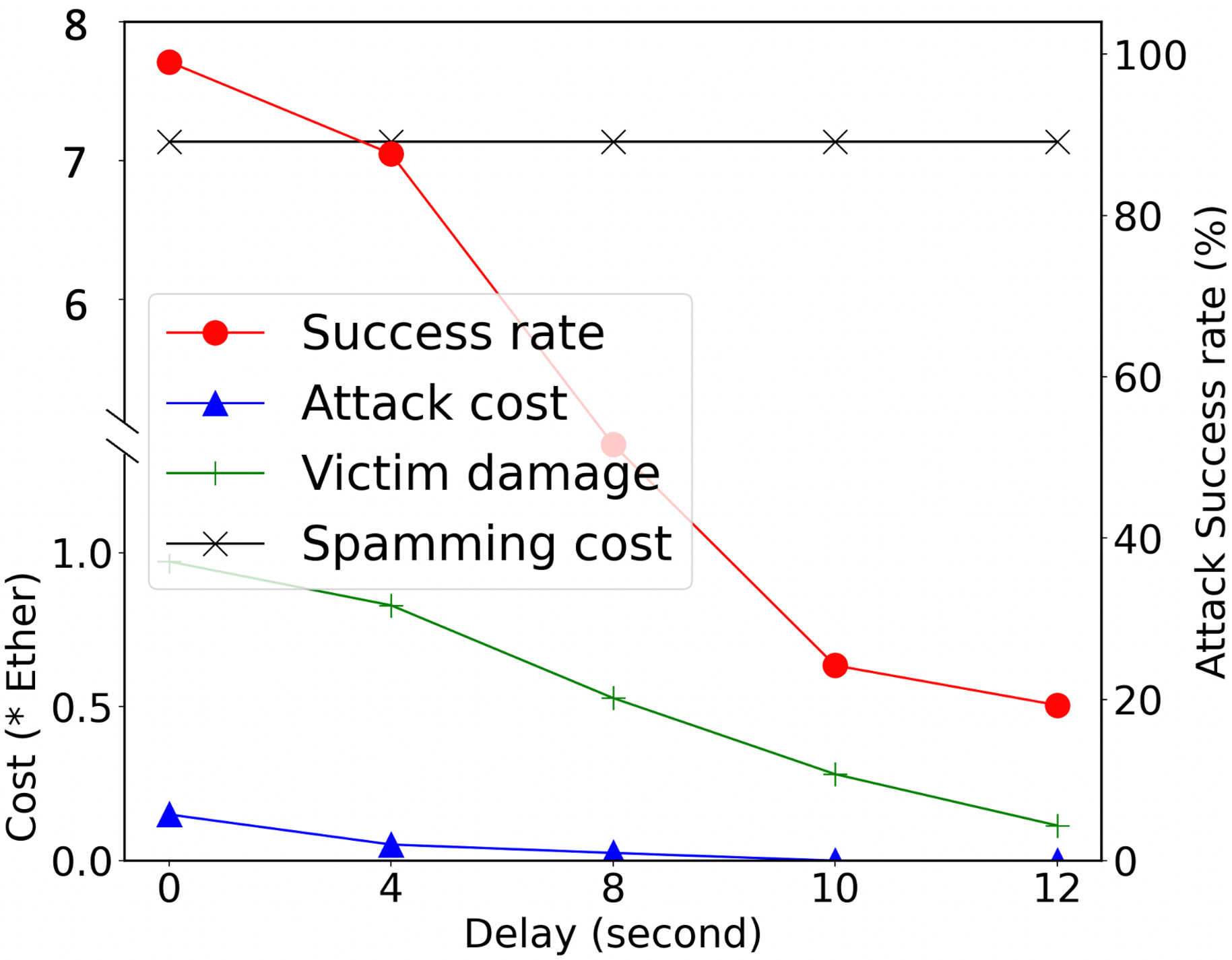}
  \label{fig:trade-off-xt8}}
  \subfloat[$XT_9$ on openEthereum w. varying \\\hspace{\textwidth}delay (single node)] {%
   \includegraphics[width=0.24\textwidth]{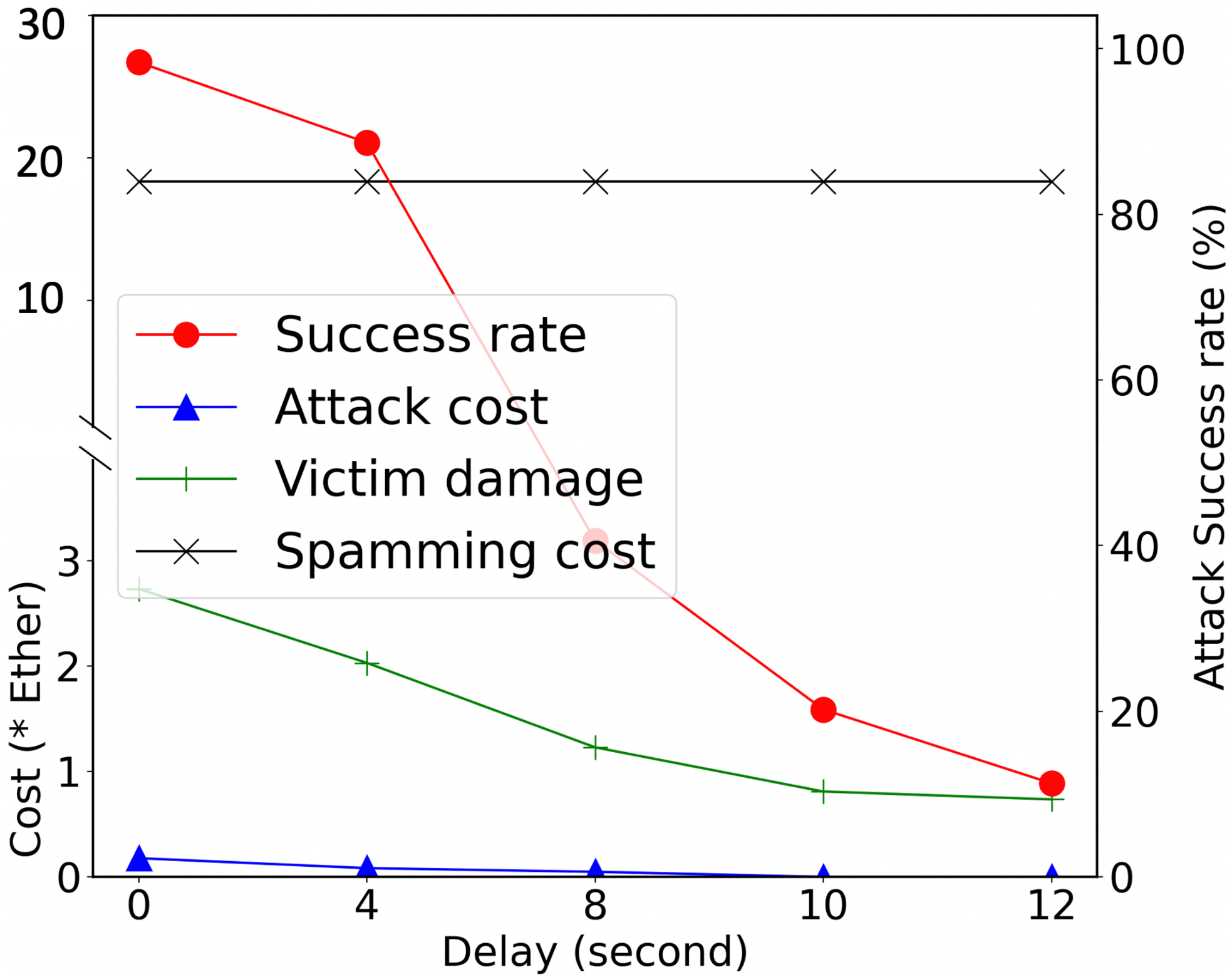}
  \label{fig:trade-off-xt9}
  
}
  \caption{Success rate and cost of turning-based eviction attack and locking attack.}%
\end{figure*}
\ignore{
Figure~\ref{fig:successrate-nethermind} reports the metrics for $XT_4$ on a victim node running Nethermind. Before the attack is launched (i.e., before the $30$-th block), the two experiment runs (attacked and no attack) produce the blocks of the same benign transaction fees, i.e., around $2-3$ Ether per block. As soon as the attack starts from the $30$-th block, the benign transaction fees quickly drop to zero Ether. The adversarial transaction fees show up and remain close to zero Ether per block. After the attack stops on $70$-th block, the benign transactions under attack gradually recover to the level of transactions under no attack. This process shows $XT_4$ with a 0-second delay has a high $100\%$ success rate, being able to evict all benign transactions in the mempool compared to the case of no attack. The $100\%$ success rate is due to that $XT_4$ is able to lock the mempool, and with $0$-second delay, the chance the transaction sequence of $XT_4$ is cut off by a block is minimal. It also shows the attack cost of $XT_4$ is extremely low, with almost zero Ether per block. The cost of spamming attack~\cite{DBLP:conf/fc/BaqerHMW16} is as high as $50$ Ether per block.

Figure~\ref{fig:successrate-geth} reports the same metrics for launching exploit $XT_6$ with $8$-second delay on a victim node running Geth-$V1.11.4$. The results are similar to those in Figure~\ref{fig:successrate-nethermind} except for the following: Instead of strictly zero-Ether fees, the adversarial-transaction fees of $XT_6$ are mostly zero Ether but also with sporadic spikes of non-zero fees under $1$ Ether. Because $XT_6$ does not lock the mempool, chances are the benign transactions sent $8$ seconds after a block is produced can be admitted into the mempool and included in the next block.
}

Figure~\ref{fig:successrate-geth} reports the metrics for $XT_6$ on a victim node running Geth $v1.11.4$. Before the attack is launched (i.e., before the $30$-th block), the two experiment runs (attacked and no attack) produce the blocks of the same benign transaction fees, i.e., around $2-3$ Ether per block. As soon as the attack starts from the $30$-th block, the benign transaction fees quickly drop to zero Ether except for some sporadic spikes (under $1$ Ether per block). The non-zero Ether cost is due to that $XT_6$ cannot lock the mempool. The adversarial transaction fees show up and remain close to zero Ether per block. After the attack stops on $70$-th block, the benign transactions under attack gradually recover to the level of transactions under no attack. 

\ignore{
Figure~\ref{fig:successrate-nethermind} reports the metrics for $XT_4$ on a victim node running Nethermind. Before the attack is launched (i.e., before the $30$-th block), the two experiment runs (attacked and no attack) produce the blocks of the same benign transaction fees, i.e., around $2-3$ Ether per block. As soon as the attack starts from the $30$-th block, the benign transaction fees quickly drop to zero Ether. The adversarial transaction fees show up and remain close to zero Ether per block. After the attack stops on $70$-th block, the benign transactions under attack gradually recover to the level of transactions under no attack. This process shows $XT_4$ with a 0-second delay has a high $100\%$ success rate, being able to evict all benign transactions in the mempool compared to the case of no attack. The $100\%$ success rate is due to that $XT_4$ is able to lock the mempool, and with $0$-second delay, the chance the transaction sequence of $XT_4$ is cut off by a block is minimal. It also shows the attack cost of $XT_4$ is extremely low, with almost zero Ether per block. The cost of a baseline spamming attack is very high, about $20$ Ether per block.
}

We also conduct experiments with varying attack delays between $0$ and $12$ seconds (note that the average time to produce a block in our collected trace is $14$ seconds). Figure~\ref{fig:delay-nethermind} shows the results of Attack $XT_4$ on Nethermind. On shorter delay than $8$ seconds, the success rates are strictly $100\%$, and the attack cost remains at $0.025$ Ether per block. When the delay grows over $8$ seconds, success rates begin to drop, and attack costs increase. At the delay of $12$ seconds, the success rate is $80\%$, and the attack cost is $1.5$ Ether per block. 
The reason is that as the delay becomes large, the transaction sequence sent in an attack can be interrupted by block production, leading to adversarial transactions included in the blockchain and normal transactions un-evicted from the mempool.

The results of Attack $XT_4$ on Geth as shown in Figure~\ref{fig:trade-off-xt4-geth} are similar to those on Nethermind. The difference is the cost of $XT_4$ on Geth is higher than that of Nethermind. In the attack sequence of $XT_4$ on Geth, it needs $384$ attack senders which means $384$ attack transactions are included in blocks on shorter delay than $8$ seconds while only $1$ attack transaction is included in $XT_4$ on Nethermind.  Figure~\ref{fig:delay-geth} shows the success rates and costs of Exploit $XT_6$ on Geth $v1.11.4$. The results are similar to those in Figure~\ref{fig:delay-nethermind} except that when the delay is short, the success rates of $XT_6$ are not $100\%$ because $XT_6$ cannot lock the attacked mempool and an attack sent too early risks admitting normal transactions sent after the attack. The results of Attack $XT_4$ on Erigon are shown in Figure~\ref{fig:trade-off-xt4-erigon}. The results are similar to those in Figure~\ref{fig:delay-geth} as the $XT_4$ on Erigon also cannot lock the attacked mempool when the delay is short. Figure~\ref{fig:trade-off-xt8} shows the success rates and cost of $XT_8$ on Reth. When the delay is $0$ second, the success rate is $99.04\%$, and the cost is $0.151$ Ether per block. The low-priced attack transactions sent with a $0$ delay occupy all the empty slots and decline all the incoming normal transactions. When the delay grows, the success rate and attack cost begin to drop. The reason is as the delay increases, more empty slots in the mempool are occupied by normal transactions. As a result, more normal transactions and less attack transactions are included in blocks. The results of Attack $XT_9$ on OpenEthereum are shown in Figure~\ref{fig:trade-off-xt9}. The results are similar to those of $XT_8$ on Reth except when the delay is short, the cost of $XT_9$ on OpenEthereum is $0.177$ per block which is higher than that of Reth. The reason is that each attack sender in the $XT_9$ attack sequence sends a high-priced child transaction for declining the incoming normal transactions and the high-priced transactions are included in blocks.

The evaluation results of all attacks on all clients are in Table~\ref{tab:attack-success-rate-cost}. On the six Ethereum clients in the public transaction path, the attack success rates are all higher than $84.63\%$, and the attack costs are all lower than $1.172$ Ether per block, which is significantly lower than the baseline.

On PBS clients and Ethereum-like clients, we also evaluate the attacks found by \textsc{mpfuzz}. Under the same experimental settings as \S~\ref{sec:setup:singlenode}, the results show similar success rates and costs with the attacks on Geth (recall these clients are Geth forks). More specifically, Table~\ref{tab:attack-success-rate-cost} shows that the success rates are higher than $92.60\%$, and attack costs are lower than $0.806$ Ether per block.

\begin{table}[!htbp]
\caption{Attack success rate and cost}
\label{tab:attack-success-rate-cost}
\centering{\footnotesize
\begin{tabularx}{0.495\textwidth}{ |X|l|X|X|X| }
\hline
Clients & Exploit& Success rate & Cost (Ether/block) & Baseline (Ether/block) \\ \hline
Geth & $XT_1$ & $99.80\%$& $0$ & $11.39$ \\  \cline{2-5}
$v1.10.25$& $XT_2$ & $99.42\%$& $0.725$ & $11.39$ \\  \cline{2-5}
& $XT_3$ & $93.02\%$& $0.0021$ & $11.39$\\  \hline
Geth & $XT_4$ & $99.42\%$& $0.806$ & $11.39$ \\  \cline{2-5}
$v1.11.4$& $XT_5$ & $92.65\%$ &$0.806$ & $11.39$ \\  \cline{2-5}
& $XT_6$ & $94.74\%$& $0.0022$ & $11.39$ \\  \hline
Erigon $v2.42.0$ & $XT_4$ & $92.53\%$& $1.172$ & $17.7$ \\  \hline
Nethermind & $XT_4$ & $99.60\%$& $0.0021$ & $10.75$ \\ \cline{2-5}
$v1.18.0$ & $XT_7$ & $84.63\%$& $0.20$ & $10.75$ \\  \hline
Besu  & $XT_2$ & $99.63\%$& $1.04$ & $17.7$\\  \cline{2-5}
$v22.7.4$ & $XT_4$ & $99.60\%$& $1.06$ & $17.7$\\  \hline
Reth $v0.1.0\-alpha.6$& $XT_4$ &$92.53\%$ &$0.672$ & $17.6$ \\  \hline
Flashbot builder $v1.11.5$& $XT_6$ & $94.74\%$& $0.0022$ & $11.39$ \\  \hline
EigenPhi & $XT_1$ & $99.80\%$& $0$ & $11.39$ \\  \cline{2-5}
builder& $XT_2$ & $99.42\%$& $0.725$ & $11.39$ \\  \cline{2-5}
 & $XT_3$ & $93.02\%$& $0.0021$ & $11.39$ \\  \cline{2-5}
& $XT_4$ & $99.42\%$& $0.806$ & $11.39$ \\  \cline{2-5}
& $XT_6$ & $94.74\%$& $0.0022$ & $11.39$ \\   \hline
bloXroute builder-ws& $XT_6$ & $94.74\%$& $0.0022$ & $11.39$ \\  \hline
go-opera & $XT_2$ & $99.12\%$& $0.201$ & $11.39$ \\  \cline{2-5}
$v1.1.3$& $XT_3$ & $92.60\%$& $0.0021$ & $11.39$ \\  \cline{2-5}
& $XT_4$ & $99.13\%$& $0.221$ & $11.39$ \\  \cline{2-5}
& $XT_6$ & $93.76\%$& $0.0022$ & $11.39$ \\  \hline
BSC $v1.3.8$& $XT_6$ & $94.74\%$& $0.0022$ & $11.39$ \\  \hline
core-geth $v1.12.18$ & $XT_6$ & $94.74\%$& $0.0022$ & $11.39$ \\  \hline
Reth $v0.1.0\-alpha.4$& $XT_8$ & $99.04\%$ &$0.151$ & $7.14$ \\  \hline
OpenEthereum & $XT_4$ &$99.56\%$ &$0.233$ & $11.39$\\  \cline{2-5}
$v3.3.5$ & $XT_9$ & $98.36\%$& $0.177$ & $18.35$\\ \hline
\end{tabularx}
}
\end{table}

\subsection{Evaluation of Found Exploits on Multi Nodes}

\label{appdx:sec:eval:local:2}
\noindent{\bf Evaluation settings}:
We extend our experiment platform to support a network of victim nodes, instead of a single victim node. Specifically, the attack node is connected to the first victim non-validator node $V_1$. Node $V_1$ is connected to the second victim non-validator node $V_2$, which is further connected to node $V_3$. The chain of victims continues until it reaches the $6$-th non-validator victim $V_6$, which is connected to the validator victim node $V_0$. The workload node is connected to the validator node $V_0$. We run the victim network among a set of geo-distributed cloud instances on the Internet. We place victim nodes $V_1, V_3, V_5$ in one geographic area in Amazon AWS cloud (a.k.a., one Availability Zone or AZ). We place the other victim nodes $V_2, V_4, V_6, V_0$ in a different geographic area, so that each message sent between $V_i$ and $V_{i+1}$ has to travel across continents on the Internet. The experiment architecture is illustrated in Figure~\ref{fig:exp:setup}.

\noindent{\bf Experiment results}:
On the multi-victim platform, we conduct similar experiments as in \S~\ref{sec:setup:singlenode}. The results are presented in Figure~\ref{fig:turning:multi}. Compared with the results on the single node, the attack cost increases faster and is much higher. For instance, when the delay is $10$/$12$ seconds, the attack cost is $8$/$8.5$ Ether per block, which is higher than the attack damage (i.e., the victim transaction fees). Under the delay, the attacks incur higher attack cost than damage and are no longer ADAMS. By contrast, from Figure~\ref{fig:delay-geth}, the attack costs on the single node with $10$/$12$-second delays are $1$/$1.4$ Ether per block, both of which are below the attack damage. The cause of the higher attack cost is the longer time to propagate the transaction sequence of an ADAMS attack in the multi-node setting than in the single node; the longer time implies the higher chance the transaction sequence is interrupted by the production of the next block and that the transactions are included to the block without being turned invalid. 
}

\section{Evaluation of \textsc{mpfuzz}}
\label{sec:eval:fuzz}
{\color{violet}
  \subsection{Ablation Study}
In this section, we describe $4$ baseline fuzzers to demonstrate the performance efficacy of the techniques proposed in \textsc{mpfuzz} as an ablation study.

\noindent{\bf Baseline B1 - stateless:} We implement a stateless fuzzer in the GoFuzz framework. 1) Given a bit string generated by GoFuzz, the fuzzer code parses it into a sequence of transactions; the length of the sequence depends on the length of the bit string. It skips a bitstring that is too short. Specifically, when parsing the bitstring into transaction sequence, every 3 bits are parsed into one transaction, corresponding to 6 adversarial symbolized transactions proposed in \textsc{mpfuzz}. 2) The fuzzer then sets up and initializes a Geth \texttt{txpool} instance with $m$ normal transactions generated from the symbolized transaction $\mathcal{N}$. 3) The fuzzer sends the transaction sequence to the initialized mempool for execution. 4) It checks the mempool state after execution: If the eviction-based test oracle (Equations~\ref{eqn:adams:evict:1}, ~\ref{eqn:adams:evict:2} and ~\ref{eqn:adams:evict:3}) is met, it emits the current transaction sequence as an exploit. 5) As a stateless fuzzer, it cannot use the same feedback technique as that of \textsc{mpfuzz}. In contrast, it takes code coverage as feedback, specifically, if this run increases the code coverage in Geth client, the fuzzer adds the current bit-string to the seeds corpus.

\noindent{\bf Baseline B2a - non-symbolized state:} Compared to \textsc{mpfuzz}, the only difference is that the baseline B2a takes concrete state in the state coverage, while \textsc{mpfuzz} takes symbolized state in the state coverage. Specifically, after sending the current transaction sequence to the mempool, it sorts the transactions in the mempool by senders and nonces. It then hashes the sorted transactions in the mempool and checks the presence of the hash digest in the seeds. If no hash exists, the current transaction sequence would increase the concrete-state coverage and be inserted into the corpus. However, the energy of a state is determined the same way as \textsc{mpfuzz}. Specifically, given the concrete state that an input reaches, the energy is determined by its symbolized state.

\noindent{\bf Baseline B2b - non-symbolized input:}  We implement the third baseline fuzzer, which is also similar to \textsc{mpfuzz} but without symbolized input mutation. In each iteration, the baseline fuzzer B2b appends to the current transaction sequence a new transaction. Given the m-slot mempool, the fuzzer tries $m$ values for senders, $m$ values for nonces, $m$ values for Gas price, and $m$ values for Ether amount. After sending to the mempool the current transaction sequence, it uses the same way to determine the feedback and energy as \textsc{mpfuzz}.

\noindent{\bf Baseline B3 - non-promising feedback \& energy} We build the last baseline fuzzer, which is the same with \textsc{mpfuzz} except that it removes the state promising-ness from its feedback and energy mechanism proposed in \textsc{mpfuzz} from the seed selection. Specifically, the feedback formula in B3 includes only state coverage ($st\_coverage$) and excludes state promisingness ($st\_promising$ as in Equation~\ref{eqn:feedback}). In addition, B3 uses as the energy the number of invalid transactions on a mempool state. The seed with more invalid transactions in the mempool has higher energy or priority to be selected for the next round of fuzzing.

\label{sec:mut:setting}
\noindent{\bf Experimental settings of fuzzing}: 
We run the four baselines and \textsc{mpfuzz} against a MUT in a Geth client reconfigured under two settings: A small setting where the mempool is resized to $6$ slots $m=6$ and all fuzzers run for two hours (if test oracle is not triggered), and a medium setting where $m=16$ and fuzzers run for $16$ hours. The fuzzing experiments are run on a local machine with an Intel i7-7700k CPU of $4$ cores and $64$ GB RAM.

\noindent{\bf
Results}: Table~\ref{appdx:tab:vs:baselines2} reports the time used that the first exploit is found under the small and medium MUT settings. For the small MUT of $6$ slots, baseline B1 cannot find any attack in two hours, B2a, B2b and B3 can find the Exploit $XT_3$ in $0.10$, $34.61$ and $1.66$ minutes respectively. By contrast, \textsc{mpfuzz} finds the same exploit ($XT_3$) in $0.03$ minutes.

For the medium setting, with $16$ slots,  baselines B1, B2b, and B3 cannot find any exploit in $16$ hours, while B2a can find Exploit $XT_3$ in $0.26$ minutes. By contrast, \textsc{mpfuzz} finds the same exploit in $0.06$ minutes.

The result that B1 has the poorest performance shows a stateless fuzzer would be ineffective in discovering ADAMS exploits. As B2a and B3 are more performant compared to B2b, it shows the symbolized input technique improves efficiency more compared to that of the promising feedback and symbolized state coverage. The comparison of results between B2a and B3 indicates that the state promisingness in feedback is more beneficial for performance than symbolized state coverage.

\begin{table}[!htbp] 
\caption{Fuzzing Geth $v1.11.3$'s mempool (in minutes) by different approaches to detect Exploit $XT_3$. OT means overtime.}
\label{appdx:tab:vs:baselines2}
\centering{\footnotesize
\begin{tabularx}{0.40\textwidth}{ |X|c|c|c|c|c| }
  \hline
Settings & B1 & B2a  & B2b & B3  &\textsc{mpfuzz} \\ \hline
$6$slot-$2$h &  OT & $0.10$  &$34.61$ &  $1.66$ &$0.03$ 
\\ \hline
$16$slot-$16$h & OT & $0.26$  & OT &OT&  $0.06$ 
\\ \hline
\end{tabularx}
}
\end{table}
}

{\color{violet}
The other experiments, including true/false positive rates and additional performance evaluation evaluation on different clients can be found in Section 7.2 and
Appendix C, available in the online supplemental material, and are identical to the original conference paper~\cite{wang2024understanding}.
}

{\color{violet}

\section{Case Study: How \textsc{mpfuzz} Finds Exploits in MUT and Extends to a Large Mempool}

We describe how \textsc{mpfuzz} finds an exploit that evades the defense. We present a case study on finding $XT_6$ in the latest Geth $\geq{}v1.11.4$. Recall that a Geth mempool has a capacity of $m'=6144$ slots, and its transaction-admission policies are characterized by three essential parameters: admitting up to $py_1'=1024$ future transactions and limiting up to $py_2'=16$ pending transactions from any senders when more than $py_3'=5120$ pending transactions are residing in the mempool. For ease of description, we set up the MUT to run the same codebase or the same admission policy but with different, smaller parameters: $m=4, py_1=1, py_2=2, py_3=2$. We first describe how \textsc{mpfuzz} finds a short exploit of $XT_6$ on the MUT. We then describe how \textsc{mpfuzz} extends the short exploit into a longer version suitable for a medium-sized mempool.

\noindent{\bf 
Fuzzing: How \textsc{mpfuzz} automatically finds exploits}:
Initially, the mempool is filled with $m=4$ normal transactions. That is, the initial symbolized state is $\mathcal{NNNN}$. The seed corpus $sdb$ initially contains an empty string. To simplify the description, we only show Symbols $\mathcal{C}$ and $\mathcal{P}$ in the input mutation for ease of illustration. \textsc{mpfuzz} retrieves the empty string and appends to it with different symbols $\mathcal{C}_0$ and $\mathcal{P}_0$. Because of the initial state $\mathcal{NNNN}$, only Symbol $\mathcal{P}_0$ is feasible. It generates the mutated input $\mathcal{P}$ and instantiates it to a parent transaction of a higher fee than normal transactions (as described in \S~\ref{sec:symbolize}). Sending the input to the mempool gets the transaction admitted, leading to transitioned state $st_1=\mathcal{NNNP}$. This is a new state that is not in the corpus, and it evicts more normal transactions $\mathcal{N}$ than the previous state $st_0$; the input produces positive feedback, and the associated input-state pair $\langle{}P,st_1=\mathcal{NNNP}\rangle{}$ is added to the corpus.

Next, \textsc{mpfuzz} retrieves from $sdb$ a seed by high energy. According to Table~\ref{tab:symbols:2}, the energy of state $\mathcal{NNNN}$ is $\frac{1}{3*4}*0$, and the energy of state $\mathcal{NNNP}$ is $\frac{1}{3*3+4}*1=1/13>0$. Hence, seed $\mathcal{NNNP}$ is selected. \textsc{mpfuzz} tries input mutation and appends to the selected input $\mathcal{P}$ one of three new symbols, that is, $\mathcal{C}_1$, $\mathcal{P}_0$ or $\mathcal{P}_1$. On state $\mathcal{NNNP}$, 1) Mutation transaction from $\mathcal{C}_1$ is admitted, transitioning the state to $\mathcal{NNPC}$, which produces positive feedback and is added to $sdb$. 2) Likewise, transaction $\mathcal{P}_0$ is admitted and produces state $\mathcal{NNPP}$ of positive feedback; the mutated input is also added to $sdb$. 3) Mutation $\mathcal{P}_1$ is admitted but produces an identical state with mutation $\mathcal{P}_0$; thus, the mutated input is not added to $sdb$.

\begin{wrapfigure}{r}{0.23\textwidth}
\centering
\includegraphics[width=0.23\textwidth]{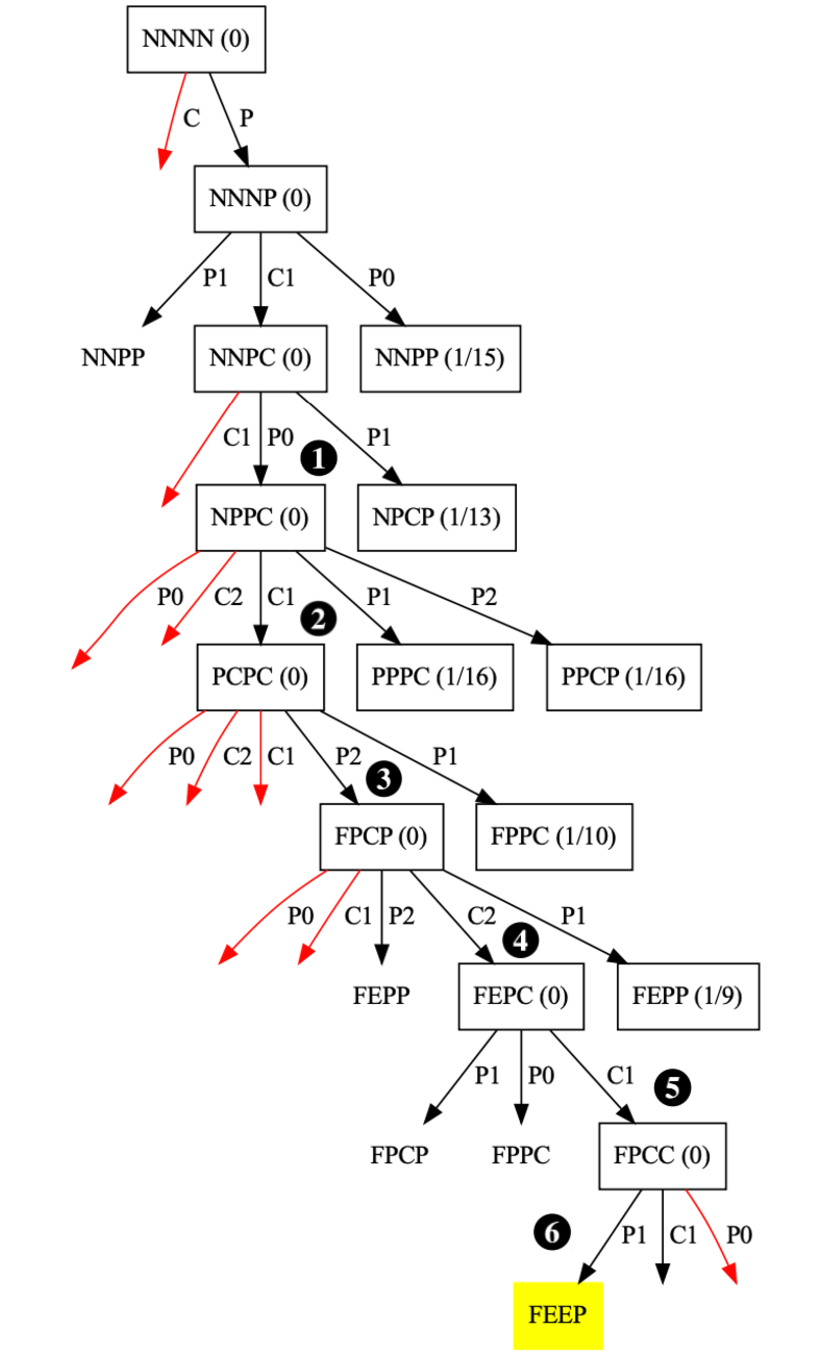}
\caption{Snapshot of the \textsc{mpfuzz} state-search tree when finding Exploit $XT_6$ on Geth $v1.11.4$.}
\label{fig:4slots-xt6}
\end{wrapfigure}

Now, there are four seeds in $sdb$: $\mathcal{NNNN} (0)$, $\mathcal{NNNP} (0)$, $\mathcal{NNPC} (1/11)$, and $\mathcal{NNPP} (1/15)$. In parentheses are their energy numbers. The seed of the highest energy $\mathcal{NNPC} (1/11)$ is selected (i.e., Step \ballnumber{1} in Figure~\ref{fig:4slots-xt6}). \textsc{mpfuzz} then mutates $\mathcal{NNPC}$ with three possibilities, that is, $\mathcal{C}_1$, $\mathcal{P}_0$, and $\mathcal{P}_1$, which produce two end states with positive-feedback, that is, $\mathcal{NPPC}$ and $\mathcal{NPCP}$. They are added to the $sdb$ with energy $\mathcal{NPPC} (1/13)$ and $\mathcal{NPCP} (1/13)$.

Let's say $\mathcal{NPPC} (1/13)$ is selected (\ballnumber{2}). \textsc{mpfuzz} then mutates input $\mathcal{P}\mathcal{C}_1\mathcal{P}_0$ with five mutation transactions, that is, $\mathcal{C}_1, \mathcal{C}_2, \mathcal{P}_0, \mathcal{P}_1, \mathcal{P}_2$, which produces three end state with positive feedback, that is, $\mathcal{PCPC} (1/11)$, $\mathcal{PPPC} (1/16)$ and $\mathcal{PPCP} (1/16)$. After that, $\mathcal{PCPC}$ is chosen for the next-round fuzzing (\ballnumber{3}). 
Mutation $\mathcal{P}_2$ is admitted and leads to state $\mathcal{FPCP}$ with positive feedback. Similarly, upon state $\mathcal{FPCP}$ (\ballnumber{4}), mutation $\mathcal{C}_2$ is admitted and leads to state $\mathcal{FEPC} (\ballnumber{5})$. Then, $\mathcal{C}_1$ transits the state to $\mathcal{FPCC}$ (\ballnumber{6}). Upon state $\mathcal{FPCC}$, mutation $\mathcal{P}_1$ is admitted and leads to state $\mathcal{FEEP}$ which 
satisfies the bug oracle of eviction attacks under $\epsilon=0.34$. The algorithm then emits the found short exploit: $\langle{}st_0=\mathcal{NNNN}, dc_0=\emptyset\rangle{}, ops=\mathcal{P}\mathcal{C}_1\mathcal{P}_0\mathcal{C}_1\mathcal{P}_2\mathcal{C}_2\mathcal{C}_1\mathcal{P}_1$ (recall Definition~\ref{def:timeline}). The snapshot of the state-search tree is depicted in Figure~\ref{fig:4slots-xt6}.

\noindent{\bf Exploit extension}: Given the short exploit automatically found on MUT (with $m=4, py_1=1, py_2=2, py_3=2$), the next step is to extend it to a longer exploit functional on larger mempool. We present a case study on extending the short exploit ($XT_6$ on MUT) to a longer exploit on a medium-size mempool (with $m=8, py_1=2, py_2=3, py_3=6$). we extend the small exploit to the longer exploit by ensuring the same admission event occurs on the actual mempool as on the smaller MUT. To do so, we first extract all the unique admission event patterns $\langle{}st_0', tx_i, st_n'\rangle{}$ in the small exploit and save them into an admission event pattern set ($aep$). Recall the admission event definition in\S~\ref{sec:extension}. In the $XT_6$ exploit found by \textsc{mpfuzz}, we extract $6$ patterns, including $\langle{}\mathcal{N}, \mathcal{P}, \mathcal{P}\rangle{}$ from \ballnumber{1}, $\langle{}\mathcal{N}, \mathcal{C}, \mathcal{C}\rangle{}$ from \ballnumber{2}, $\langle{}\mathcal{PC}, \mathcal{P'}, \mathcal{P'F}\rangle{}$ from \ballnumber{3}, $\langle{}\mathcal{PC}, \mathcal{C'}, \mathcal{C'E}\rangle{}$ from \ballnumber{4}, $\langle{}\mathcal{EP}, \mathcal{C}, \mathcal{PC}\rangle{}$ from \ballnumber{5}, and $\langle{}\mathcal{PCC}, \mathcal{P'}, \mathcal{P'EE}\rangle{}$ from \ballnumber{6}.

\begin{figure}[!ht]
  \centering
  \includegraphics[width=0.375\textwidth]{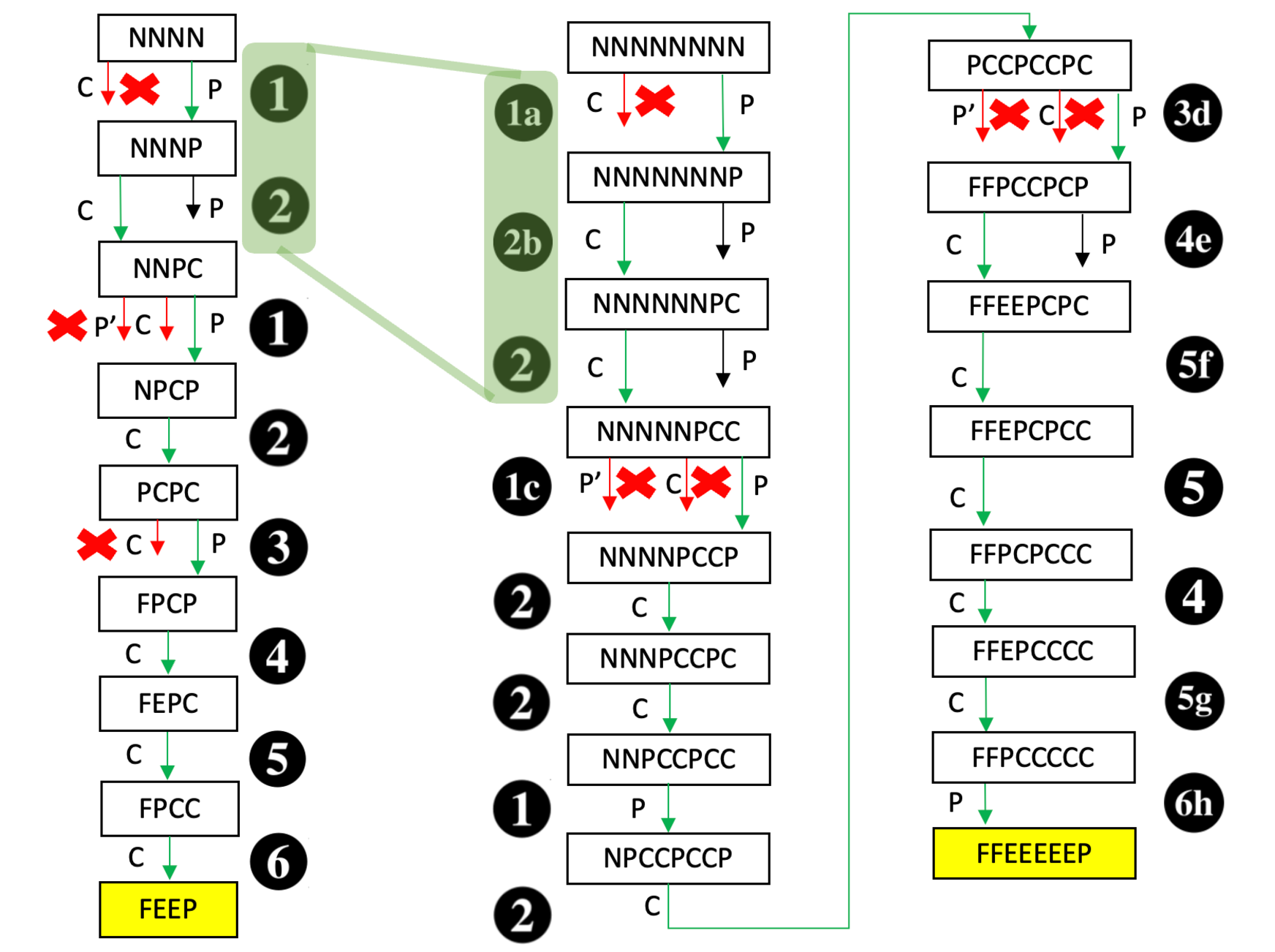}
  \caption{Exploit $XT_6$ extension on Geth $v1.11.4$ with a medium-size mempool.}
  \label{fig:extension}
  \end{figure}

We then construct the attack transaction sequence on the larger-size mempool by aligning the patterns identified from the small exploit with corresponding admission events. Specifically, this entails verifying that the admission event pattern of a given event $f(st_0,tx_i)\Rightarrow{}st_n$ on the larger-size mempool is present within the $aep$. Initially, the mempool is filled with $m=8$ normal transactions and the symbolized state is $\mathcal{NNNNNNNN}$. There are two patterns in $aep$ whose initial state $st_0'$ matches the mempool state, that is $\langle{}\mathcal{N}, \mathcal{P}, \mathcal{P}\rangle{}$ and $\langle{}\mathcal{N}, \mathcal{C}, \mathcal{C}\rangle{}$ (\ballnumber{1a}). Between these two patterns, only sending transaction $\mathcal{P}$ is feasible. Therefore, \textsc{mpfuzz} sends $\mathcal{P}$ upon the initial state which leads to transitioned state $\mathcal{NNNNNNNP}$. In this admission event, the submission of transaction $\mathcal{P}$ results in its admission in the mempool by evicting a $\mathcal{N}$ in the initial state. Thus, it aligns the pattern of $\langle{}\mathcal{N}, \mathcal{P}, \mathcal{P}\rangle{}$. This positive outcome prompts \textsc{mpfuzz} to proceed with the extension process accordingly. 

Given the mempool state $\mathcal{NNNNNNNP}$, there are two patterns in $aep$ whose initial state $st_0'$ matches the mempool state, that is $\langle{}\mathcal{N}, \mathcal{P}, \mathcal{P}\rangle{}$ and $\langle{}\mathcal{N}, \mathcal{C}, \mathcal{C}\rangle{}$ (\ballnumber{2b}). Both of the $tx_i$ in these two patterns are feasible. In this case, \textsc{mpfuzz} applies the pattern that decreases the most $opcost()$ first. Recall the specification of the $opcost()$ in Table~\ref{tab:symbols:2}. Thus, \textsc{mpfuzz} applies the pattern $\langle{}\mathcal{N}, \mathcal{C}, \mathcal{C}\rangle{}$ that has the smaller $opcost()$ of the end state. Sending a $\mathcal{C}$ leads to the admission of $\mathcal{C}$ by evicting a $\mathcal{N}$, which aligns the end state of the pattern. Similarly, \textsc{mpfuzz} extends another $\mathcal{C}$ in the input transaction sequence and transitions the mempool state to $\mathcal{NNNNNPCC}$. At this time, there are $4$ patterns whose initial state $st_0'$ matches the mempool state and is also feasible. In the $opcost()$ decreasing order, \textsc{mpfuzz} first try to apply pattern $\langle{}\mathcal{PCC}, \mathcal{P'}, \mathcal{P'EE}\rangle{}$ by sending a $\mathcal{P}$. However, the mempool state led by admitting $\mathcal{P}$ does not match the end state $st_n'$ of the pattern. This negative outcome results \textsc{mpfuzz} backtrack by rollback the mempool state before admitting $\mathcal{P}$, which is $\mathcal{NNNNNPCC}$. Similarly, the second pattern $\langle{}\mathcal{N}, \mathcal{C}, \mathcal{C}\rangle{}$ fails to transition the mempool state to correspond with the end state of the pattern. Consequently, \textsc{mpfuzz} applies the next pattern $\langle{}\mathcal{N}, \mathcal{P}, \mathcal{P}\rangle{}$ transitions the mempool state to $\mathcal{NNNNPCCP}$ (\ballnumber{1c}).

Continuing the aforementioned process, \textsc{mpfuzz} leads the mempool state to $\mathcal{PCCPCCPC}$ (\ballnumber{3d}). Among the three patterns whose initial state matches the mempool state, only applying $\langle{}\mathcal{PC}, \mathcal{P'}, \mathcal{P'F}\rangle{}$ can get a corresponding end state in the state transition. That is, sending a $\mathcal{P}$ results to the mempool state as $\mathcal{FFPCCPCP}$. Subsequently, \textsc{mpfuzz} selects the pattern $\langle{}\mathcal{PC}, \mathcal{C'}$, due to its lower $opcost()$ (\ballnumber{4e}) and then applies the only pattern that aligns the mempool state transition, which is $\langle{}\mathcal{EP}, \mathcal{C}, \mathcal{PC}\rangle{}$, resulting in the mempool state of $\mathcal{FFEPCPCC}$ (\ballnumber{5f}). Similar to the aforementioned process, \textsc{mpfuzz} leads the mempool state to $FFPCCCCC$ (\ballnumber{5g}). Upon that, \textsc{mpfuzz} applies the only pattern that aligns the mempool state transition, which is $\langle{}\mathcal{PCC}, \mathcal{P'}, \mathcal{P'EE}\rangle{}$, and transitions the mempool state to $FFEEEEEP$ (\ballnumber{6h}). The mempool state satisfies the bug oracle of eviction attacks within the small mempool of MUT, satisfying the condition that $\epsilon \leq 0.34$. The algorithm then emits the extend exploit against the medium-size mempool, which is $ops = PCCPCCPCPCCCP$.

}

\section{Discussions}

\noindent{\bf Responsible bug disclosure}: 
We have disclosed the discovered ADAMS vulnerabilities to the Ethereum Foundation, which oversees the bug bounty program across major Ethereum clients, including Geth, Besu, Nethermind, and Erigon. 
\tangSide{R5}
{\color{blue}
We also reported the found bugs to the developers of Reth, Flashbot, EigenPhi, bloXroute, BSC, Ethereum Classic and Fantom. Besides $7$ DETER bugs that \textsc{mpfuzz} rediscovered, $24$ newly discovered ADAMS bugs are reported. As of March 2024, $15$ bugs are confirmed: $XT_1$ (Nethermind, EigenPhi builder), $XT_2$ (EigenPhi builder), $XT_3$ (EigenPhi builder), $XT_4$ (Geth, Erigon, Nethermind, Besu, EigenPhi builder), $XT_5$ (Geth), $XT_6$ (Geth, Flashbot Builder, EigenPhi builder), $XT_7$ (Nethermind), and $XT_8$ (Reth).} 
After our reporting, $XT_4$/$XT_7$/$XT_8$ have been fixed on Geth $v1.11.4$/Nethermind $v1.21.0$/Reth $v0.1.0-alpha.6$. Bug reporting is documented~\cite{me:mpfuzz:report}. 

\noindent{\bf Ethical concerns}: 
When evaluating attacks, we only mounted attacks on the testnet and did so with minimal impacts on the tested network. For instance, our attack lasts a short period of time, say no more than $4$ blocks produced. We did not test our attack on the Ethereum mainnet. Additionally, we obscure the first few digits of the block number in the attack screenshot, i.e., Figure~\ref{fig:testnet:rinkeby_ed4}.
When reporting bugs, we disclose to the developers the mitigation design tradeoff and the risk of fixing one attacks by enabling other attacks (described above). To prevent introducing new bugs, we did not suggest fixes against turning-based locking ($XT_5$, $XT_6$) and locking ($XT_8$).

  \section{Conclusion and Future Works}
 
\noindent{\bf Conclusion}: This paper presents \textsc{mpfuzz}, the first mempool fuzzer to find asymmetric DoS bugs by exploring symbolized mempool states and optimistically estimating the promisingness of an intermediate state in reaching bug oracles. Running \textsc{mpfuzz} on popular Ethereum clients discovers new mempool-DoS bugs, which exhibit various sophisticated patterns, including stealthy mempool eviction and mempool locking. 

{\tangSide{R2, R5, R7} \color{blue} 
\noindent{\bf Limitations and future works}: The exploit generation in this work is not fully automated. Manual tasks include mempool reduction, \textsc{mpfuzz} setup (configuring $\epsilon$ and symbols), etc. Automating these tasks is the future work.

\textsc{mpfuzz}'s bug oracles neither captures complete dependencies among concrete transactions nor guarantees completeness in finding vulnerabilities. Certified mempool security with completeness is also an open problem.

\textsc{mpfuzz} targets a victim mempool of limited size: the vulnerabilities found in this work are related to transaction eviction, which does not occur in a mempool of infinite capacity. Thus, the \textsc{mpfuzz} workflow cannot find exploits in the mempool of infinite capacity. It is an open problem whether a mempool of large or infinite capacity has DoS bugs.

\section{Acknowledgments}
The authors thank the anonymous reviewers in USENIX security’24. The authors appreciate the discussion with Kangjie Lu in the early stage of this work.  All authors but Kai Li are partially supported by two Ethereum Foundation academic grants and NSF grants CNS-2139801, CNS-1815814, DGE2104532. Kai Li is supported by NSF grant CNS2347486 and one Ethereum Academic Grant.
}

\ignore{
{{  \section{Discussion}
At present, the transaction symbolization of \textsc{mpfuzz} is designed by a heuristic method. Through generating one concrete transaction in each symbol, \textsc{mpfuzz} covers all the input space of transactions and the state of mempool. By prioritizing the seed with the energy designed by the heuristic method, \textsc{mpfuzz} triggers the bug oracle efficiently. A more practical method is to automatically extract the symbolized transactions by statically analyzing the control flow code path. In the control flow graph of the mempool source code, each unique code path is triggered by a group of concrete transactions. The key idea is to map each group of concrete transactions triggering the same code path into a distinct symbol. It is sufficient to cover all the code paths in mempool source code. However, whether it covers all the state of mempool needs to be studied in future. In addition, how to generate the concrete transaction by analyzing the code path as well as assigning energy score needs to be explored in the future.
}}

}


}



%

\bibliographystyle{IEEEtran}
\bibliography{bkc,fuzz,txtbk,yuzhetang,bug}


\vfill

\end{document}